\documentclass[11pt,letter]{article}
\pdfoutput=1 
\usepackage{jheppub}

\usepackage[dvipsnames]{xcolor}

\usepackage{cleveref}
\usepackage{amsmath}
\usepackage{mathtools}
\usepackage{amssymb}
\usepackage{graphicx}
\usepackage{color}
\usepackage{colordvi}
\usepackage{colortbl}
\usepackage{hyperref}
\usepackage{slashed}
\usepackage{subcaption}
\usepackage{physics}
\usepackage{upgreek}
\usepackage{cases}
\usepackage{mathtools,tabu}
\usepackage{slashed}
\usepackage{stackengine}
\stackMath
\usepackage{graphicx}

\hypersetup{hyperfootnotes=false}

\allowdisplaybreaks

\usepackage{tikz}
\usetikzlibrary{positioning, shapes.geometric, arrows.meta,calc,backgrounds}

\newcommand{\be}{\begin{equation}}
\newcommand{\ee}{\end{equation}}
\newcommand{\ba}{\begin{array}}
\newcommand{\ea}{\end{array}}
\newcommand{\balg}{\begin{align}}
\newcommand{\ealg}{\end{align}}
\newcommand{\bit}{\begin{itemize}}
\newcommand{\eit}{\end{itemize}}
\newcommand{\trm}[1]{\textrm{#1}}

\newcommand{\Mpc}{\trm{\Mpc}}
\newcommand{\yr}{\trm{\yr}}
\newcommand{\eV}{\trm{\eV}}

\newcommand{\nn}{\nonumber}

\newcommand{\boldit}[1]{ {\mathbf #1}}

\newcommand{\cs}[2]{ {\boldit{c}^{(#1#2)}_s}}
\newcommand{\css}[2]{ {\boldit{c}^{(#1#2,ss)}_s}}
\newcommand{\cnlsm}[2]{ {\boldit{c}^{(#1#2,d)}_s}}

\newcommand{\ap}{{\alpha'}}

\def\be{\begin{equation}}
\def\ee{\end{equation}}
\newcommand\bea{\begin{align}}
\newcommand\eea{\end{align}}
\def\nn{\nonumber}
\def\eqn#1{Equation~(\ref{#1})}

\def\tab#1{Tab.~{\ref{#1}}}

\def\midrule{\hline}
\def\bottomrule{\hline}
\def\toprule{\hline}

\definecolor{NUpurple}{RGB}{078,042,132}

\definecolor{SZblue}{RGB}{38,100,153}

\colorlet{alex}{blue!50!green}
\colorlet{ref-link}{green!70!black!70!blue}
\colorlet{cite-link}{red!60!blue}
\colorlet{fc}{green!50!black}

\definecolor{ucp-color}{RGB}{0,100,144}
\definecolor{cut-color}{RGB}{203, 4, 31}
\definecolor{hgrey0}{RGB}{175,175,175}
\colorlet{blobcolor}{CadetBlue}

\colorlet{n1-arrow}{ProcessBlue}
\colorlet{n2-arrow}{Plum}
\colorlet{n3-arrow}{orange}

\hypersetup{
  citecolor = cite-link,
  linkcolor = ref-link,
}

\newcommand{\pcub}{(s t u)}
\newcommand{\psq}{(s^2 + t^2 + u^2)}

\def\sect#1{Section~\ref{#1}}

\def\tab#1{Tab.~{\ref{#1}}}

\newcommand{\compAdj}[2]{\left(#1\,\textcircled{$\mathfrak{a}$}\,#2\right)}

\usepackage[normalem]{ulem}

\definecolor{hgreen}{rgb}{0,0.545,0}
\definecolor{hblue}{rgb}{0,0,0.545}
\definecolor{hred}{rgb}{0.475,0.0,0.15}

\def\midrule{\hline}
\def\bottomrule{\hline}
\def\toprule{\hline}

\definecolor{nhpRed}{RGB}{161,0,0}
\definecolor{nhp4}{RGB}{203, 4, 31}
\definecolor{nhp3}{RGB}{244,99,30}
\definecolor{nhp2}{RGB}{255,159,0}
\definecolor{nhp1}{RGB}{48,152,152}
\definecolor{nhpBlue}{RGB}{0,100,144}
\definecolor{cutred}{RGB}{219,56,49}
\definecolor{hgreen}{RGB}{25,176,146}
\definecolor{hgreen1}{RGB}{175,230,175}
\definecolor{hblue}{RGB}{52,152,219}
\definecolor{hbluedark}{RGB}{36, 106, 160}
\definecolor{hblue1}{RGB}{255,255,166}
\definecolor{hred}{RGB}{139,1,0}
\definecolor{hreddark}{RGB}{151, 58, 81}
\definecolor{hred1}{RGB}{255,155,155}
\definecolor{cutred}{RGB}{219,56,49}
\definecolor{hgrey4}{RGB}{75,75,75}
\definecolor{hgrey5}{RGB}{50,50,50}
\definecolor{hgrey3}{RGB}{100,100,100}
\definecolor{hgrey}{RGB}{125,125,125}
\definecolor{hgrey2}{RGB}{125,125,125}
\definecolor{hgrey1}{RGB}{150,150,150}
\definecolor{hgrey0}{RGB}{175,175,175}
\definecolor{hgreyLight}{RGB}{242,242,242}
\definecolor{darkgreen}{RGB}{59,126,108}
\definecolor{jaxoblue}{HTML}{0086FF}

\newcommand\oop[1]{ \hat{#1} }

\usepackage{mathtools} 
\newcommand{\stringKLT}[1][]{\mathop{\otimes}\limits^{\alpha'}\!{}^{#1}}
\newcommand\justD[1]{ \mathcal{D}_{#1}}
\newcommand\mD[1]{ \int \mathcal{D}_{#1}}

\newcommand\Lag[0]{\mathcal{L}}

\newcommand{\prodList}[1]{{#1}_{\times}}
\newcommand{\ce}{\varepsilon}

\newcommand{\fieldsBold}[4]{{\boldsymbol{#1}^{\boldsymbol{#2}}_{\boldsymbol{#3}}\boldsymbol(\boldsymbol{#4}\boldsymbol)}}

\usetikzlibrary{
    decorations.pathmorphing, 
    decorations.markings,    
    positioning,
    arrows.meta,
    bending,intersections,
    shapes.misc,
    calc}

\tikzset{
    step/.style={rectangle, draw=blue!40, thick, rounded corners, fill=blue!5, text width=4.2cm, align=center, minimum height=1.4cm},
    feyn vertex/.style={inner sep=0pt, outer sep=0pt, minimum size=0pt},
    feyn empty vertex/.style={inner sep=1pt, minimum size=2pt, fill=white, circle, draw},
    feyn cross vertex/.style={cross out, draw, minimum size=4pt, inner sep=0pt},
    feyn propagator/.style={thick, draw=black},
    dotted_ellipsis_line/.style={ 
        feyn propagator, 
        dotted
    },
    ellipsis_points_label/.style args={[#1]#2}{
    postaction={
        decoration={
            markings,
            mark=at position 0.5 with { 
                \node[font=\Large, black, #1] {#2}; 
            }
        },
        decorate
    }
   },
    fermion/.style={
        feyn propagator,
        decoration={markings, mark=at position 0.55 with {\arrow{Latex[length=1.5mm, width=1.2mm]}}},
        postaction={decorate}
    },
    antifermion/.style={
        feyn propagator,
        decoration={markings, mark=at position 0.55 with {\arrowreversed{Latex[length=1.5mm, width=1.2mm]}}},
        postaction={decorate}
    },
    photon/.style={
        feyn propagator,
        decorate,
        decoration={snake, segment length=4pt, amplitude=1pt, pre length=1pt, post length=1pt}
    },
    gluon/.style={
        feyn propagator,
        decorate,
        decoration={coil, segment length=4pt, amplitude=2.5pt, aspect=0.4, pre length=1pt, post length=1pt}
    },
    scalar/.style={
        feyn propagator,
        dashed,
        dash pattern=on 3pt off 2pt
    },
    ghost/.style={
        feyn propagator,
        dotted,
        decoration={markings, mark=at position 0.55 with {\arrow{Latex[length=1.5mm, width=1.2mm]}}},
        postaction={decorate}
    },
    graviton/.style={
        feyn propagator,
        decorate, 
        decoration={snake, amplitude=.4mm, segment length=1.5mm, pre length=.5mm, post length=.5mm}, 
        double},
    graviton simple/.style={
        feyn propagator,
        double distance=1.2pt,
    },
    graviton wavy/.style={
        feyn propagator,
        decorate,
        decoration={
         pre length=.5mm, post length=.5mm,
            bumps,         
            segment length=1.5mm, 
            amplitude=.4mm    
        },
        double
    },
    momentum/.style={sloped, midway, font=\scriptsize, auto=false},
    momentum above/.style={momentum, above},
    momentum below/.style={momentum, below},
    momentum left/.style={momentum, left},
    momentum right/.style={momentum, right},
    feyn loop/.style={feyn propagator, loop, min distance=10mm, looseness=10},
    crossed dot/.style={
        cross out,
        draw,
        minimum size=4pt,
        inner sep=0pt,
        outer sep=0pt
    },
    dot/.style={
        circle,
        fill,
        minimum size=4pt,
        inner sep=0pt,
        outer sep=0pt
    },
    empty dot/.style={
        circle,
        draw,
        minimum size=4pt,
        inner sep=0pt,
        outer sep=0pt
    },
    blob/.style={
        circle,
        fill=hgreen,
        minimum size=12pt,
        inner sep=0pt,
        outer sep=0pt,
        draw=hgreen,
        text=white
    },
    mygluon/.style={
        feyn propagator,
        decorate,
        decoration={
            coil,
            segment length=6pt,
            amplitude=3.5pt,     
            aspect=.8,
            pre length=.5pt,
            post length=.5pt,
            mirror              
        }
    },
    momentumlabel/.style args={[#1]#2}{
        postaction={
            decoration={
                markings,
                mark=at position 0.55 with {
                    \node[momentum, #1] {#2};
                }
            },
            decorate 
        }
    }
}

\makeatletter
\def\convertto#1#2{\strip@pt\dimexpr #2*65536/\number\dimexpr 1#1}
\makeatother

\newcommand{\pgfextractangle}[3]{
  \pgfmathanglebetweenpoints{\pgfpointanchor{#2}{center}}
  {\pgfpointanchor{#3}{center}}
  \global\let#1\pgfmathresult  
}

\pgfmathsetmacro{\extL}{0.7}

\definecolor{cutred}{RGB}{219,56,49}
\definecolor{hgreen}{RGB}{25,176,146}
\definecolor{hgreen1}{RGB}{175,230,175}
\definecolor{hblue}{RGB}{52,152,219}
\definecolor{hblue1}{RGB}{255,255,166}
\definecolor{hred}{RGB}{216,83,117}
\definecolor{hred1}{RGB}{255,155,155}
\definecolor{cutred}{RGB}{219,56,49}
\definecolor{hgrey4}{RGB}{75,75,75}
\definecolor{hgrey5}{RGB}{50,50,50}
\definecolor{hgrey3}{RGB}{100,100,100}
\definecolor{hgrey}{RGB}{125,125,125}
\definecolor{hgrey2}{RGB}{125,125,125}
\definecolor{hgrey1}{RGB}{150,150,150}
\definecolor{hgrey0}{RGB}{175,175,175}
\definecolor{darkgreen}{RGB}{59,126,108}

\newenvironment{myfeynmandiagram}[1][]{%
    \begin{tikzpicture}[#1]%
}{%
    \end{tikzpicture}%
}

\newcommand{\myvertex}[4][]{
    \node[#1] #2 #3 {#4};
}

\newcommand{\mypropagator}[3][]{
    \draw[#1] #2 #3;
}

\newcommand{\contactFour}[4]{%
  {%
    \begin{myfeynmandiagram}[
        scale=1.4, 
        baseline=(current bounding box.center)
      ]
      \myvertex {(a1)} {at (-1, -0.6)} {\(#1\)}
      \myvertex {(a2)} {at (-1, 0.6)} {\(#2\)}
      \myvertex {(a3)} {at (0.3, 0.6)} {\(#3\)}
      \myvertex {(a4)} {at (0.3, -0.6)} {\(#4\)}

      \myvertex[feyn vertex] {(mid1)} {at (-0.35,0)} {}

      \mypropagator[feyn propagator] {(mid1)} {-- (a1)}
      \mypropagator[feyn propagator] {(mid1)} {-- (a2)}
      \mypropagator[feyn propagator] {(mid1)} {-- (a3)}
      \mypropagator[feyn propagator] {(mid1)} {-- (a4)}
    \end{myfeynmandiagram}%
  }%
}

\newcommand{\mgraph}{%
  {%
    \begin{myfeynmandiagram}[
        scale=1.5,
        baseline=(current bounding box.center)
      ]
      \myvertex[feyn vertex] {(mid1)} {at (-1,0)} {}
      \myvertex[feyn vertex] {(midAb)} {at (-0.35,0)} {}
      \myvertex {(midAt)} {at (-0.35,0.8)} {\(\phantom{3}\)}
      \myvertex[feyn vertex] {(midB)} {at (-0.18,0.35)} {}
      \myvertex[feyn vertex] {(midC)} {at (0.18,0.35)} {}
      \myvertex[feyn vertex] {(midDb)} {at (0.35,0)} {}
      \myvertex {(midDt)} {at (0.35,0.8)} {\(\phantom{4}\)}
      \myvertex[feyn vertex] {(mid2)} {at (1,0)} {}
      
      \myvertex {(a1)} {at (-1, -0.8)} {\(1\)}
      \myvertex {(a2)} {at (-1, 0.8)} {\(2\)}
      \myvertex {(a3)} {at (1, 0.8)} {\(m-1\)}
      \myvertex {(a4)} {at (1, -0.8)} {\(m\)}

      \mypropagator[feyn propagator] {(midAb)} {-- (midAt)}
      \mypropagator[feyn propagator] {(midDb)} {-- (midDt)}
      \mypropagator[feyn propagator] {(mid2)} {-- (a3)}
      \mypropagator[feyn propagator] {(mid2)} {-- (a4)}
      \mypropagator[feyn propagator] {(mid2)} {-- (midDb)}
      \mypropagator[feyn propagator] {(midDb)} {-- (midAb)}
      \mypropagator[feyn propagator] {(midAb)} {-- (mid1)}
      \mypropagator[feyn propagator] {(a1)} {-- (mid1)}
      \mypropagator[feyn propagator] {(a2)} {-- (mid1)}

      \mypropagator[dotted_ellipsis_line]%
                           {(midB)} {-- (midC)};

    \end{myfeynmandiagram}%
  }%
}

\newcommand{\sixtrim}[6]{%
  {%
    \begin{myfeynmandiagram}[
        scale=1.5,
        baseline=(current bounding box.center)
      ]
      \myvertex {(a1)} {at (-1, -0.8)} {\(#1\)}
      \myvertex {(a2)} {at (-1, 0.8)} {\(#2\)}
      \myvertex[feyn vertex] {(mid1)} {at (-1,0)} {}
      \myvertex[feyn vertex] {(midDb)} {at (0,0)} {}
      \myvertex[feyn vertex] {(midAb)} {at (0,.5)} {}
      \myvertex {(midAt)} {at (-.45,.8)} {\(#3\)}
      \myvertex {(midDt)} {at (.45,.8)} {\(#4\)}
      \myvertex[feyn vertex] {(mid2)} {at (1,0)} {}
      \myvertex {(a3)} {at (1, 0.8)} {\(#5\)}
      \myvertex {(a4)} {at (1,- 0.8)} {\(#6\)}

      \mypropagator[feyn propagator] {(midAb)} {-- (midAt)}
      \mypropagator[feyn propagator] {(midAb)} {-- (midDt)}
      \mypropagator[feyn propagator] {(midDb)} {-- (midAb)}
      \mypropagator[feyn propagator] {(mid2)} {-- (a3)}
      \mypropagator[feyn propagator] {(mid2)} {-- (a4)}
      \mypropagator[feyn propagator] {(mid2)} {-- (midDb)}
      \mypropagator[feyn propagator] {(midDb)} {-- (mid1)}
      \mypropagator[feyn propagator] {(a1)} {-- (mid1)}
      \mypropagator[feyn propagator] {(a2)} {-- (mid1)}
    \end{myfeynmandiagram}%
  }%
}

\newcommand{\schannel}{%
  {%
    \begin{myfeynmandiagram}[%
        baseline=(current bounding box.center)
      ]

      \myvertex[feyn vertex] {(mid1)} {at (-0.5, 0)} {} %
      \myvertex[feyn vertex] {(mid2)} {at (0.5, 0)} {} %

      \myvertex {(a1)} {at (-1.5, 0)} {$a_1$}
      \myvertex {(a2)} {at (-0.5, 1)} {$a_2$}
      \myvertex {(a3)} {at (0.5, 1)} {$a_3$}
      \myvertex {(a4)} {at (1.5, 0)} {$a_4$}

      \mypropagator[feyn propagator] {(mid1)} {-- (a1)}
      \mypropagator[feyn propagator] {(mid1)} {-- (a2)}
      \mypropagator[feyn propagator] {(mid2)} {-- (a4)}
      \mypropagator[feyn propagator] {(mid2)} {-- (a3)}

      \mypropagator[ultra thick, draw=hred] {(mid1)} {-- (mid2)}

    \end{myfeynmandiagram}%
  }%
}
\newcommand{\tchannel}{%
  {%
    \begin{myfeynmandiagram}[%
        baseline=(current bounding box.center)
      ]

      \myvertex[feyn vertex] {(mid1)} {at (0, 0)} {}   %
      \myvertex[feyn vertex] {(mid2)} {at (0, 1)} {}   %

      \myvertex {(a1)} {at (-1, 0)} {$a_1$}
      \myvertex {(a2)} {at (-0.75, 1.75)} {$a_2$} %
      \myvertex {(a3)} {at (0.75, 1.75)} {$a_3$}
      \myvertex {(a4)} {at (1, 0)} {$a_4$}

      \mypropagator[feyn propagator] {(mid1)} {-- (a1)}
      \mypropagator[feyn propagator] {(mid1)} {-- (a4)}
      \mypropagator[feyn propagator] {(mid2)} {-- (a2)}
      \mypropagator[feyn propagator] {(mid2)} {-- (a3)}

      \mypropagator[ultra thick, draw=hred] {(mid1)} {-- (mid2)}

    \end{myfeynmandiagram}%
  }%
}

\newcommand{\crossedUchannel}{%
  {%
    \begin{myfeynmandiagram}[scale=.85] %

      \def\Radius{2cm}
      \def\MidOffset{0.5cm}
      \def\Gap{3.5pt}

      \myvertex {(a1)} {at (210:\Radius)} {$a_1$}  %
      \myvertex {(a2)} {at (150:\Radius)} {$a_2$} %
      \myvertex {(a3)} {at (30:\Radius)} {$a_3$}   %
      \myvertex {(a4)} {at (-30:\Radius)} {$a_4$}  %

      \myvertex[feyn vertex] {(mid1)} {at (-\MidOffset,0)} {} %
      \myvertex[feyn vertex] {(mid2)} {at (\MidOffset,0)} {}  %

      \mypropagator[feyn propagator] {(mid1)} {-- (a2)}
      \mypropagator[feyn propagator] {(mid2)} {-- (a3)}

      \mypropagator[ultra thick, draw=hred] {(mid1)} {-- (mid2)}

      \mypropagator[name path global=OverPath,feyn propagator] { (a1) } {-- (mid2)}
      \mypropagator[name path global=UnderPath,feyn propagator] { (a4) } {-- (mid1)}
      \fill[white,name intersections={of=OverPath and UnderPath, total=\numInts}]
      \foreach \s in {1,...,\numInts}{(intersection-\s) circle (2pt) node {}};
      \mypropagator[name path global=OverPath,feyn propagator] { (a1) } {-- (mid2)}
    \end{myfeynmandiagram}%
  }%
}

\newsavebox{\sChanBox}
\newsavebox{\tChanBox}
\newsavebox{\uChanBox}

\savebox{\sChanBox}{\schannel}
\savebox{\tChanBox}{\tchannel}
\savebox{\uChanBox}{\crossedUchannel}

\newcommand{\fourgraphT}[4]{%
  {%
    \begin{myfeynmandiagram}[
        scale=1.4, %
        baseline=(current bounding box.center)
      ]
      \myvertex {(a1)} {at (-0.9, -0.6)} {\(#1\)}
      \myvertex {(a2)} {at (-0.9, 1.6)} {\(#2\)}
      \myvertex {(a3)} {at (0.9, 1.6)} {\(#3\)}
      \myvertex {(a4)} {at (0.9, -0.6)} {\(#4\)}

      \myvertex[feyn vertex] {(mid1)} {at (0,0)} {}
      \myvertex[feyn vertex] {(mid2)} {at (0,1.0)} {}

      \mypropagator[feyn propagator] {(mid2)} {-- (a3)}
      \mypropagator[feyn propagator] {(mid2)} {-- (a2)}
      \mypropagator[feyn propagator] {(mid2)} {-- (mid1)}
      \mypropagator[feyn propagator] {(a1)} {-- (mid1)}
      \mypropagator[feyn propagator] {(a4)} {-- (mid1)}
    \end{myfeynmandiagram}%
  }%
}

\newcommand{\fourgraphAct}[4]{%
  {%
    \begin{myfeynmandiagram}[
        scale=1.4, %
        baseline=(current bounding box.center)
      ]
      \myvertex {(a1)} {at (-1.5, -0.8)} {\(#1\)}
      \myvertex {(a2)} {at (-1.5, 0.8)} {\(#2\)}
      \myvertex {(a3)} {at (0.8, 0.8)} {\(#3\)}
      \myvertex {(a4)} {at (0.8, -0.8)} {\(#4\)}

      \myvertex[feyn vertex] {(mid1)} {at (-1,0)} {}
      \myvertex[feyn vertex] {(mid2)} {at (0.3,0)} {}

      \mypropagator[feyn propagator] {(mid2)} {-- (a3)}
      \mypropagator[feyn propagator] {(mid2)} {-- (a4)}
      \mypropagator[feyn propagator] {(mid2)} {-- (mid1)}
      \mypropagator[feyn propagator] {(a1)} {-- (mid1)}
      \mypropagator[feyn propagator] {(a2)} {-- (mid1)}
    \end{myfeynmandiagram}%
  }%
}

\newcommand{\fourgraphScalarProp}[4]{%
  {%
    \begin{myfeynmandiagram}[
        scale=1.4, %
        baseline=(current bounding box.center)
      ]
      \myvertex {(a1)} {at (-1.5, -0.8)} {\(#1\)}
      \myvertex {(a2)} {at (-1.5, 0.8)} {\(#2\)}
      \myvertex {(a3)} {at (0.8, 0.8)} {\(#3\)}
      \myvertex {(a4)} {at (0.8, -0.8)} {\(#4\)}

      \myvertex[feyn vertex] {(mid1)} {at (-1,0)} {}
      \myvertex[feyn vertex] {(mid2)} {at (0.3,0)} {}

      \mypropagator[scalar,thick] {(mid2)} {-- (a3)}
      \mypropagator[mygluon,thick] {(mid2)} {-- (a4)}
      \mypropagator[scalar,thick] {(mid2)} {-- (mid1)}
      \mypropagator[mygluon,thick] {(a1)} {-- (mid1)}
      \mypropagator[scalar,thick] {(a2)} {-- (mid1)}
    \end{myfeynmandiagram}%
  }%
}

\newcommand{\fourgraphVectorProp}[4]{%
  {%
    \begin{myfeynmandiagram}[
        scale=1.4, %
        baseline=(current bounding box.center)
      ]
      \myvertex {(a1)} {at (-1.5, -0.8)} {\(#1\)}
      \myvertex {(a2)} {at (-1.5, 0.8)} {\(#2\)}
      \myvertex {(a3)} {at (0.8, 0.8)} {\(#3\)}
      \myvertex {(a4)} {at (0.8, -0.8)} {\(#4\)}

      \myvertex[feyn vertex] {(mid1)} {at (-1,0)} {}
      \myvertex[feyn vertex] {(mid2)} {at (0.3,0)} {}

      \mypropagator[mygluon,thick] {(mid2)} {-- (a3)}
      \mypropagator[mygluon,thick] {(mid2)} {-- (a4)}
      \mypropagator[mygluon,thick] {(mid2)} {-- (mid1)}
      \mypropagator[scalar,thick] {(a1)} {-- (mid1)}
      \mypropagator[scalar,thick] {(a2)} {-- (mid1)}
    \end{myfeynmandiagram}%
  }%
}

\newcommand{\fourgraphExtScalar}[4]{%
  {%
    \begin{myfeynmandiagram}[
        scale=1.4, %
        baseline=(current bounding box.center)
      ]
      \myvertex {(a1)} {at (-1.5, -0.8)} {\(#1\)}
      \myvertex {(a2)} {at (-1.5, 0.8)} {\(#2\)}
      \myvertex {(a3)} {at (0.8, 0.8)} {\(#3\)}
      \myvertex {(a4)} {at (0.8, -0.8)} {\(#4\)}

      \myvertex[feyn vertex] {(mid1)} {at (-1,0)} {}
      \myvertex[feyn vertex] {(mid2)} {at (0.3,0)} {}

      \mypropagator[scalar,thick] {(mid2)} {-- (a3)}
      \mypropagator[scalar,thick] {(mid2)} {-- (a4)}
      \mypropagator[mygluon,thick] {(mid2)} {-- (mid1)}
      \mypropagator[scalar,thick] {(a1)} {-- (mid1)}
      \mypropagator[scalar,thick] {(a2)} {-- (mid1)}
    \end{myfeynmandiagram}%
  }%
}

\newcommand{\fourgraphExtScalarSmall}[4]{%
  {%
    \begin{myfeynmandiagram}[
        scale=1.4, %
        baseline=(current bounding box.center)
      ]
      \myvertex {(a1)} {at (-1.4, -0.6)} {\(#1\)}
      \myvertex {(a2)} {at (-1.4, 0.6)} {\(#2\)}
      \myvertex {(a3)} {at (0.4, 0.6)} {\(#3\)}
      \myvertex {(a4)} {at (0.4, -0.6)} {\(#4\)}

      \myvertex[feyn vertex] {(mid1)} {at (-1,0)} {}
      \myvertex[feyn vertex] {(mid2)} {at (0,0)} {}

      \mypropagator[scalar,thick] {(mid2)} {-- (a3)}
      \mypropagator[scalar,thick] {(mid2)} {-- (a4)}
      \mypropagator[mygluon,thick] {(mid2)} {-- (mid1)}
      \mypropagator[scalar,thick] {(a1)} {-- (mid1)}
      \mypropagator[scalar,thick] {(a2)} {-- (mid1)}
    \end{myfeynmandiagram}%
  }
}

\newcommand{\fivegraphAct}[5]{%
  {%
    \begin{myfeynmandiagram}[
        scale=1.4, %
        baseline=(current bounding box.center)
      ]
      \myvertex {(a1)} {at (-1, -0.8)} {\(#1\)}
      \myvertex {(a2)} {at (-1, 0.8)} {\(#2\)}
      \myvertex {(a3)} {at (1, 0.8)} {\(#4\)}
      \myvertex {(a4)} {at (1, -0.8)} {\(#5\)}

      \myvertex[feyn vertex] {(mid1)} {at (-1,0)} {}
      \myvertex[feyn vertex] {(midAb)} {at (0,0)} {}
      \myvertex {(midAt)} {at (0,0.8)} {\(#3\)}
      \myvertex[feyn vertex] {(mid2)} {at (1,0)} {}

      \mypropagator[feyn propagator] {(midAb)} {-- (midAt)}
      \mypropagator[feyn propagator] {(mid2)} {-- (a3)}
      \mypropagator[feyn propagator] {(mid2)} {-- (a4)}
      \mypropagator[feyn propagator] {(mid2)} {-- (midAb)}
      \mypropagator[feyn propagator] {(midAb)} {-- (mid1)}
      \mypropagator[feyn propagator] {(a1)} {-- (mid1)}
      \mypropagator[feyn propagator] {(a2)} {-- (mid1)}
    \end{myfeynmandiagram}%
  }%
}

\newcommand{\fivegraphNMcutLeftUnique}[5]{%
  {%
    \begin{myfeynmandiagram}[
        scale=1.4, %
        baseline=(current bounding box.center)
      ]
      \myvertex {(a1)} {at (-1, -0.8)} {\(#1\)}
      \myvertex {(a2)} {at (-1, 0.8)} {\(#2\)}
      \myvertex {(a3)} {at (1, 0.8)} {\(#4\)}
      \myvertex {(a4)} {at (1, -0.8)} {\(#5\)}

      \myvertex[feyn vertex] {(mid1)} {at (-1,0)} {}
      \myvertex[feyn vertex] {(midAb)} {at (0,0)} {}
      \myvertex {(midAt)} {at (0,0.8)} {\(#3\)}
      \myvertex[feyn vertex] {(mid2)} {at (1,0)} {}

      \myvertex {(a5)} {at (-0.5, -0.7)} {}
      \myvertex {(a6)} {at (-0.5, 0.7)} {}
      \myvertex[dot,fill=darkgreen] {(a7)} {at (0.45, -0.5)} {}
      \myvertex[dot,fill=darkgreen] {(a8)} {at (0.45, 0.5)} {}

      \mypropagator[feyn propagator] {(midAb)} {-- (midAt)}
      \mypropagator[feyn propagator] {(mid2)} {-- (a3)}
      \mypropagator[feyn propagator] {(mid2)} {-- (a4)}
      \mypropagator[feyn propagator] {(mid2)} {-- (midAb)}
      \mypropagator[feyn propagator] {(midAb)} {-- (mid1)}
      \mypropagator[feyn propagator] {(a1)} {-- (mid1)}
      \mypropagator[feyn propagator] {(a2)} {-- (mid1)}

      \mypropagator[scalar,ultra thick,draw=cutred] {(a5)} {-- (a6)}
      \mypropagator[ultra thick,draw=darkgreen] {(a7)} {to[bend right=45] (a8)}
    \end{myfeynmandiagram}%
  }%
}

\newcommand{\fivegraphNMcutRightUnique}[5]{%
  {%
    \begin{myfeynmandiagram}[
        scale=1.4, %
        baseline=(current bounding box.center)
      ]
      \myvertex {(a1)} {at (-1, -0.8)} {\(#1\)}
      \myvertex {(a2)} {at (-1, 0.8)} {\(#2\)}
      \myvertex {(a3)} {at (1, 0.8)} {\(#4\)}
      \myvertex {(a4)} {at (1, -0.8)} {\(#5\)}

      \myvertex[feyn vertex] {(mid1)} {at (-1,0)} {}
      \myvertex[feyn vertex] {(midAb)} {at (0,0)} {}
      \myvertex {(midAt)} {at (0,0.8)} {\(#3\)}
      \myvertex[feyn vertex] {(mid2)} {at (1,0)} {}

      \myvertex {(a5)} {at (0.5, -0.7)} {}
      \myvertex {(a6)} {at (0.5, 0.7)} {}
      \myvertex[dot,fill=darkgreen] {(a7)} {at (-0.45, -0.5)} {}
      \myvertex[dot,fill=darkgreen] {(a8)} {at (-0.45, 0.5)} {}

      \mypropagator[feyn propagator] {(midAb)} {-- (midAt)}
      \mypropagator[feyn propagator] {(mid2)} {-- (a3)}
      \mypropagator[feyn propagator] {(mid2)} {-- (a4)}
      \mypropagator[feyn propagator] {(mid2)} {-- (midAb)}
      \mypropagator[feyn propagator] {(midAb)} {-- (mid1)}
      \mypropagator[feyn propagator] {(a1)} {-- (mid1)}
      \mypropagator[feyn propagator] {(a2)} {-- (mid1)}

      \mypropagator[scalar,ultra thick,draw=cutred] {(a5)} {-- (a6)}
      \mypropagator[ultra thick,draw=darkgreen] {(a7)} {to[bend right=45] (a8)}
    \end{myfeynmandiagram}%
  }%
}

\newcommand{\fivegraphContactLeftRightUnique}[5]{%
  {%
    \begin{myfeynmandiagram}[
        scale=1.4, %
        baseline=(current bounding box.center)
      ]
      \myvertex {(a1)} {at (-1, -0.8)} {\(#1\)}
      \myvertex {(a2)} {at (-1, 0.8)} {\(#2\)}
      \myvertex {(a3)} {at (1, 0.8)} {\(#4\)}
      \myvertex {(a4)} {at (1, -0.8)} {\(#5\)}

      \myvertex[feyn vertex] {(mid1)} {at (-1,0)} {}
      \myvertex[feyn vertex] {(midAb)} {at (0,0)} {}
      \myvertex {(midAt)} {at (0,0.8)} {\(#3\)}
      \myvertex[feyn vertex] {(mid2)} {at (1,0)} {}

      \myvertex[dot,fill=darkgreen] {(a5)} {at (0.5, -0.5)} {}
      \myvertex[dot,fill=darkgreen] {(a6)} {at (0.5, 0.5)} {}
      \myvertex[dot,fill=darkgreen] {(a7)} {at (-0.45, -0.5)} {}
      \myvertex[dot,fill=darkgreen] {(a8)} {at (-0.45, 0.5)} {}

      \mypropagator[feyn propagator] {(midAb)} {-- (midAt)}
      \mypropagator[feyn propagator] {(mid2)} {-- (a3)}
      \mypropagator[feyn propagator] {(mid2)} {-- (a4)}
      \mypropagator[feyn propagator] {(mid2)} {-- (midAb)}
      \mypropagator[feyn propagator] {(midAb)} {-- (mid1)}
      \mypropagator[feyn propagator] {(a1)} {-- (mid1)}
      \mypropagator[feyn propagator] {(a2)} {-- (mid1)}

      \mypropagator[ultra thick,draw=darkgreen] {(a5)} {to[bend right=45] (a6)}
      \mypropagator[ultra thick,draw=darkgreen] {(a7)} {to[bend right=45] (a8)}
    \end{myfeynmandiagram}%
  }%
}

\newcommand{\fivegraphMcut}[5]{%
  {%
    \begin{myfeynmandiagram}[
        scale=1.4, %
        baseline=(current bounding box.center)
      ]
      \myvertex {(a1)} {at (-1, -0.8)} {\(#1\)}
      \myvertex {(a2)} {at (-1, 0.8)} {\(#2\)}
      \myvertex {(a3)} {at (1, 0.8)} {\(#4\)}  %
      \myvertex {(a4)} {at (1, -0.8)} {\(#5\)} %

      \myvertex[feyn vertex] {(mid1)} {at (-1,0)} {}
      \myvertex[feyn vertex] {(midAb)} {at (0,0)} {}
      \myvertex {(midAt)} {at (0,0.8)} {\(#3\)}
      \myvertex[feyn vertex] {(mid2)} {at (1,0)} {}

      \myvertex {(a5)} {at (-0.5, -0.7)} {}
      \myvertex {(a6)} {at (-0.5, 0.7)} {}
      \myvertex {(a7)} {at (0.5, -0.7)} {}
      \myvertex {(a8)} {at (0.5, 0.7)} {}

      \mypropagator[feyn propagator] {(midAb)} {-- (midAt)}
      \mypropagator[feyn propagator] {(mid2)} {-- (a3)}
      \mypropagator[feyn propagator] {(mid2)} {-- (a4)}
      \mypropagator[feyn propagator] {(mid2)} {-- (midAb)}
      \mypropagator[feyn propagator] {(midAb)} {-- (mid1)}
      \mypropagator[feyn propagator] {(a1)} {-- (mid1)}
      \mypropagator[feyn propagator] {(a2)} {-- (mid1)}

      \mypropagator[scalar,ultra thick,draw=cutred] {(a5)} {-- (a6)} %
      \mypropagator[scalar,ultra thick,draw=cutred] {(a7)} {-- (a8)} %
    \end{myfeynmandiagram}%
  }%
}

\newcommand{\sixgraph}[6]{%
  {%
    \begin{myfeynmandiagram}[
        scale=1.4, %
        baseline=(current bounding box.center)
      ]
      \myvertex {(a1)} {at (-1, -0.7)} {\(#1\)} %
      \myvertex {(a2)} {at (-1, 0.7)} {\(#2\)}  %
      \myvertex {(a3)} {at (-0.33, 0.7)} {\(#3\)} %
      \myvertex {(a4)} {at (0.33, 0.7)} {\(#4\)}  %
      \myvertex {(a5)} {at (1, 0.7)} {\(#5\)}    %
      \myvertex {(a6)} {at (1, -0.7)} {\(#6\)}   %

      \myvertex[feyn vertex] {(mid1)} {at (-1,0)} {}    %
      \myvertex[feyn vertex] {(mid2)} {at (-0.33,0)} {}  %
      \myvertex[feyn vertex] {(mid3)} {at (0.33,0)} {}   %
      \myvertex[feyn vertex] {(mid4)} {at (1,0)} {}     %

      \mypropagator[feyn propagator] {(mid1)} {-- (a1)}  %
      \mypropagator[feyn propagator] {(mid1)} {-- (a2)}  %
      \mypropagator[feyn propagator] {(mid2)} {-- (a3)}  %
      \mypropagator[feyn propagator] {(mid3)} {-- (a4)}  %
      \mypropagator[feyn propagator] {(mid4)} {-- (a5)}  %
      \mypropagator[feyn propagator] {(mid4)} {-- (a6)}  %
      \mypropagator[feyn propagator] {(mid1)} {-- (mid2)} %
      \mypropagator[feyn propagator] {(mid2)} {-- (mid3)} %
      \mypropagator[feyn propagator] {(mid3)} {-- (mid4)} %
    \end{myfeynmandiagram}%
  }%
}

\newcommand{\sixGraphMC}[6]{%
  {%
    \begin{myfeynmandiagram}[
        scale=1.4, %
        baseline=(current bounding box.center)
      ]
      \myvertex {(a1)} {at (-1, -0.7)} {\(#1\)} %
      \myvertex {(a2)} {at (-1, 0.7)} {\(#2\)}  %
      \myvertex {(a3)} {at (-0.33, 0.7)} {\(#3\)} %
      \myvertex {(a4)} {at (0.33, 0.7)} {\(#4\)}  %
      \myvertex {(a5)} {at (1, 0.7)} {\(#5\)}    %
      \myvertex {(a6)} {at (1, -0.7)} {\(#6\)}   %

      \myvertex[feyn vertex] {(mid1)} {at (-1,0)} {}    %
      \myvertex[feyn vertex] {(mid2)} {at (-0.33,0)} {}  %
      \myvertex[feyn vertex] {(mid3)} {at (0.33,0)} {}   %
      \myvertex[feyn vertex] {(mid4)} {at (1,0)} {}     %

      \myvertex {(a7)} {at (-0.67, -0.7)} {}  %
      \myvertex {(a8)} {at (-0.67, 0.7)} {}   %
      \myvertex {(a9)} {at (0, -0.7)} {}      %
      \myvertex {(a10)} {at (0, 0.7)} {}      %
      \myvertex {(a11)} {at (0.67, -0.7)} {}  %
      \myvertex {(a12)} {at (0.67, 0.7)} {}   %

      \mypropagator[feyn propagator] {(mid1)} {-- (a1)}  %
      \mypropagator[feyn propagator] {(mid1)} {-- (a2)}  %
      \mypropagator[feyn propagator] {(mid2)} {-- (a3)}  %
      \mypropagator[feyn propagator] {(mid3)} {-- (a4)}  %
      \mypropagator[feyn propagator] {(mid4)} {-- (a5)}  %
      \mypropagator[feyn propagator] {(mid4)} {-- (a6)}  %
      \mypropagator[feyn propagator] {(mid1)} {-- (mid2)} %
      \mypropagator[feyn propagator] {(mid2)} {-- (mid3)} %
      \mypropagator[feyn propagator] {(mid3)} {-- (mid4)} %

      \mypropagator[scalar,ultra thick,draw=cutred] {(a7)} {-- (a8)}   %
      \mypropagator[scalar,ultra thick,draw=cutred] {(a9)} {-- (a10)}  %
      \mypropagator[scalar,ultra thick,draw=cutred] {(a11)} {-- (a12)} %
    \end{myfeynmandiagram}%
  }%
}

\newcommand{\sixGraphNNMCright}[6]{%
  {%
    \begin{myfeynmandiagram}[
        scale=1.4, %
        baseline=(current bounding box.center)
      ]
      \myvertex {(a1)} {at (-1, -0.7)} {\(#1\)} %
      \myvertex {(a2)} {at (-1, 0.7)} {\(#2\)}  %
      \myvertex {(a3)} {at (-0.33, 0.7)} {\(#3\)} %
      \myvertex {(a4)} {at (0.33, 0.7)} {\(#4\)}  %
      \myvertex {(a5)} {at (1, 0.7)} {\(#5\)}    %
      \myvertex {(a6)} {at (1, -0.7)} {\(#6\)}   %

      \myvertex[feyn vertex] {(mid1)} {at (-1,0)} {}    %
      \myvertex[feyn vertex] {(mid2)} {at (-0.33,0)} {}  %
      \myvertex[feyn vertex] {(mid3)} {at (0.33,0)} {}   %
      \myvertex[feyn vertex] {(mid4)} {at (1,0)} {}     %

      \myvertex {(a7)} {at (0.67, -0.7)} {}  %
      \myvertex {(a8)} {at (0.67, 0.7)} {}   %
      \myvertex[dot,fill=darkgreen] {(a9)} {at (-0.67, -0.5)} {}  %
      \myvertex[dot,fill=darkgreen] {(a10)} {at (-0.67, 0.5)} {}  %
      \myvertex[dot,fill=darkgreen] {(a11)} {at (0, -0.5)} {}  %
      \myvertex[dot,fill=darkgreen] {(a12)} {at (0, 0.5)} {}  %

      \mypropagator[feyn propagator] {(mid1)} {-- (a1)}  %
      \mypropagator[feyn propagator] {(mid1)} {-- (a2)}  %
      \mypropagator[feyn propagator] {(mid2)} {-- (a3)}  %
      \mypropagator[feyn propagator] {(mid3)} {-- (a4)}  %
      \mypropagator[feyn propagator] {(mid4)} {-- (a5)}  %
      \mypropagator[feyn propagator] {(mid4)} {-- (a6)}  %
      \mypropagator[feyn propagator] {(mid1)} {-- (mid2)} %
      \mypropagator[feyn propagator] {(mid2)} {-- (mid3)} %
      \mypropagator[feyn propagator] {(mid3)} {-- (mid4)} %

      \mypropagator[scalar,ultra thick,draw=cutred] {(a7)} {-- (a8)}   %
      \mypropagator[ultra thick,draw=darkgreen] {(a9)} {to[bend right=45] (a10)} %
      \mypropagator[ultra thick,draw=darkgreen] {(a11)} {to[bend right=45] (a12)} %
    \end{myfeynmandiagram}%
  }%
}

\newcommand{\sixGraphUniqueContact}[6]{%
  {%
    \begin{myfeynmandiagram}[
        scale=1.4, %
        baseline=(current bounding box.center)
      ]
      \myvertex {(a1)} {at (-1, -0.7)} {\(#1\)} %
      \myvertex {(a2)} {at (-1, 0.7)} {\(#2\)}  %
      \myvertex {(a3)} {at (-0.33, 0.7)} {\(#3\)} %
      \myvertex {(a4)} {at (0.33, 0.7)} {\(#4\)}  %
      \myvertex {(a5)} {at (1, 0.7)} {\(#5\)}    %
      \myvertex {(a6)} {at (1, -0.7)} {\(#6\)}   %

      \myvertex[feyn vertex] {(mid1)} {at (-1,0)} {}    %
      \myvertex[feyn vertex] {(mid2)} {at (-0.33,0)} {}  %
      \myvertex[feyn vertex] {(mid3)} {at (0.33,0)} {}   %
      \myvertex[feyn vertex] {(mid4)} {at (1,0)} {}     %

      \myvertex[dot,fill=darkgreen] {(a7)} {at (-0.67, -0.35)} {}  %
      \myvertex[dot,fill=darkgreen] {(a8)} {at (-0.67, 0.35)} {}   %
      \myvertex[dot,fill=darkgreen] {(a9)} {at (0, -0.35)} {}     %
      \myvertex[dot,fill=darkgreen] {(a10)} {at (0, 0.35)} {}     %
      \myvertex[dot,fill=darkgreen] {(a11)} {at (0.67, -0.35)} {} %
      \myvertex[dot,fill=darkgreen] {(a12)} {at (0.67, 0.35)} {}  %

      \mypropagator[feyn propagator] {(mid1)} {-- (a1)}  %
      \mypropagator[feyn propagator] {(mid1)} {-- (a2)}  %
      \mypropagator[feyn propagator] {(mid2)} {-- (a3)}  %
      \mypropagator[feyn propagator] {(mid3)} {-- (a4)}  %
      \mypropagator[feyn propagator] {(mid4)} {-- (a5)}  %
      \mypropagator[feyn propagator] {(mid4)} {-- (a6)}  %
      \mypropagator[feyn propagator] {(mid1)} {-- (mid2)} %
      \mypropagator[feyn propagator] {(mid2)} {-- (mid3)} %
      \mypropagator[feyn propagator] {(mid3)} {-- (mid4)} %

      \mypropagator[ultra thick,draw=darkgreen] {(a7)} {to[bend right=45] (a8)}   %
      \mypropagator[ultra thick,draw=darkgreen] {(a9)} {to[bend right=45] (a10)}  %
      \mypropagator[ultra thick,draw=darkgreen] {(a11)} {to[bend right=45] (a12)} %
    \end{myfeynmandiagram}%
  }%
}

\newcommand{\threeGraph}[3]{%
  {%
    \begin{myfeynmandiagram}[
        scale=1.4, %
        baseline=(current bounding box.center)
      ]
      \myvertex {(a1)} {at (-0.9, -0.9)} {\(#1\)}
      \myvertex {(a2)} {at (-0.9, 0.9)} {\(#2\)}
      \myvertex {(a3)} {at (1.2, 0)} {\(#3\)}

      \myvertex[feyn vertex] {(mid1)} {at (0,0)} {}

      \mypropagator[scalar,thick] {(mid1)} {-- (a1)}
      \mypropagator[scalar,thick] {(mid1)} {-- (a2)}
      \mypropagator[mygluon,thick] {(a3)} {-- (mid1)}
    \end{myfeynmandiagram}%
  }%
}

\newcommand{\fourOp}[5]{%
  {%
    \begin{myfeynmandiagram}[
        scale=1.4, %
        baseline=(current bounding box.center)
      ]
      \myvertex {(a1)} {at (-1.2, -1.2)} {\(#1\)}
      \myvertex {(a2)} {at (-1.2, 1.2)} {\(#2\)}
      \myvertex {(a3)} {at (1.2, 1.2)} {\(#3\)}
      \myvertex {(a4)} {at (1.2, -1.2)} {\(#4\)}

      \myvertex[blob,draw=hgreen] {(mid1)} {at (0,0)} {\(\,#5\,\)}

      \mypropagator[scalar,ultra thick,draw=hblue] {(mid1)} {-- (a1)}
      \mypropagator[scalar,ultra thick,draw=hblue] {(mid1)} {-- (a2)}
      \mypropagator[mygluon,ultra thick,draw=hred] {(a3)} {-- (mid1)}
      \mypropagator[mygluon,ultra thick,draw=hred] {(a4)} {-- (mid1)}
    \end{myfeynmandiagram}%
  }%
}

\newcommand{\fourgraphScalarPropOp}[6]{%
  {%
    \begin{myfeynmandiagram}[
        scale=1.4, %
        baseline=(current bounding box.center)
      ]
      \myvertex {(a1)} {at (-1.9, -1.2)} {\(#1\)}
      \myvertex {(a2)} {at (-1.9, 1.2)} {\(#2\)}
      \myvertex {(a3)} {at (1.6, 1.2)} {\(#3\)}
      \myvertex {(a4)} {at (1.6, -1.2)} {\(#4\)}

      \myvertex[blob,draw=hgreen] {(mid1)} {at (-1,0)} {\(\,#5\,\)}
      \myvertex[blob,draw=hgreen] {(mid2)} {at (0.7,0)} {\(\,#6\,\)}

      \mypropagator[scalar,ultra thick,draw=hblue] {(mid2)} {-- (a3)}
      \mypropagator[mygluon,ultra thick,draw=hred] {(mid2)} {-- (a4)}
      \mypropagator[scalar,ultra thick,draw=hblue] {(mid2)} {-- (mid1)}
      \mypropagator[mygluon,ultra thick,draw=hred] {(a1)} {-- (mid1)}
      \mypropagator[scalar,ultra thick,draw=hblue] {(a2)} {-- (mid1)}
    \end{myfeynmandiagram}%
  }%
}

\newcommand{\fourgraphVectorPropOp}[6]{%
  {%
    \begin{myfeynmandiagram}[
        scale=1.4,
        baseline=(current bounding box.center)
      ]
      \myvertex {(a1)} {at (-1.9, -1.2)} {\(#1\)}
      \myvertex {(a2)} {at (-1.9, 1.2)} {\(#2\)}
      \myvertex {(a3)} {at (1.6, 1.2)} {\(#3\)}
      \myvertex {(a4)} {at (1.6, -1.2)} {\(#4\)}

      \myvertex[blob, hgreen] {(mid1)} {at (-1, 0)} {\(\,#5\,\)}
      \myvertex[white] {(mid3)} {at (-1, 0)} {\(\,#5\,\)}
      \myvertex[blob, hgreen] {(mid2)} {at (0.7, 0)} {\(\,#6\,\)}
      \myvertex[white] {(mid4)} {at (0.7, 0)} {\(\,#6\,\)}

      \mypropagator[mygluon, ultra thick, draw=hred] {(mid2)} {-- (a3)}
      \mypropagator[mygluon, ultra thick, draw=hred] {(mid2)} {-- (a4)}
      \mypropagator[mygluon, ultra thick, draw=hred] {(mid2)} {-- (mid1)}
      \mypropagator[scalar, ultra thick, draw=hblue] {(a1)} {-- (mid1)}
      \mypropagator[scalar, ultra thick, draw=hblue] {(a2)} {-- (mid1)}
    \end{myfeynmandiagram}%
  }%
}

\newcommand{\fourgraphVectorPropCut}[6]{%
  {%
    \begin{myfeynmandiagram}[
        scale=1.4, %
        baseline=(current bounding box.center)
      ]
      \myvertex {(a1)} {at (-1.9, -1.2)} {\(#1\)}
      \myvertex {(a2)} {at (-1.9, 1.2)} {\(#2\)}
      \myvertex {(a3)} {at (1.6, 1.2)} {\(#3\)}
      \myvertex {(a4)} {at (1.6, -1.2)} {\(#4\)}

      \myvertex[blob, hgreen] {(mid1)} {at (-1, 0)} {\(\,#5\,\)}
      \myvertex[white] {(mid3)} {at (-1, 0)} {\(\,#5\,\)}
      \myvertex[blob, hgreen] {(mid2)} {at (0.7, 0)} {\(\,#6\,\)}
      \myvertex[white] {(mid4)} {at (0.7, 0)} {\(\,#6\,\)}

      \myvertex {(a5)} {at (-0.15, -0.7)} {}
      \myvertex {(a6)} {at (-0.15, 0.7)} {}

      \mypropagator[mygluon,thick] {(mid2)} {-- (a3)}
      \mypropagator[mygluon,thick] {(mid2)} {-- (a4)}
      \mypropagator[mygluon,thick] {(mid2)} {-- (mid1)}
      \mypropagator[scalar,thick] {(a1)} {-- (mid1)}
      \mypropagator[scalar,thick] {(a2)} {-- (mid1)}

      \mypropagator[scalar,ultra thick,draw=cutred] {(a5)} {-- (a6)}
    \end{myfeynmandiagram}%
  }%
}

\newcommand{\fourgraphScalarPropCut}[6]{%
  {%
    \begin{myfeynmandiagram}[
        scale=1.4,
        baseline=(current bounding box.center)
      ]
      \myvertex {(a1)} {at (-1.9, -1.2)} {\(#1\)}
      \myvertex {(a2)} {at (-1.9, 1.2)} {\(#2\)}
      \myvertex [blob,hgreen] {(mid1)} {at (-1,0)} {\(\,#5\,\)}
      \myvertex [white] {(mid3)} {at (-1,0)} {\(\,#5\,\)}
      \myvertex [blob,hgreen] {(mid2)} {at (0.7,0)} {\(\,#6\,\)}
      \myvertex [white] {(mid4)} {at (0.7,0)} {\(\,#6\,\)}
      \myvertex {(a3)} {at (1.6, 1.2)} {\(#3\)}
      \myvertex {(a4)} {at (1.6, -1.2)} {\(#4\)}
      \myvertex {(a5)} {at (-0.15, -0.7)} {}
      \myvertex {(a6)} {at (-0.15, 0.7)} {}

      \mypropagator [scalar,thick,hgrey] {(mid2)} {-- (a3)}
      \mypropagator [mygluon,thick,hgrey] {(mid2)} {-- (a4)}
      \mypropagator [scalar,thick,hgrey] {(mid2)} {-- (mid1)}
      \mypropagator [mygluon,thick,hgrey] {(a1)} {-- (mid1)}
      \mypropagator [scalar,thick,hgrey] {(a2)} {-- (mid1)}
      \mypropagator [scalar,ultra thick,cutred] {(a5)} {-- (a6)}
    \end{myfeynmandiagram}
  }%
}

\newcommand{\fivegraphNMcutLeft}[5]{%
  {%
    \begin{myfeynmandiagram}[
        scale=1.4, %
        baseline=(current bounding box.center)
      ]
      \myvertex {(a1)} {at (-1, -0.8)} {\(#1\)}
      \myvertex {(a2)} {at (-1, 0.8)} {\(#2\)}
      \myvertex {(a3)} {at (1, 0.8)} {\(#4\)}
      \myvertex {(a4)} {at (1, -0.8)} {\(#5\)}

      \myvertex[feyn vertex] {(mid1)} {at (-1,0)} {}
      \myvertex[feyn vertex] {(midAb)} {at (0,0)} {}
      \myvertex {(midAt)} {at (0,0.8)} {\(#3\)}
      \myvertex[feyn vertex] {(mid2)} {at (1,0)} {}

      \myvertex {(a5)} {at (-0.5, -0.7)} {}
      \myvertex {(a6)} {at (-0.5, 0.7)} {}

      \mypropagator[feyn propagator] {(midAb)} {-- (midAt)}
      \mypropagator[feyn propagator] {(mid2)} {-- (a3)}
      \mypropagator[feyn propagator] {(mid2)} {-- (a4)}
      \mypropagator[feyn propagator] {(mid2)} {-- (midAb)}
      \mypropagator[feyn propagator] {(midAb)} {-- (mid1)}
      \mypropagator[feyn propagator] {(a1)} {-- (mid1)}
      \mypropagator[feyn propagator] {(a2)} {-- (mid1)}

      \mypropagator[scalar,ultra thick,draw=cutred] {(a5)} {-- (a6)}
    \end{myfeynmandiagram}%
  }%
}

\newcommand{\fivegraphNMcutRight}[5]{%
  {%
    \begin{myfeynmandiagram}[
        scale=1.4, %
        baseline=(current bounding box.center)
      ]
      \myvertex {(a1)} {at (-1, -0.8)} {\(#1\)}
      \myvertex {(a2)} {at (-1, 0.8)} {\(#2\)}
      \myvertex {(a3)} {at (1, 0.8)} {\(#4\)}
      \myvertex {(a4)} {at (1, -0.8)} {\(#5\)}

      \myvertex[feyn vertex] {(mid1)} {at (-1,0)} {}
      \myvertex[feyn vertex] {(midAb)} {at (0,0)} {}
      \myvertex {(midAt)} {at (0,0.8)} {\(#3\)}
      \myvertex[feyn vertex] {(mid2)} {at (1,0)} {}

      \myvertex {(a5)} {at (0.5, -0.7)} {}
      \myvertex {(a6)} {at (0.5, 0.7)} {}

      \mypropagator[feyn propagator] {(midAb)} {-- (midAt)}
      \mypropagator[feyn propagator] {(mid2)} {-- (a3)}
      \mypropagator[feyn propagator] {(mid2)} {-- (a4)}
      \mypropagator[feyn propagator] {(mid2)} {-- (midAb)}
      \mypropagator[feyn propagator] {(midAb)} {-- (mid1)}
      \mypropagator[feyn propagator] {(a1)} {-- (mid1)}
      \mypropagator[feyn propagator] {(a2)} {-- (mid1)}

      \mypropagator[scalar,ultra thick,draw=cutred] {(a5)} {-- (a6)}
    \end{myfeynmandiagram}%
  }%
}

\newcommand{\sixGraphNMCleftU}[6]{%
  {%
    \begin{myfeynmandiagram}[
        scale=1.4, %
        baseline=(current bounding box.center)
      ]
      \myvertex {(a1)} {at (-1, -0.7)} {\(#1\)} %
      \myvertex {(a2)} {at (-1, 0.7)} {\(#2\)}  %
      \myvertex {(a3)} {at (-0.33, 0.7)} {\(#3\)} %
      \myvertex {(a4)} {at (0.33, 0.7)} {\(#4\)}  %
      \myvertex {(a5)} {at (1, 0.7)} {\(#5\)}    %
      \myvertex {(a6)} {at (1, -0.7)} {\(#6\)}   %

      \myvertex[feyn vertex] {(mid1)} {at (-1,0)} {}    %
      \myvertex[feyn vertex] {(mid2)} {at (-0.33,0)} {}  %
      \myvertex[feyn vertex] {(mid3)} {at (0.33,0)} {}   %
      \myvertex[feyn vertex] {(mid4)} {at (1,0)} {}     %

      \myvertex {(a7)} {at (0, -0.7)} {}  %
      \myvertex {(a8)} {at (0, 0.7)} {}   %
      \myvertex {(a9)} {at (0.67, -0.7)} {}  %
      \myvertex {(a10)} {at (0.67, 0.7)} {}  %
      \myvertex[dot,fill=darkgreen] {(a11)} {at (-0.665, -0.35)} {}  %
      \myvertex[dot,fill=darkgreen] {(a12)} {at (-0.665, 0.35)} {}  %

      \mypropagator[feyn propagator] {(mid1)} {-- (a1)}  %
      \mypropagator[feyn propagator] {(mid1)} {-- (a2)}  %
      \mypropagator[feyn propagator] {(mid2)} {-- (a3)}  %
      \mypropagator[feyn propagator] {(mid3)} {-- (a4)}  %
      \mypropagator[feyn propagator] {(mid4)} {-- (a5)}  %
      \mypropagator[feyn propagator] {(mid4)} {-- (a6)}  %
      \mypropagator[feyn propagator] {(mid1)} {-- (mid2)} %
      \mypropagator[feyn propagator] {(mid2)} {-- (mid3)} %
      \mypropagator[feyn propagator] {(mid3)} {-- (mid4)} %

      \mypropagator[scalar,ultra thick,draw=cutred] {(a7)} {-- (a8)}   %
      \mypropagator[scalar,ultra thick,draw=cutred] {(a9)} {-- (a10)}  %
      \mypropagator[ultra thick,draw=darkgreen] {(a11)} {to[bend right=45] (a12)} %
    \end{myfeynmandiagram}%
  }%
}

\newcommand{\sixGraphNMCmiddleU}[6]{%
  {%
    \begin{myfeynmandiagram}[
        scale=1.4, %
        baseline=(current bounding box.center)
      ]
      \myvertex {(a1)} {at (-1, -0.7)} {\(#1\)} %
      \myvertex {(a2)} {at (-1, 0.7)} {\(#2\)}  %
      \myvertex {(a3)} {at (-0.33, 0.7)} {\(#3\)} %
      \myvertex {(a4)} {at (0.33, 0.7)} {\(#4\)}  %
      \myvertex {(a5)} {at (1, 0.7)} {\(#5\)}    %
      \myvertex {(a6)} {at (1, -0.7)} {\(#6\)}   %

      \myvertex[feyn vertex] {(mid1)} {at (-1,0)} {}    %
      \myvertex[feyn vertex] {(mid2)} {at (-0.33,0)} {}  %
      \myvertex[feyn vertex] {(mid3)} {at (0.33,0)} {}   %
      \myvertex[feyn vertex] {(mid4)} {at (1,0)} {}     %

      \myvertex {(a7)} {at (-0.67, -0.7)} {}  %
      \myvertex {(a8)} {at (-0.67, 0.7)} {}   %
      \myvertex {(a9)} {at (0.67, -0.7)} {}  %
      \myvertex {(a10)} {at (0.67, 0.7)} {}  %
      \myvertex[dot,fill=darkgreen] {(a11)} {at (0, -0.35)} {}  %
      \myvertex[dot,fill=darkgreen] {(a12)} {at (0, 0.35)} {}  %

      \mypropagator[feyn propagator] {(mid1)} {-- (a1)}  %
      \mypropagator[feyn propagator] {(mid1)} {-- (a2)}  %
      \mypropagator[feyn propagator] {(mid2)} {-- (a3)}  %
      \mypropagator[feyn propagator] {(mid3)} {-- (a4)}  %
      \mypropagator[feyn propagator] {(mid4)} {-- (a5)}  %
      \mypropagator[feyn propagator] {(mid4)} {-- (a6)}  %
      \mypropagator[feyn propagator] {(mid1)} {-- (mid2)} %
      \mypropagator[feyn propagator] {(mid2)} {-- (mid3)} %
      \mypropagator[feyn propagator] {(mid3)} {-- (mid4)} %

      \mypropagator[scalar,ultra thick,draw=cutred] {(a7)} {-- (a8)}   %
      \mypropagator[scalar,ultra thick,draw=cutred] {(a9)} {-- (a10)}  %
      \mypropagator[ultra thick,draw=darkgreen] {(a11)} {to[bend right=45] (a12)} %
    \end{myfeynmandiagram}%
  }%
}

\newcommand{\sixGraphNMCrightU}[6]{%
  {%
    \begin{myfeynmandiagram}[
        scale=1.4, %
        baseline=(current bounding box.center)
      ]
      \myvertex {(a1)} {at (-1, -0.7)} {\(#1\)} %
      \myvertex {(a2)} {at (-1, 0.7)} {\(#2\)}  %
      \myvertex {(a3)} {at (-0.33, 0.7)} {\(#3\)} %
      \myvertex {(a4)} {at (0.33, 0.7)} {\(#4\)}  %
      \myvertex {(a5)} {at (1, 0.7)} {\(#5\)}    %
      \myvertex {(a6)} {at (1, -0.7)} {\(#6\)}   %

      \myvertex[feyn vertex] {(mid1)} {at (-1,0)} {}    %
      \myvertex[feyn vertex] {(mid2)} {at (-0.33,0)} {}  %
      \myvertex[feyn vertex] {(mid3)} {at (0.33,0)} {}   %
      \myvertex[feyn vertex] {(mid4)} {at (1,0)} {}     %

      \myvertex {(a7)} {at (-0.67, -0.7)} {}  %
      \myvertex {(a8)} {at (-0.67, 0.7)} {}   %
      \myvertex {(a9)} {at (0, -0.7)} {}  %
      \myvertex {(a10)} {at (0, 0.7)} {}  %
      \myvertex[dot,fill=darkgreen] {(a11)} {at (.66, -0.35)} {}  %
      \myvertex[dot,fill=darkgreen] {(a12)} {at (.66, 0.35)} {}  %

      \mypropagator[feyn propagator] {(mid1)} {-- (a1)}  %
      \mypropagator[feyn propagator] {(mid1)} {-- (a2)}  %
      \mypropagator[feyn propagator] {(mid2)} {-- (a3)}  %
      \mypropagator[feyn propagator] {(mid3)} {-- (a4)}  %
      \mypropagator[feyn propagator] {(mid4)} {-- (a5)}  %
      \mypropagator[feyn propagator] {(mid4)} {-- (a6)}  %
      \mypropagator[feyn propagator] {(mid1)} {-- (mid2)} %
      \mypropagator[feyn propagator] {(mid2)} {-- (mid3)} %
      \mypropagator[feyn propagator] {(mid3)} {-- (mid4)} %

      \mypropagator[scalar,ultra thick,draw=cutred] {(a7)} {-- (a8)}   %
      \mypropagator[scalar,ultra thick,draw=cutred] {(a9)} {-- (a10)}  %
      \mypropagator[ultra thick,draw=darkgreen] {(a11)} {to[bend right=45] (a12)} %
    \end{myfeynmandiagram}%
  }%
}

\newcommand{\sixGraphNNMCleftU}[6]{%
  {%
    \begin{myfeynmandiagram}[
        scale=1.4, %
        baseline=(current bounding box.center)
      ]
      \myvertex {(a1)} {at (-1, -0.7)} {\(#1\)} %
      \myvertex {(a2)} {at (-1, 0.7)} {\(#2\)}  %
      \myvertex {(a3)} {at (-0.33, 0.7)} {\(#3\)} %
      \myvertex {(a4)} {at (0.33, 0.7)} {\(#4\)}  %
      \myvertex {(a5)} {at (1, 0.7)} {\(#5\)}    %
      \myvertex {(a6)} {at (1, -0.7)} {\(#6\)}   %

      \myvertex[feyn vertex] {(mid1)} {at (-1,0)} {}    %
      \myvertex[feyn vertex] {(mid2)} {at (-0.33,0)} {}  %
      \myvertex[feyn vertex] {(mid3)} {at (0.33,0)} {}   %
      \myvertex[feyn vertex] {(mid4)} {at (1,0)} {}     %

      \myvertex {(a7)} {at (-0.67, -0.7)} {}  %
      \myvertex {(a8)} {at (-0.67, 0.7)} {}   %
      \myvertex[dot,fill=darkgreen] {(a9)} {at (0, -0.35)} {}  %
      \myvertex[dot,fill=darkgreen] {(a10)} {at (0, 0.35)} {}  %
      \myvertex[dot,fill=darkgreen] {(a11)} {at (0.67, -0.35)} {}  %
      \myvertex[dot,fill=darkgreen] {(a12)} {at (0.67, 0.35)} {}  %

      \mypropagator[feyn propagator] {(mid1)} {-- (a1)}  %
      \mypropagator[feyn propagator] {(mid1)} {-- (a2)}  %
      \mypropagator[feyn propagator] {(mid2)} {-- (a3)}  %
      \mypropagator[feyn propagator] {(mid3)} {-- (a4)}  %
      \mypropagator[feyn propagator] {(mid4)} {-- (a5)}  %
      \mypropagator[feyn propagator] {(mid4)} {-- (a6)}  %
      \mypropagator[feyn propagator] {(mid1)} {-- (mid2)} %
      \mypropagator[feyn propagator] {(mid2)} {-- (mid3)} %
      \mypropagator[feyn propagator] {(mid3)} {-- (mid4)} %

      \mypropagator[scalar,ultra thick,draw=cutred] {(a7)} {-- (a8)}   %
      \mypropagator[ultra thick,draw=darkgreen] {(a9)} {to[bend right=45] (a10)} %
      \mypropagator[ultra thick,draw=darkgreen] {(a11)} {to[bend right=45] (a12)} %
    \end{myfeynmandiagram}%
  }%
}

\newcommand{\sixGraphNNMCmiddleU}[6]{%
  {%
    \begin{myfeynmandiagram}[
        scale=1.4, %
        baseline=(current bounding box.center)
      ]
      \myvertex {(a1)} {at (-1, -0.7)} {\(#1\)} %
      \myvertex {(a2)} {at (-1, 0.7)} {\(#2\)}  %
      \myvertex {(a3)} {at (-0.33, 0.7)} {\(#3\)} %
      \myvertex {(a4)} {at (0.33, 0.7)} {\(#4\)}  %
      \myvertex {(a5)} {at (1, 0.7)} {\(#5\)}    %
      \myvertex {(a6)} {at (1, -0.7)} {\(#6\)}   %

      \myvertex[feyn vertex] {(mid1)} {at (-1,0)} {}    %
      \myvertex[feyn vertex] {(mid2)} {at (-0.33,0)} {}  %
      \myvertex[feyn vertex] {(mid3)} {at (0.33,0)} {}   %
      \myvertex[feyn vertex] {(mid4)} {at (1,0)} {}     %

      \myvertex {(a7)} {at (0, -0.7)} {}  %
      \myvertex {(a8)} {at (0, 0.7)} {}   %
      \myvertex[dot,fill=darkgreen] {(a9)} {at (-0.67, -0.35)} {}  %
      \myvertex[dot,fill=darkgreen] {(a10)} {at (-0.67, 0.35)} {}  %
      \myvertex[dot,fill=darkgreen] {(a11)} {at (0.67, -0.35)} {}  %
      \myvertex[dot,fill=darkgreen] {(a12)} {at (0.67, 0.35)} {}  %

      \mypropagator[feyn propagator] {(mid1)} {-- (a1)}  %
      \mypropagator[feyn propagator] {(mid1)} {-- (a2)}  %
      \mypropagator[feyn propagator] {(mid2)} {-- (a3)}  %
      \mypropagator[feyn propagator] {(mid3)} {-- (a4)}  %
      \mypropagator[feyn propagator] {(mid4)} {-- (a5)}  %
      \mypropagator[feyn propagator] {(mid4)} {-- (a6)}  %
      \mypropagator[feyn propagator] {(mid1)} {-- (mid2)} %
      \mypropagator[feyn propagator] {(mid2)} {-- (mid3)} %
      \mypropagator[feyn propagator] {(mid3)} {-- (mid4)} %

      \mypropagator[scalar,ultra thick,draw=cutred] {(a7)} {-- (a8)}   %
      \mypropagator[ultra thick,draw=darkgreen] {(a9)} {to[bend right=45] (a10)} %
      \mypropagator[ultra thick,draw=darkgreen] {(a11)} {to[bend right=45] (a12)} %
    \end{myfeynmandiagram}%
  }%
}

\newcommand{\sixGraphNNMCrightU}[6]{%
  {%
    \begin{myfeynmandiagram}[
        scale=1.4, %
        baseline=(current bounding box.center)
      ]
      \myvertex {(a1)} {at (-1, -0.7)} {\(#1\)} %
      \myvertex {(a2)} {at (-1, 0.7)} {\(#2\)}  %
      \myvertex {(a3)} {at (-0.33, 0.7)} {\(#3\)} %
      \myvertex {(a4)} {at (0.33, 0.7)} {\(#4\)}  %
      \myvertex {(a5)} {at (1, 0.7)} {\(#5\)}    %
      \myvertex {(a6)} {at (1, -0.7)} {\(#6\)}   %

      \myvertex[feyn vertex] {(mid1)} {at (-1,0)} {}    %
      \myvertex[feyn vertex] {(mid2)} {at (-0.33,0)} {}  %
      \myvertex[feyn vertex] {(mid3)} {at (0.33,0)} {}   %
      \myvertex[feyn vertex] {(mid4)} {at (1,0)} {}     %

      \myvertex {(a7)} {at (0.67, -0.7)} {}  %
      \myvertex {(a8)} {at (0.67, 0.7)} {}   %
      \myvertex[dot,fill=darkgreen] {(a9)} {at (-0.67, -0.35)} {}  %
      \myvertex[dot,fill=darkgreen] {(a10)} {at (-0.67, 0.35)} {}  %
      \myvertex[dot,fill=darkgreen] {(a11)} {at (0, -0.35)} {}  %
      \myvertex[dot,fill=darkgreen] {(a12)} {at (0, 0.35)} {}  %

      \mypropagator[feyn propagator] {(mid1)} {-- (a1)}  %
      \mypropagator[feyn propagator] {(mid1)} {-- (a2)}  %
      \mypropagator[feyn propagator] {(mid2)} {-- (a3)}  %
      \mypropagator[feyn propagator] {(mid3)} {-- (a4)}  %
      \mypropagator[feyn propagator] {(mid4)} {-- (a5)}  %
      \mypropagator[feyn propagator] {(mid4)} {-- (a6)}  %
      \mypropagator[feyn propagator] {(mid1)} {-- (mid2)} %
      \mypropagator[feyn propagator] {(mid2)} {-- (mid3)} %
      \mypropagator[feyn propagator] {(mid3)} {-- (mid4)} %

      \mypropagator[scalar,ultra thick,draw=cutred] {(a7)} {-- (a8)}   %
      \mypropagator[ultra thick,draw=darkgreen] {(a9)} {to[bend right=45] (a10)} %
      \mypropagator[ultra thick,draw=darkgreen] {(a11)} {to[bend right=45] (a12)} %
    \end{myfeynmandiagram}%
  }%
}

 \newcommand{\OpPromotion}{\oop{\mathcal{O}}}

\declareslashed{}{\big/}{0.15}{0}{\mathcal{A}}
\declareslashed{}{\big/}{0.15}{0}{\mathcal{L}}

\title{The double copy effective action: a quantum (chromodynamics) approach to space-time}
\author[a,b]{John Joseph M. Carrasco}
\author[a]{Suna Zekio\u{g}lu}

\affiliation[a]{The Amplitudes and Insights Group, Department of Physics \& Astronomy, Northwestern University, Evanston, Illinois 60208, USA}
\affiliation[b]{Center for Interdisciplinary Exploration and Research in Astrophysics (CIERA), Northwestern University, 1800 Sherman Ave, Evanston, IL 60201, USA}

\date{\today}
\abstract{
Conventional Lagrangian formulations of gauge and gravity theories emphasize compactness and off-shell symmetry.  This often obscures the structure of on-shell physical observables. In this work, we present a constructive framework that elevates gauge-invariant scattering amplitudes to  the defining data for quantum field theory actions, including effective field theories. Focusing on  double-copy theories, we promote color-dual amplitude numerators to quantum operators. This enables the systematic identification of novel local operator content at each multiplicity and the construction of double-copy-compatible actions. 
By applying this framework to the well-established double-copy relationship between Einstein gravity and Yang-Mills theory, which holds for all-multiplicity tree-level amplitudes, we demonstrate a systematic path to constructing the operator expansion of $\sqrt{-g}R$ from factorized gauge-theory components. This clarifies how gravitational interactions can be understood as emerging from simpler gauge-theoretic structures at the action level.
This formalism extends color-kinematics duality from amplitude data to operator constructions, naturally realizing the double copy at the level of actions and asymptotic quantum states. We illustrate the method with Yang-Mills theory, Einstein gravity, and its application to generating higher-derivative operators inspired by Z-theory and open superstring amplitudes.
This work provides a concrete bridge between structured amplitudes and effective actions, offering a physically grounded alternative to traditional EFT basis-building. It reveals at the operator level deep structural connections between gauge theory and gravity (connections long recognized in scattering amplitudes) from fundamental interactions to their quantum state descriptions and higher-derivative extensions.}

\begin{document}
\maketitle
\newpage

\section{Introduction}

Recent advances in explicit scattering amplitude constructions~\cite{Carrasco:2019yyn,Carrasco:2021ptp,Carrasco:2023wib} have shown that it is straightforward to bootstrap color-dual higher-derivative corrections to all orders in mass dimension --- at least at low multiplicity. But what local operators should be added to the action to generate these bootstrapped amplitudes?

The conventional route is to postulate field-level ans\"atze and fix their coefficients by matching to known amplitude data. While this approach can succeed at low multiplicity or when guided by symmetry constraints, it becomes increasingly inefficient and opaque at higher orders. Previous efforts to incorporate color-kinematics duality at the Lagrangian level have often focused on constructing effective Yang-Mills Lagrangians whose Feynman rules are designed to directly yield BCJ-satisfying kinematic numerators for all graph contributions (e.g., \cite{Bern:2010yg,Tolotti:2013caa, Cheung:2016prv}). Such approaches typically involve adding either auxiliary fields or higher-valency interaction terms to the standard Lagrangian, which may be non-local and are often defined recursively such that the full Lagrangian remains on-shell equivalent to Yang-Mills but is expressed in a form that makes the duality manifest at the level of Feynman rules. 

Additionally significant insights into the origins and off-shell validity of color-kinematics duality also come from analyzing specific action principles in various frameworks. Examples include the pure spinor formalism for supersymmetric Yang-Mills theories~\cite{Ben-Shahar:2021doh}, where dual numerators emerge from the action's structure, and studies of Chern-Simons theory~\cite{Ben-Shahar:2021zww}, where kinematic algebras satisfying Jacobi identities can be identified directly from the action, leading to off-shell CK duality for currents and correlators. These formalism-centric approaches further our understanding of how such dualities can be inherent, even off-shell, properties of particular fundamental actions.

Our approach differs significantly in its philosophy and goals. We take well-structured, color-dual (or double-copy) amplitudes as the primary input. Our framework then provides a systematic method to derive the minimal set of local operators in an effective action that are necessary to reproduce these physical observables. We achieve this by sharply distinguishing novel $m$-point contact interactions from contributions reconstructible via unitarity from lower-multiplicity amplitudes. While color-kinematics duality is a crucial guiding principle that ensures the consistency and structural integrity of our input amplitudes (and is essential for applications like the gravitational double copy), our method does not aim to generate specific numerator forms from Feynman rules. Instead, it translates the physical information encoded in these already-structured numerators directly into local operator terms in the action. This allows us to maintain locality in the fundamental operator additions at each order and provides a direct bridge from on-shell S-matrix data to a standard effective action. 

\begin{figure}[h]
\centering
\includegraphics[width=0.8\textwidth]{"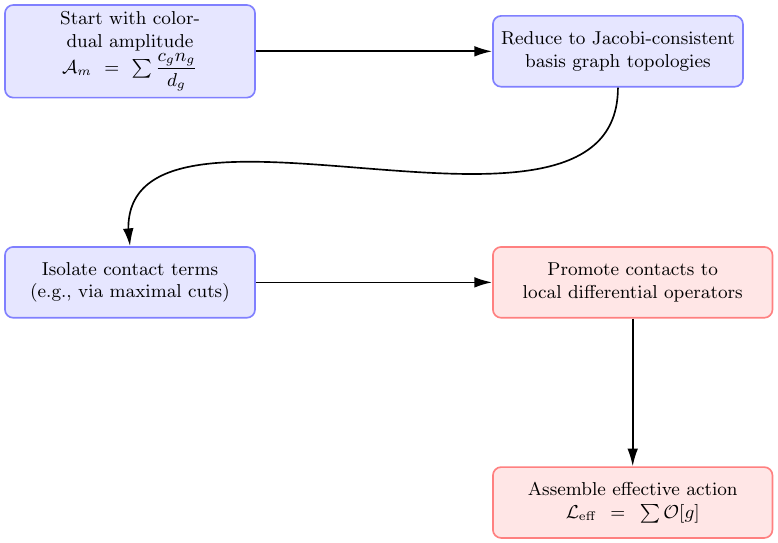"}
\caption{
\textbf{From amplitudes to operators.} The operator promotion procedure begins with a color-dual amplitude, reduces to a basis of cubic graphs using Jacobi identities, isolates local contact terms (e.g., via maximal cut 2), promotes each numerator to a field-space operator, and assembles the resulting operators into the effective action. The structure is preserved under double copy.
}
\label{fig:introFlow}
\end{figure}

Specifically, we describe a promotion procedure, \cref{fig:introFlow}, that maps each graph contribution in an \(m\)-point amplitude to a corresponding local \(m\)-field operator. This approach preserves color-kinematics duality and, when applicable, maintains manifest double-copy structure. While the method is particularly well-suited to constructing higher-derivative color-dual operators, it also provides a path to rewriting entire double-copy theories in terms of their color-dual graph basis.   This S-matrix-driven approach to constructing actions has important historical precedents. Of related prior work, we find the closest in spirit and approach to be the landmark work of Bern and Grant~\cite{Bern:1999ji} which, preceding the convenience of color-dual double-copy, still constructed higher-multiplicity gravity amplitudes from Yang-Mills amplitudes  via the $\alpha' \to 0$ KLT relations~\cite{Kawai:1985xq, Bern:1998sv} and subtracted all cuts manually to achieve incredibly compact representations of the Lagrangian level graviton contact terms through  five points.   

The operator promotion procedure we present utilizes a systematic method,
closely related to a generalization of maximal cuts, to isolate the novel
contact information contained in amplitude data at each multiplicity.  In
doing so, it enforces locality and removes redundancy associated with
lower-order physics.  This streamlines operator construction while making
the double-copy structure of the resulting actions manifest.  Operators
derived in this way inherit the symmetries of the input amplitudes.

It is important to emphasize that this construction does not define a new
S-matrix prescription.  Rather, it provides an inverse map from on-shell
amplitudes to an action with manifest double-copy structure.  The resulting
action is fully compatible with standard methods for extracting physical
predictions, including perturbiner expansions of the equations of motion,
identifying Berends-Giele currents, as well as conventional path-integral 
approaches.

This framework finds a particularly compelling application in relating Yang-Mills theory to Einstein gravity. Given that tree-level gravitational scattering amplitudes are understood to be constructible as a double copy of Yang-Mills amplitudes to all multiplicities, and that color-dual representations for Yang-Mills numerators are systematically available, our method provides a direct bridge to the corresponding operator structure. Specifically, it furnishes a constructive algorithm for deriving the operator expansion of the Einstein-Hilbert action, $\sqrt{-g}R$, from factorized Yang-Mills building blocks. This process reveals how the complete tower of gravitational interactions can be systematically assembled from gauge-theoretic components at the level of the action.  This finds a particular resonance in the class of twofold-symmetry representations of the Einstein-Hilbert action of Cheung and Remmen~\cite{Cheung:2016say}.  Reference~\cite{Cheung:2016say} presents a well-structured set of gauge transformations and gauge fixing so that `left'-indices only contract with `left'-indices, and `right'-indices only contract with `right'-indices.  While motivated by double-copy they did not establish `left' and `right' with distinct Yang-Mills-type operators.  Here such a structure is clear --- although we identify gravitational states in a more traditional gauge involving a symmetrization and projection out of any dilatonic trace, and explicitly use propagators that enforce such a projection to physical graviton states.

Beyond this result for fundamental forces, our approach readily extends to constructing higher-derivative operators, and we illustrate its application to structures found in Z-theory and open superstring amplitudes. Furthermore, we explore how this operator-level double copy informs a consistent state-level encoding for quantum gravity.

This work finds itself at the intersection of a number of fields. For amplitudes practitioners, it offers a concrete and structurally faithful map from bootstrapped amplitude data to local operators. For effective field theorists, it provides an efficient alternative to traditional ansatz- and basis-building methods, yielding manifestly gauge-invariant, double-copy-compatible operators imported from their natural habitat. For researchers interested in the foundational structure of QFT and gravity, the method opens a path toward reconstructing semi-classical gravitational actions from gauge-theoretic constituents --- an open invitation to flat-space holography. Finally, we hope to present the method with sufficient procedural clarity and pedagogical grounding to serve as a practical and easily automatable tool.

This paper is organized as follows. In \cref{sec:review}, we review the duality between color and kinematics and the graph-based double-copy structure shared by many theories. In \cref{sec:simpleScalar}, we introduce a pedagogical covariantized free-scalar bootstrap to illustrate our method in a transparent setting. \Cref{sec:contactExtraction} describes how contact terms are isolated from amplitude data using a generalized maximal-cut strategy. \Cref{sec:promotion} presents the core operator promotion procedure, mapping Jacobi-consistent graph numerators to local field-space operators. In \cref{sec:examples}, we apply the method across a range of theories --- gauge, string-inspired, and gravitational --- highlighting its efficiency and generality. In \cref{sec:doublecopyGrav} we lay out the path to quantum gravity via double copy, first highlighting how every YM tree-level amplitude encodes the necessary Yang-Mills gravity contact required by $\sqrt{-g}R$ at that multiplicity, then laying out a dictionary for gravity states as double-copied Yang-Mills states. We conclude in \cref{sec:outlook} with discussion of broader applications and structural implications.

\section{Review}
\label{sec:review}
\subsection{Color-dual representations}
The duality between color-and-kinematics originally identified at tree-level~\cite{Bern:2008qj} and soon thereafter generalized to the multiloop integrand level~\cite{Bern:2010ue} as well as perturbative and complete classical solutions to the equation of motion~\cite{Monteiro:2014cda, Luna:2015paa,Luna:2016due, Kosower:2018adc} now bridges many aspects of physics from particle physics, to string theory, to mathematical physics, and from gravitational wave astrophysics, to inflationary cosmology.  We would not be able to do justice to the field here, and instead defer to a number of tutorials and reviews~\cite{Elvang:2013cua, Carrasco:2015iwa, Bern:2019prr, Borsten:2020bgv,Adamo:2022dcm,  Bern:2022wqg, Bern:2023zkg}.

In this paper we will focus on the most familiar type of color-dual representations --- adjoint antisymmetric where kinematics obey the same structural relations as color-weights dressed with antisymmetric adjoint color factors ($f^{abc}$).  Namely this will mean satisfying Jacobi and antisymmetry.  This is sufficient to bootstrap Yang-Mills and the Nonlinear Sigma Model to all multiplicity, and when allowing additional color-structures to combine with kinematics allows for many higher derivative interactions including the open and closed bosonic and superstring theories at tree-level.  The methods presented here generalize trivially to double-copies that require symmetric structure constants ($d^{abc}$)~\cite{Carrasco:2022jxn}, but that will not be the main focus of this paper. 

All scattering amplitudes can be expressed in terms of cubic (trivalent) graphs. Higher point contact contributions can be absorbed into cubic graph dressings by including relevant inverse propagators.  At tree level for $m$ particles scattering, we need consider a maximum of $(2m-5)!!$ distinct cubic graphs, $\Gamma^m_3$.  For the Yang-Mills theory we map these graphs to color-weights (dressing all vertices with $f^{abc}$ structure constants) $c_g$, kinematic-weights (functions of momenta and polarizations) $n_g$, and propagators $1/d_g$.  As such the full amplitude is given by:
\begin{equation}
\label{cubicAmp}
\mathcal{A}_m = \sum_{g\in \Gamma^m_3}  \frac{c_g n_g}{d_g}\,
\end{equation}
where we have suppressed the coupling whose power goes as $m-2$ with multiplicity $m$.
This set grows factorially with  $m$ because of all the different labels each topology can have.  While it is possible to choose individual mappings for each distinctly labeled graph, this is in some sense artificial for gluons whose polarizations have not yet been specified.

Fortunately there are only an exponential number\footnote{\href{https://oeis.org/A000672}{A000672}.} of distinct cubic topologies, $\mathcal{T}^m_3$, at each multiplicity. Hence we will consider  a single functional dressings for  each graph's topology.  We can therefore dress each topology and sum over permutations of labels to capture all $(2m-5)!!$ channels,
\begin{equation}
\mathcal{A}_m = \sum_{g\in \mathcal{T}^m_3}  S_g \frac{  c_g n_g}{d_g} + \text{permutations} \,.
\end{equation}
We have introduced symmetry factors $S_n$ to account for the overcounting of relabeling due to each topology's symmetry. 

\begin{figure}[t]
\centering
\includegraphics[width=0.8\textwidth]{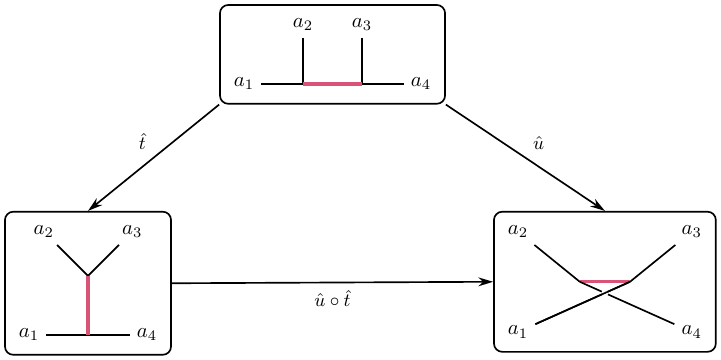}
\caption{
\textbf{Jacobi moves at four points.}
Each trivalent graph at four-point multiplicity is related to the others by a local graph transformation, shown as an edge in this graph-of-graphs. These moves correspond to the algebraic Jacobi identities in both color and kinematics. The cyclic ordering of the arrows corresponds to choosing an orientation convention for the antisymmetric vertex structure.
}
\label{fig:whitehead4pt}
\end{figure}

Each one of the distinct topologies in $\mathcal{T}^m_3$ can be expressed as a finite number of Jacobi moves\footnote{Sometimes called Whitehead moves~\cite{rafi2013,Carrasco:2015iwa,Bern:2019prr}.} (see \cref{fig:whitehead4pt}) from a single basis graph topology, the so called half-ladder $m$-point graph,
\begin{equation}
\label{aGraphDef}
b_{m} = \mgraph
\end{equation}
 This is a familiar concept for color-weights and was used by Dixon, Del Duca, and Maltoni to prove the Kleiss-Kuijif field theory relations.

By way of example, consider the first occasion for a non-half-ladder graph, the trimerous topology that contributes at 6-points:
\begin{equation}
g_{\text{tri}}=\sixtrim{1}{2}{3}{4}{5}{6}
\end{equation}

Antisymmetric adjoint structure constant ($f^{abc}$) based color weights $c(g_{\text{tri}})$ satisfy a Jacobi relation with the color weights of the two differently labeled half-ladder graphs:
\begin{equation}
c \circ \sixtrim{a}{b}{c}{d}{e}{f} = c \circ \sixgraph{a}{b}{c}{d}{e}{f} -  c \circ \sixgraph{a}{b}{d}{c}{e}{f} \,.
\end{equation}

Color-dual representations of Yang-Mills amplitudes simply have the kinematic weights satisfy exactly the self-same Jacobi and antisymmetry rules, so the kinematic numerators $n(g_{\text{tri}})$ will be given as the difference between the relabeled basis graphs. 
\begin{equation}
n \circ \sixtrim{a}{b}{c}{d}{e}{f} = n \circ \sixgraph{a}{b}{c}{d}{e}{f} -  n \circ \sixgraph{a}{b}{d}{c}{e}{f} \,.
\end{equation}
The fact that adjoint color-dual kinematic numerators also satisfy Jacobi imposes additional relations on ordered amplitudes known as BCJ relations to a basis of $(n-3)!$ orderings.

For theories color-dual to $f^{abc}$ color, we write every tree-level amplitude as:
\begin{equation}
\label{scatteringGauge}
\mathcal{A}_m =  S(b_{m})  \frac {c(b_{m}) n(b_{m}) }{d(b_{m})} + \text{Jacobi moves} + \text{permutations} \,.
\end{equation}
We will give various explicit kinematic weights $n(b_{m})$ for a variety of theories in later sections.  At this time perhaps it can help ground the discussion to generically offer the propagator and adjoint color weights associated with $b_{m}$. 

For theories like Yang-Mills the color-weights are simply given by dressing every vertex with a gauge theory structure constant labelled according to the graph,
\begin{equation}
c(b_{m}) = f^{e_1 e_2 i_1} f^{i_1 e_3 i_2} \cdots f^{i_{m-3} e_{m-1} e_m}\,.
\end{equation}
For convenience we distinguish between external color labels $e_j$ and internal color-labels $i_k$.

For massless fields of any spin, with an all outgoing momentum convention, we can dress the propagators of the half-ladder, $b_m$,
\begin{equation}
d(b_{m}) = (k_1+k_2)^2 (k_1+k_2+k_3)^2 \cdots (k_{m-2}+k_{m-1}+k_m)^2 (k_{m-1} +k_m)^2 \,.
\end{equation}

Famously, local individual kinematic weights $n_g$ are not generically gauge invariant.  There must be cancelations between distinct channels.  Algebraic relations between color-weights $c_g$ associated with distinct channels ensure the gauge invariance of the full amplitude.  Therefore we can preserve this gauge invariance by replacing the $c_g$ with any other graph weights $\tilde{n}_g$ that obey the same algebraic relations --- this is known as taking the double-copy between two theories.  Critically that means removing a weight in the numerator from each (usually a color-weight) and then combining the remaining kinematic weights,
\begin{equation}
\label{scatteringDoubleCopy}
\mathcal{A}_m =  S(b_{m})  \frac {\tilde{n}(b_{m}) n(b_{m}) }{d(b_{m})} + \text{Jacobi moves} + \text{permutations} \,.
\end{equation}
Taking both copies to be vector (spin-1) kinematic weights we arrive at amplitudes for gravitational (spin-2) theories. Linearized diffeomorphism invariance emerges from the double-copy of linearized gauge invariance~\cite{Bern:2019prr}. 

Generically we will write tree-level amplitudes in theories that participate in the {\em double-copy} web of theories as follows:
\begin{equation}
\mathcal{A}^{A\otimes B}_m  = S(b_{m})  \frac {n_A(b_{m}) n_B(b_{m}) }{d(b_{m})} + \text{Jacobi moves} + \text{permutations}
\end{equation}
In each copy $n_A$, every external field contributes a particular little-group weight corresponding to its representation of the Lorentz group.  This is manifested in arbitrary $D$-dimensions via formal polarization vectors and formal spinors.   Here we consider double-copy amplitudes for theories of maximum spin-2, so each copy every external leg can contribute maximum spin-1 little-group weight.

\subsection{Higher-derivative compositional bootstrap}

The compositional bootstrap of \cite{Carrasco:2019yyn,Carrasco:2021ptp,Carrasco:2022jxn} provides a systematic framework for constructing and classifying the predictions of higher derivative operators in gauge and gravity theories. This approach leverages color-kinematics duality to efficiently encode higher derivative corrections to all orders while maintaining manifest double-copy structure. The key insight is that the functional forms associated with various graphical algebraic structures can be composed to generate additional functional forms of higher mass dimension.  The algebraic properties of color and kinematic numerators, such as Jacobi identities and vertex symmetries, are crucial for ensuring the overall consistency and physical properties of the resulting amplitudes, including Bose symmetry where applicable.

With a scalar graph weight of linear order one can create a ladder to generate all higher order scalar interactions all the way to the ultra-violet. Antisymmetric $f^{abc}$ structure constants generate color factors that participate in the algebraic relations that characterize the original duality between color-and kinematics. 

We summarize some of the results that will refer to directly, but refer the interested reader to the above references for more details and context.  We require three antisymmetric and Jacobi-satisfying building blocks to exhaust all such higher-derivative modifications to single-trace color-weights at four-points --- up to products of scalar permutation invariants.  As such any arbitrary higher derivative generalization of $c_s\equiv f^{a_1 a_2 e} f^{e a_3 a_4}$, is spanned by,
\begin{equation} 
\label{gaugeSoln}
 \boldit{c}_s  = \sum_{i} {\alpha'}^i (a_{XY} \cs{X}{Y} +  a^{ss}_{XY} \css{X}{Y}+a^{\rm d}_{XY} \cnlsm{X}{Y} )\,.
\end{equation} 
The three building blocks are as follows\,
\begin{align}
 \boldit{c}^{(X,Y)}_s &\equiv c_s   \pcub^X \psq^Y\ap^{3X+2Y}\,,
 \label{trivialColor} \\
\boldit{c}^{(X,Y,ss)}_s &\equiv  \left( c_t (u-s) + c_u (s-t) \right)  \pcub^X \psq^Y  \ap^{1+3X+2Y}\, .
\label{colorAndKinJac} \\
\cnlsm{X}{Y} {} &\equiv d^{abcd} s(u-t) \pcub^X \psq^Y \ap^{(2+3X+2Y)}\, .
\label{cNLSM}
\end{align}
We introduce the mass-dimension carrying $\alpha'$, the other antisymmetric channel adjoint weights via relabeling,  $c_t = f^{a_4 a_1 e} f^{a_2 a_3 e}$ and $c_u = f^{a_3 a_1 e} f^{a_4 a_2 e}$,  and admit the fully permutation invariant color $d^{abcd} = \frac{1}{3!} \sum_{\sigma \in S_3(b,c,d)} \rm{Tr}(T^{a} T^{\sigma_1}T^{\sigma_2}T^{\sigma_3})\,$.  The other higher derivative channels follow $\boldit{c}_s$ via relabeling, and all higher-derivative color-numerators satisfy antisymmetry about each vertex, and Jacobi as written,
\be
\boldit{c}_s = \boldit{c}_t + \boldit{c}_u \,.
\ee
This allows us to lift the eight adjoint vector color-weights to all orders in Mandelstams spanning every higher-derivative (parity preserving) vector weight that can be written as an antisymmetric adjoint double-copy.   One can complete all higher derivative vector weights by admitting kinematic and color weights that obey symmetric-adjoint double-copy.  One can of course carry out a similar program for parity odd higher derivative operators by fixing in dimensions and admitting dual vector dressings.  

Having established how tree-level scattering amplitudes in a wide web of theories can be expressed in a color-dual or double-copy form using a basis of graph numerators, including all-orders higher-derivative corrections, we now turn to the central challenge: constructing the corresponding local operators in an effective action that generate these structured amplitudes.

\section{An Invitation: The Simple Scalar}
\label{sec:simpleScalar}
We will illustrate our method for writing down actions using the example of the covariantized free scalar, a theory of adjoint scalars and gluons that we will define in two ways: an amplitudes-based approach that defines the theory by its physical properties and a traditional approach of writing down gauge-invariant operators. This will serve as a pedagogic invitation to our method of operator promotion, introducing the ideas one by one within the example, later to be generalized.

\subsection{Defining the simple scalar}
\subsubsection{Bootstrapping the theory from physical principles}
We define the simple scalar to be a \textit{gauge-invariant, color-dual theory of a real adjoint scalar minimally coupled to Yang-Mills}, at the lowest mass dimension possible. These basic physical principles immediately allow us to write down three-point and four-point amplitudes in this theory, which tell us what operators need to be present in the theory.

We can write down the three-point amplitude of two scalars and one gluon by constraining an ansatz to satisfy the Ward identity on the gluon leg and Bose invariance: 
\begin{equation}
\mathcal{A}_{\varphi\varphi g} \equiv \mathcal{A}\left(\varphi^a(k_1), \varphi^b(k_2), g^c(k_3) \right) = \alpha f^{abc} \left( k_2 \cdot \varepsilon_3  \right)
\end{equation}
\begin{figure}[htbp]
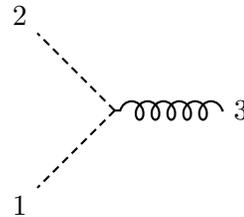
 
    \centering
    \threeGraph{1}{2}{3}
    \caption{$\varphi \varphi g$ vertex} 
    \label{figure:3ptvertex}
\end{figure}
where $\alpha$ is a free parameter that will later be fixed from considerations of gauge invariance and consistent factorization at higher multiplicity. An ansatz at lowest mass dimension for an amplitude of one scalar interacting with two gluons admits no gauge invariant solutions, so we do not consider such a three-point interaction as part of the simple scalar theory. 

\begin{figure}
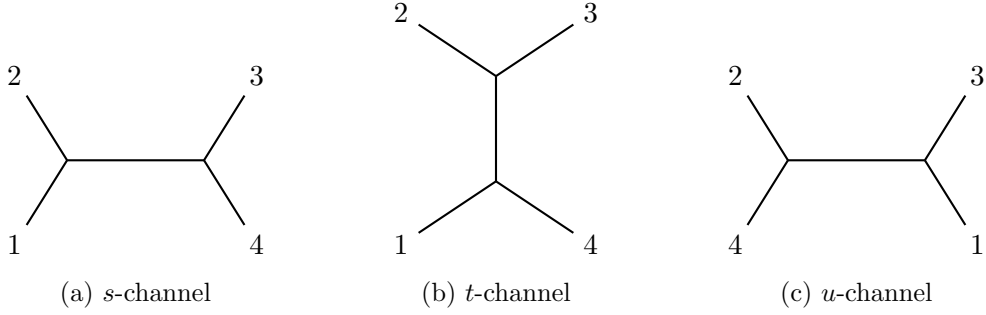

\captionsetup[subfigure]{justification=centering}
\centering
\begin{subfigure}[b]{0.3\textwidth}
         \centering
         \fourgraphAct{1}{2}{3}{4}
         \caption{$s$-channel}
     \end{subfigure}
     \begin{subfigure}[b]{0.3\textwidth}
         \centering
         \fourgraphT{1}{2}{3}{4}
         \caption{$t$-channel}
     \end{subfigure}
     \begin{subfigure}[b]{0.3\textwidth}
         \centering
         \fourgraphAct{4}{2}{3}{1}
         \caption{$u$-channel}
     \end{subfigure}
\caption{Standard definitions of $s$, $t$, and $u$ channel cubic four-point graph labelings.} 
\label{figure:standard4ptcubic}
\end{figure}

From the existence of three gluon interactions (from pure Yang-Mills) and two scalar, one gluon interactions, we can consider two distinct four-point amplitudes: two external scalars and two external gluons, or four external scalars. We write the four-scalar case as a sum over the three cubic graph channels (as defined in Figure~\ref{figure:standard4ptcubic}), dressing each graph with adjoint color $c$, a kinematic numerator $n$, and the associated massless cubic propagator $d$: 
\begin{equation} 
\mathcal{A}_{\varphi\varphi\varphi\varphi} = \sum_{g \in \{s,t,u\}} \frac{c_g n_g}{d_g}
\end{equation}
The color factor associated with a graph $g$ with leg labels $(abcd)$ is given simply by the adjoint factor built from structure constant contractions, $c_g = f^{abe}f^{ecd}$. Each graph is dressed with a functional numerator $n$, which we constrain to obey adjoint color-kinematics duality (antisymmetry around each vertex and a Jacobi identity on the internal leg): 
\begin{equation}
\begin{split}
n(abcd) &= -n(bacd)\\
n(abcd) &= -n(abdc)\\
n(abcd) &= +n(dcba)
\end{split}
\end{equation}
\begin{equation}
n(abcd) = n(dabc) + n(dbca) 
\end{equation} 
These conditions fix the form of the numerator to be: 
\begin{equation}
n(abcd) = \beta \left( s_{bc} - s_{ac} \right)
\end{equation} 
where the free parameter $\beta$ can be fixed on factorization of the full four scalar amplitude to be $\beta = \alpha^2 /2$, according to: 
\begin{equation}
\begin{split}
\lim_{s \rightarrow 0} s \mathcal{A}\left(\varphi^a(k_1), \varphi^b(k_2), \varphi^c(k_3), \varphi^d(k_4) \right)  = \sum_{states}  \mathcal{A}\left(\varphi^a(k_1), \varphi^b(k_2), g^{e,(h)}(k_3 + k_4) \right) \\ \times \mathcal{A}\left(\varphi^c(k_3), \varphi^d(k_4), g^{e,(\bar{h})}(k_3 + k_4) \right) 
\end{split}
\end{equation}

Computation of the full amplitude indicates that no new information is present in the four scalar amplitude that was not already contained within the three point vertex (i.e., there is no four scalar contact term necessary to ensure the theory satisfies color-kinematics duality at four points): 
\begin{equation}
\label{eqn:fourScalarAmp} 
\mathcal{A}_{\varphi\varphi\varphi\varphi} = \frac{\alpha^2}{2} \left( \frac{c_s (u - t)}{s} + \frac{c_t (u-s)}{t} + \frac{c_u (s-t)}{u}  \right) \, . 
\end{equation}

We can write down the two scalar, two gluon amplitude $\mathcal{A}(\varphi(k_1),\varphi(k_2),g(k_3),g(k_4))$ from similar considerations. There are now two cubic graph topologies, one with a scalar propagator and one with a vector propagator, defined as follows: 
\begin{equation}
\fourgraphScalarProp{d}{a}{b}{c} \quad \quad  \quad n_{\varphi}(abcd)
\end{equation}
\begin{equation}
\fourgraphVectorProp{a}{b}{c}{d} \quad \quad  \quad n_{\text{v}}(abcd)
\end{equation}
We will proceed in a standard manner: assign each topology an ansatz; constrain each ansatz on its graph symmetries, factorization, and gauge invariance of the constructed full amplitude; and finally relate the topologies using the kinematic Jacobi relation. As is typical in adjoint color-kinematics considerations, this construction does not exclude the possibility of a four point contact term of the form: 
\begin{equation}
\contactFour{a}{b}{c}{d} 
\end{equation}
Rather, if such a contact term is demanded by gauge invariance, it will be generated by this ansatz procedure as assigned to the cubic graphs in a manner consistent with color-dual functional representations, just as the Yang-Mills four point contact can be absorbed onto the cubic graphs by multiplying by appropriate factors of unity ($s/s$, $t/t$, $u/u$). 

The full amplitude can then be written as a sum over the dressings of the relevant cubic graphs --- represented schematically, 
\begin{equation}
\label{eqn:schematic4pt}
\mathcal{A}_{\varphi\varphi g g}  \sim \fourgraphScalarProp{k_4}{k_1}{k_2}{k_3} 
+ \fourgraphScalarProp{k_3}{k_1}{k_2}{k_4} + \fourgraphVectorProp{k_1}{k_2}{k_3}{k_4} \, . 
\end{equation}
In terms of our functional dressings and adjoint color factors, the amplitude is given by: 
\begin{equation}
\mathcal{A}_{\varphi\varphi g g}  = \frac{c_t n_{\varphi}(1234)}{t} - \frac{c_u n_{\varphi}(1243)}{u} + \frac{c_s n_{\text{v}}(1234)}{s} \, .
\end{equation}

Graph symmetries provide the following constraints on the $n_{\varphi}$ and $n_{\text{v}}$ functional numerators (these can be understood from drawing the automorphisms of each topology and considering vertex antisymmetry, in line with color-kinematics duality): 
\begin{equation}
n_{\varphi}(abcd) = +n_{\varphi}(badc)
\end{equation}
\begin{equation}
\begin{split} 
n_{\text{v}}(abcd) &= -n_{\text{v}}(bacd) \\
n_{\text{v}}(abcd) &= -n_{\text{v}}(abdc)
\end{split}
\end{equation}
Such partially-fixed numerator functions are then further constrained to consistently factorize down to appropriate products of three point amplitudes; for example, the $s$-channel cut of our amplitude, described by the appropriate limit of the $n_{\text{v}}$ numerator dressing, can be fixed as follows: 
\begin{equation}
\begin{split}
\lim_{s \rightarrow 0} s \mathcal{A}\left(\varphi^a(k_1), \varphi^b(k_2), g^c(k_3), g^d(k_4) \right)  &= \lim_{s \rightarrow 0} \, c_s n_{\text{v}}(1234) \\ &= \sum_{states}  \mathcal{A}\left(\varphi^a(k_1), \varphi^b(k_2), g^{e,(h)}(k_3 + k_4) \right) \\ &\quad \quad \quad \,\, \times \mathcal{A}\left(g^c(k_3), g^d(k_4), g^{e,(\bar{h})}(k_3 + k_4) \right) 
\end{split}
\end{equation}
where the first three point amplitude arises from our simple scalar theory, and the second from pure Yang-Mills theory, 
\begin{align}
\mathcal{A}_3^{\text{YM}} &= g f^{abc} \left[ (k_1-k_2)\cdot{\varepsilon_3} (\varepsilon_1 \cdot \varepsilon_2) +\text{cyclic}\right]\\
&= -2g f^{abc} \left[
  (k_2\cdot \varepsilon_3)(\varepsilon_1 \cdot \varepsilon_2) +
  (k_3\cdot \varepsilon_1) (\varepsilon_2 \cdot \varepsilon_3)+
  (k_1\cdot \varepsilon_2)( \varepsilon_3 \cdot \varepsilon_1 )  
  \right]
\end{align}
Upon fixing on all distinct cuts, the two scalar, two gluon amplitude $\mathcal{A}_{\varphi\varphi g g}$ contains just two parameters: $g$, the Yang-Mills coupling, and $\alpha$, the unfixed overall coefficient of the three point simple scalar amplitude $\mathcal{A}_{\varphi\varphi g}$. By demanding this amplitude be gauge invariant, $\alpha$ is fixed to $-2g$, consistent with the knowledge that the couplings of particles to non-abelian gauge bosons are determined entirely by the pure gauge theory coupling and the relevant group representation. The fully fixed numerators are as follows: 
\begin{equation}
n_{\varphi}(1234) = - 4 g^2 (k_2\cdot\varepsilon_3)(k_1 \cdot\varepsilon_4) - 2 g^2 (k_2\cdot k_3)(\varepsilon_3\cdot\varepsilon_4)
\end{equation}
\begin{equation}
\begin{split}
n_{\text{v}}(1234) &= \, 4 g^2 (k_1\cdot \varepsilon_3) (k_2\cdot \varepsilon_4) - 
 4 g^2 (k_1\cdot \varepsilon_4) (k_2\cdot \varepsilon_3) \\ &\, \,- 
 4 g^2 (k_2\cdot k_3) (\varepsilon_3\cdot \varepsilon_4) - 
 2 g^2 (k_1\cdot k_2) (\varepsilon_3\cdot \varepsilon_4)
\end{split}
\end{equation}
One can note that each dressing contains a term proportional to its associated cubic propagator, confirming that this theory does indeed require a two scalar, two gluon contact term to preserve gauge invariance that we have assigned to the cubic graphs in a notion consistent with color-kinematics duality. 

These fully fixed numerator dressings automatically satisfy the kinematic Jacobi identity which relates the two distinct graph topologies: 
\begin{equation}
\fourgraphScalarProp{d}{a}{b}{c}  =  \fourgraphScalarProp{c}{a}{b}{d} + \fourgraphVectorProp{a}{b}{c}{d} 
\end{equation}
\begin{equation}
\label{eqn:svJacobi}
n_{\varphi}(abcd) = n_{\varphi}(abdc) + n_{\text{v}}(abcd) 
\end{equation}
This confirms our ability to define this simple scalar theory as a consistent, color-dual, gauge-invariant theory of real adjoint scalars coupled to Yang-Mills.

In summary, by demanding a gauge-invariant, antisymmetric-adjoint color-dual, consistently factorizing theory, we have arrived at the information sufficient to define the simple scalar: 
\begin{equation}
n_{3}(abc) \equiv n \circ \threeGraph{a}{b}{c} = 2 g \left( k_b \cdot \varepsilon_c \right) = g \left(k_b-k_a \right ) \cdot \varepsilon_c
\end{equation} 
\begin{equation}
n_{\varphi}(abcd) \equiv n \circ \fourgraphScalarProp{d}{a}{b}{c} = - 2 g^2 \left[ 2 (k_b\cdot\varepsilon_c)(k_a \cdot\varepsilon_d) + (k_b\cdot k_c)(\varepsilon_c\cdot\varepsilon_d) \right]
\end{equation}
The dressing for graphs with four external scalars is determined completely by the three point numerators (there is no four scalar contact information); the dressing for the two scalar, two gluon graph with a gluon propagator, $n_{\text{v}}$, is determined entirely in terms of the scalar propagator graph dressing $n_{\varphi}$ by the Jacobi relation specified in equation~\ref{eqn:svJacobi}; and the two scalar, two gluon contact term has been absorbed in a color-dual manner onto the aforementioned cubic graph dressings. We can conclude that all amplitudes in this simple scalar theory can be constructed from just these two pieces of information.

\subsubsection{Traditional form of the action}
The simple scalar is a real adjoint scalar minimally coupled to Yang-Mills. We can write down the action as follows, 
\begin{equation}
\label{eqn:traditionalAction}
\mathcal{S} = \int d^d x \left( -\frac{1}{4}\Tr \left( F^2 \right) + \frac{1}{2}\left(D_{\mu}\varphi \right)^{a}\!\left(D^{\mu}\varphi \right)^{a}  \right) 
\end{equation}
where the covariant derivative can be written in terms of the real structure constants $f^{abc}$ of the Yang-Mills SU(N) gauge group, 
\begin{equation}
\left(D_{\mu} \varphi \right)^a = \partial_{\mu} \varphi^a + g f^{abc}  \varphi^b A_{\mu}^c \,
\end{equation}
yielding the familiar Lagrangian density for the covariantized free scalar theory,
\begin{equation} 
\mathcal{L} = -\frac{1}{4}\Tr \left( F^2 \right) + \frac{1}{2} (\partial \varphi)^2 + g f^{abc} (\partial^\mu \varphi^a)  A^b_\mu  \varphi^{c}   
  + \frac{g^2}{2} f^{abe}f^{ecd} \varphi^a A_{\mu}^b  \varphi^c A^{\mu\, ^d} 
\end{equation}
From this definition of the theory, it is clear that the scalar-scalar-gluon and scalar-scalar-gluon-gluon vertices are given explicitly in terms of the Yang-Mills coupling $g$; however, color-kinematics duality is obscured by this form of the action. 

\subsection{Constructing operators}
From what we learned building the amplitudes for this theory, we expect the following schematic terms to appear in the Lagrangian for the simple scalar: 
\begin{equation}
\mathcal{L} = \mathcal{L}_{\text{Yang-Mills}} + \mathcal{L}_{\text{free-scalar}} + \mathcal{L}_{\varphi\varphi g}
+ \mathcal{L}_{\varphi\varphi g g} \, ,
\end{equation}
consistent with the types of terms seen when the traditional action formalism for this theory is expanded. We will take a different approach to constructing our action, by instead using the content of the three-point amplitude $\mathcal{A}_{\varphi\varphi g}$ to write down $\mathcal{L}_{\varphi\varphi g}$, and, likewise, the content of the four-point amplitude $\mathcal{A}_{\varphi\varphi g g}$ to write down $\mathcal{L}_{\varphi\varphi g g}$. 

We will here introduce a method for promoting an amplitude to a corresponding operator in this simplified case, and provide the more general prescription in \cref{sec:promotion}.  This will work by promoting kinematic numerators $n$ and propagators $d$ to operators, denoted symbolically as $\oop{n}$, $\oop{d}$, given in terms of fields and derivatives.  Recall that the amplitude for particle content $\mathcal{P}$ is taken generically to be written a sum over all relevant cubic graphs, each dressed with a kinematic numerator $n_g$, a color factor $c_g$, and the graph's associated cubic propagators $d_g$, 
\begin{equation}
\mathcal{A}_{\mathcal{P}} = \sum_{g \in \Gamma^{(3)}_{\mathcal{P}}} \frac{c_g n_g}{d_g} \, .
\end{equation} 
Promoting  will allow us to write down the relevant contribution to the Lagrangian density by constructing a sum over \textit{distinct four-point trivalent-graph topologies} $\uptau \in  \mathcal{T}^4_3$, each dressed with a corresponding numerator operator, color factor, propagator operator, and symmetry factor (which will be explained shortly): 
\begin{equation}
\label{eqn:Lform}
\mathcal{L}_{\mathcal{P}} = \int \mathcal{D}_{|\mathcal{P}|}{ \sum_{\uptau \in \mathcal{T}^4_3{}_{\mathcal{P}}}  s_{\uptau} \frac{c_{\uptau} \oop{n}_{\uptau}}{\oop{d}_{\uptau}} {\mathcal{P}}}
\end{equation}
where the sum runs over all distinct cubic graph topologies $ \mathcal{T}^4_3{}_{\mathcal{P}}$ available given the relevant field content.  We have the promoted operators acting on field operators represented here by $\mathcal{P}$, and we introduced an integration as a convenience to localize any labeling coordinates we may have given our fields to make it easy for derivatives to land in the right spot.  Color-kinematics duality can then be exploited to write these topology dressings in terms of those corresponding to the minimal basis topologies that arise from solving the Jacobi relations.

This operator $\mathcal{L}_{\mathcal{P}}$ encodes the relevant information in the theory concerning particle content $\mathcal{P}$ in the following sense: dressing \textit{only the contact graph} for this particle content with the vertex rule derived from $\mathcal{L}_{\mathcal{P}}$ reproduces the entire amplitude $\mathcal{A}_{\mathcal{P}}$, even if the theory contains no such contact interaction! This works because the denominators $\oop{d}_{\uptau}$ in our operator construction encode the propagator structure associated with all graph contributions to the amplitude (cubic, contact, or otherwise). Simply put, the Feynman vertex rule already contains all the information about graph connectivity and causal structure in the theory, allowing us to assign everything to the contact diagram. 

Of course, this is not how Feynman rule calculations are meant to proceed; rather, all possible graph topologies are drawn and dressed with all the rules obtained from the full Lagrangian of the theory. Following such a prescription naively using our operators will result in some redundancy (in which our contact diagrams are dressed to include cubic graph information, and then, separately, cubic graphs are dressed with the lower-point vertices). We will remove such redunancy systematically by constructing additional operators corresponding to the cut contributions to the amplitudes of the theory. 

The generation of Feynman rules considers all the possible ways that fields in the action could line up with the fields in initial and final states, effectively constituting a sum over all relevant permutations. This will lead to each topology's contribution being overcounted by precisely its number of automorphisms, necessitating a symmetry factor to compensate. Each symmetry factor $s_{\uptau}$ is simply the \textit{reciprocal of the number of automorphisms} of the graph topology $\uptau$. For example, for four external scalars in our theory, there is only one topology, 
\begin{equation}
\fourgraphExtScalar{1}{2}{3}{4} \,
\end{equation}
and thus only one term in our candidate operator. In calculating the corresponding Feynman rule, all permutations of $(1234)$ will be generated, but these 24 permutations contain only three distinct variations (the $s$, $t$, and $u$ channels, as visualized in Figure~\ref{figure:standard4ptcubic}). Each physically distinguishable channel appears 8 total times, meaning we must divide the result by 8 to correct for this overcounting. We see this factor of 8 precisely because it is the number of automorphisms of this graph topology --- the total number of ways we could have labeled it that are, by Bose symmetry, indistinguishable from the original: 
\begin{equation}
\begin{split}
\left\{ \small \fourgraphExtScalarSmall{1}{2}{3}{4} \, , \, \fourgraphExtScalarSmall{2}{1}{3}{4} \, , \, \fourgraphExtScalarSmall{1}{2}{4}{3} \, , \, \fourgraphExtScalarSmall{2}{1}{4}{3} \, , \, \right. \\ \small \left.
\fourgraphExtScalarSmall{4}{3}{2}{1} \, , \, \fourgraphExtScalarSmall{3}{4}{2}{1} \, , \, \fourgraphExtScalarSmall{4}{3}{1}{2} \, , \, \fourgraphExtScalarSmall{3}{4}{1}{2} \right\}
\end{split}
\end{equation}
So, for this topology, the appropriate symmetry factor would be $s = 1/8$. It is also precisely for this reason that Feynman rules generate all permutations that we do not specify particular orders of arguments to $c_{\uptau}$ and $\oop{n}_{\uptau}$ in our schematic formula for $\mathcal{L}$; choosing (1234) will not yield a different result than (2431), or any other allowed permutation, as long as it is chosen the same way for $c$, $\oop{n}$, and $\oop{d}$.

\subsubsection{Three field operator}
At three point in our simple scalar theory the only non-vanishing interaction involving scalars relates two external scalars and one external gluon.  This amplitude corresponds to dressing the only topology, as seen in Figure~\ref{figure:3ptvertex}. We found this amplitude using bootstrap methods to be given by: 
\begin{equation}
\label{eqn:a3pt}
\mathcal{A}_{\varphi\varphi g} =  g f^{abc} \left( k_2- k_1 \right) \cdot \varepsilon_3  = 2 g f^{abc} \left( k_2 \cdot \varepsilon_3 \right) 
\end{equation}
The color factor is simply the adjoint structure constant $f^{abc}$, there are no propagators (thus $d_g$ is simply unity, and we need not worry about any propagator operators in the denominator at three points), and the kinematic numerator is given by: 
\begin{equation}
n_{\varphi\varphi g}(123) = n \circ \threeGraph{1}{2}{3} = 2 g \left( k_2 \cdot \varepsilon_3 \right)
\end{equation} 
This graph has just two automorphisms, so the symmetry factor is $1/2$. Hence, in line with the procedure sketched in equation~\ref{eqn:Lform}, we want to write down an operator whose integrand takes the following form to encode the three-point content of this simple scalar theory,
\begin{equation}
\label{eqn:schematicL}
\mathcal{L}_{\varphi\varphi g} =  \mD{3} \,  \frac{1}{2} \, f^{abc} \oop{n}^{\mu}_{\varphi\varphi g}\, \varphi^a \varphi^b A_{\mu}^c 
\end{equation}

We will now proceed to detail how we promote kinematic numerators to operators. As is standard, we interpret momenta as arising from derivatives acting upon the fields, and polarizations from the field's corresponding vector index.  We are aiming for a setup that allows: 
\begin{align}
\frac{1}{2} \mD{3} \oop{n}_{\varphi\varphi g} \varphi^a \varphi^b A_{\mu}^c  &= g\, \varphi^a(x) \left( \partial^{\mu}\varphi^b(x) \right) A_{\mu}^c (x)\, , \qquad \text{with} \\
 \mD{3} &\equiv \int d^4 x_1 \, d^4 x_2 \,  d^4 x_3 \delta^4(x_1 - x)\delta^4(x_2- x)\delta^4(x_3 - x) \,. 
\end{align}
We will find it crucial in our method to be able to separate derivative and field operators in these expressions, so that, schematically, we can write such an operator as: 
\begin{equation}
\oop{n}^\mu_{\varphi\varphi g} \varphi^a \varphi^b A_{\mu}^c  \sim 2g\, \partial^{\mu}_2 \left[ \varphi^a(x) \varphi^b_2(x) A_{\mu}{}^c (x) \right] \, ,
\end{equation}
where the label $2$ has been introduced as temporary notation that instructs the reader that this derivative only acts on the field $\varphi^b_2(x)$ we've labeled with this same subscript, and not, say, on the vector field $A_{\mu}^c (x)$. How can we encode this sort of selective derivative mathematically? 

We can achieve this by assigning each field in the operator its own dummy position $x_i$, which is set by a Dirac delta function to be evaluated only at the spacetime point $x$ that is universal to all fields in the operator, so as not to corrupt the locality of the interaction. Essentially, our procedure will always be to write an arbitrary field $\phi$ as follows, 
\begin{equation}
\phi(x) = \int d^d x_i \delta^{(d)}(x - x_i) \phi(x_i)
\end{equation}
with a unique label $i$ for each field in the operator. This allows us to rewrite the operator as follows, 
\begin{equation}
\oop{n}^\mu_{\varphi\varphi g} = 2g\, \partial^{\mu}_2\,,
\end{equation}
so that the full contribution to the Lagrangian $\mathcal{L}_{\varphi\varphi g} =\frac{1}{2} \mD{3} \left( f^{abc} \oop{n}^\mu_{\varphi\varphi g} \right) \phi^a \phi^b A^c_\mu$ will hence take the form: 
\begin{equation}
\mathcal{L}_{\varphi\varphi g} =  g f^{abc} \int \left(\prod_i^3 d^d x_i \delta^{(d)}(x - x_i) \right) \frac{\partial}{\partial x_2{}_{\mu}} \left[ \varphi^a(x_1) \varphi^b(x_2) A^c_{\mu} (x_3) \right] \, .
\end{equation}
To write this more succinctly, we will introduce two pieces of notation: first, the shorthand $\partial^{\mu}_2 \equiv \partial / \partial x_2{}_{\mu}$ denotes derivatives with respect to the different spacetime position labels.  In general, we define a convenient measure to encode the dummy spacetime variables,
\begin{equation}
\justD{n}^d x \equiv \left(\prod_i^n d^d x_i \delta^{(d)}(x - x_i) \right)
\end{equation}
so that our $\varphi \varphi g$ contribution to the Lagrangian can be expressed in the compact form: 
\begin{align}\label{3ptAmpAsOp}
\mathcal{L}_{\varphi\varphi g} &=  g f^{abc} \mD{3}^d x \, \partial_{2}^{\mu} \left[ \varphi^a(x_1) \varphi^b(x_2) A^{c}_{\mu} (x_3) \right] \, \\
&= g f^{abc} \varphi^a (\partial^\mu \varphi^b) A^c_\mu 
\end{align}

We see we got there by basically taking $k^\mu_i \to \partial^\mu_i$, and $ \eta_{\mu\nu} \epsilon^\nu \to \eta_{\mu\nu}$, a procedure to be elaborated on in more detail in \cref{sec:promotion}. 

Now that we have our operator, we will confirm it gives rise to the correct amplitude by explicitly computing its corresponding vertex rule. The operator can be rewritten in momentum-space via a Fourier transform (all particles are taken to be outgoing to match the conventions used in our amplitudes-based approach): 
\begin{equation}
\mathcal{L}_{\varphi\varphi g} =  g f^{abc} \int \left(\prod_i^3 d^d k_i e^{i k_i \cdot x} \right) (ik_2^{\mu}) \left[ \tilde{\varphi}^a(k_1) \tilde{\varphi}^b(k_2) \tilde{A}_{\mu}^c (k_3) \right]
\end{equation} 
We now proceed to calculate its contribution to the action $\mathcal{S} = \int d^d x \, \mathcal{L}$ and relabel dummy variables and indices: 
\begin{equation}
\mathcal{S}_{\varphi\varphi g} =  g f^{xyz} \int d^d x \left(\prod_i^3 d^d p_i  \right) e^{i (p_1 + p_2 + p_3)\cdot x}  (ip_2^{\nu}) \left[ \tilde{\varphi}^x(p_1) \tilde{\varphi}^y(p_2) \tilde{A}_{\nu}^z (p_3) \right] 
\end{equation} 
\begin{align}
\mathcal{S}_{\varphi\varphi g} &=  g f^{xyz} \int \left(\prod_i^3 d^d p_i  \right) (2\pi)^d \delta^{(d)}(p_1 + p_2 + p_3) (ip_2^{\nu}) \left[ \tilde{\varphi}^x(p_1) \tilde{\varphi}^y(p_2) \tilde{A}_{\nu}^z (p_3) \right] \\
\end{align} 
We can write down the associated Feynman rule for the $\varphi\varphi g$ vertex by taking the functional derivative of the action: 
\begin{equation}
\mathcal{V}^{abc}_{\mu} \equiv \frac{\delta \mathcal{S}}{\delta \tilde{\varphi}^a(k_1) \delta \tilde{\varphi}^b(k_2) \delta \tilde{A}^{\mu,c} (k_3)}
\end{equation}
\begin{equation}
\begin{split}
\mathcal{V}^{abc}_{\mu} =  g f^{xyz} \int \left(\prod_i^3 d^d p_i  \right) (2\pi)^d \delta^{(d)}(p_1 + p_2 + p_3) (ip_{2,\nu}) \delta^{cz} \delta^{\nu}_{\mu} \delta^{(d)}(k_3 - p_3) \\ \left[ \delta^{ax}\delta^{by}  \delta^{(d)}(k_1 - p_1) \delta^{(d)}(k_2 - p_2) + \delta^{bx} \delta^{ay} \delta^{(d)}(k_2 - p_1) \delta^{(d)}(k_1 - p_2) \right]
\end{split}
\end{equation} 
\begin{equation}
\mathcal{V}^{abc}_{\mu} =  i g (2\pi)^d \delta^{(d)}(k_1 + k_2 + k_3) f^{abc} \left(k_2 - k_1\right)_{\mu}
\end{equation} 
The full on-shell all-outgoing three-point $S$-matrix element $\langle \varphi \varphi g | S | 0 \rangle$ can be computed directly from this vertex rule by contracting it with the external gluon's polarization vector:  
\begin{equation}
\langle \varphi \varphi g | S | 0 \rangle = \mathcal{V}^{abc}_{\mu} \varepsilon^{\mu}_3 = 2 i g (2\pi)^d \delta^{(d)}(k_1 + k_2 + k_3) f^{abc}  \left(k_2 \cdot \varepsilon_3 \right)
\end{equation} 
Finally, from the standard definition $\langle f | S | i \rangle = (2\pi)^d \delta^{(d)}(k_f - k_i) i \mathcal{A}$, we can read off the scattering amplitude, 
\begin{equation}
\mathcal{A} =  2 g f^{abc}  \left(k_2 \cdot \varepsilon_3 \right) \, .
\end{equation} 
This result is in perfect agreement with the result found using amplitudes-based considerations, as desired. 

\subsubsection{Four field operator}
Now we would like to encode the information from the two scalar, two gluon amplitude $\mathcal{A}_{\varphi \varphi g g}$ in a (admittedly nonlocal) four-field operator $\mathcal{L}_{\varphi \varphi g g}$.  Of course this will introduce redundancies as we discuss and remove below, but for now we want a four-field operator that reproduces the four-field amplitude in its entirety.

This amplitude is expressed as the following sum over graphs, depicted in equation~\ref{eqn:schematic4pt} and written functionally as:
\begin{equation}
\mathcal{A}_{\varphi\varphi g g}  = \frac{c_t n_{\varphi}(1234)}{t} - \frac{c_u n_{\varphi}(1243)}{u} + \frac{c_s n_{\text{v}}(1234)}{s} \, ,
\end{equation}

In this case, we have two topologies, labeled $\varphi$ and v (according to the particle-type of their propagators). We must then construct our Lagrangian in terms of the associated dressings of both topologies: 
\begin{equation}
\mathcal{L}_{\varphi\varphi g g} = \mD{4} \left( s_{\varphi} \frac{c^{abcd}_{\varphi} \oop{n}^{\mu\nu}_{\varphi}}{\oop{d}_{\varphi}} + s_{\text{v}} \frac{c^{abcd}_{\text{v}} \oop{n}^{\mu\nu}_{\text{v}}}{\oop{d}_{\text{v}}} \right)  \varphi^a \varphi^b A^c_{\mu} A^d_{\nu}  \equiv \mathcal{O}_{\varphi} + \mathcal{O}_{\text{v}}
\end{equation}
Again, we arrive at the symmetry factors by counting automorphisms for each topology: 
\begin{equation}
\begin{split}
\fourgraphScalarProp{k_4}{k_1}{k_2}{k_3}  \quad &\Rightarrow \quad s_{\varphi} = \frac{1}{2} \\
\fourgraphVectorProp{k_1}{k_2}{k_3}{k_4}  \quad &\Rightarrow \quad s_{\text{v}} = \frac{1}{4}
\end{split}
\end{equation}
The numerator operators can be written in the same manner as at three points. The scalar propagator graph dressing $n_{\varphi}$ is translated into the operator $\oop{n}_{\varphi}$ as follows: 
\begin{equation}
n_{\varphi}(1234) = - 2 g^2 \left[ 2 (k_2\cdot\varepsilon_3)(k_1 \cdot\varepsilon_4) + (k_2\cdot k_3)(\varepsilon_3\cdot\varepsilon_4) \right]
\end{equation}
\begin{equation}
\oop{n}_{\varphi}^{\mu\nu} = -2 g^2  \,  \big( 2\, \partial_2^{\mu} \,\partial_1^{\nu} + \eta^{\mu\nu} \, \partial_2^{\rho} \, \partial_{3,\rho} \big) 
\end{equation}
Finally, we encode the propagator structure of the graphs using derivatives in the denominator, so the $d_{\varphi} = s_{23}=(k_2+k_3)^2=2 k_2\cdot k_3$ propagator is promoted to $\oop{d}_{\varphi} = 2 \left( \partial_2 \cdot \partial_3 \right) = 2 \left( \partial_{2,\mu} \partial_3^{\mu} \right)$. We can then arrive at this topology's contribution to the Lagrangian as: 
\begin{equation}
\label{eqn:Os}
\mathcal{O}_{\varphi} = - g^2 f^{dae}f^{ebc} \mD{4} \, \frac{\big( 2\, \partial_2^{\mu} \,\partial_1^{\nu} + \eta^{\mu\nu} \, \partial_2^{\rho} \, \partial_{3,\rho} \big)}{2 \left( \partial_2 \cdot \partial_3 \right)} \varphi^a(x_1) \varphi^b(x_2) A^{c}_{\mu} (x_3) A^{d}_{\nu} (x_4) 
\end{equation}
While the non-local appearance of this operator, with derivatives in the denominator, can ring alarm bells signaling action at a distance, computation of the associated vertex rule will show that this non-locality is actually precisely encoding the causal structure of our theory: the presence of a denominator in this expression, which has been written as a four-field contact term, gives rise to the cubic propagators appropriate for this operator's contribution to the consistently-factorizing amplitude. 

The vector propagator graph dressing $n_{\text{v}}$ follows from the kinematic Jacobi relation: 
\begin{equation}
n_{\text{v}}(abcd)  = n_{\varphi}(abcd) - n_{\varphi}(abdc)
\end{equation}
We opt to simply promote this Jacobi identity between functional numerators $n$ to one between operators $\oop{n}$, and then construct $\mathcal{O}_{\text{v}}$ in the same way as the scalar-propagator case, with $c_{\text{v}} = f^{abe}f^{ecd}$, $\oop{d}_{\text{v}} = 2 \left( \partial_1 \cdot \partial_2 \right)$, and $s_{\text{v}} = 1/4$.
\begin{equation}
\begin{split}
\label{eqn:Ov}
\mathcal{O}_{\text{v}} = - \frac{1}{2} g^2 f^{abe}f^{ecd} \mD{4} \, & \frac{\big( 2\, \partial_2^{[\mu ,} \partial_1^{\nu]} + \eta^{\mu\nu} \, \partial_2^{\rho} \, \partial_{3,\rho}  - \eta^{\mu\nu} \, \partial_2^{\rho} \, \partial_{4,\rho} \big)}{2 \left( \partial_1 \cdot \partial_2 \right)} \times \\ & \quad \, \,  \varphi^a(x_1) \varphi^b(x_2) A^{c}_{\mu} (x_3) A^{d}_{\nu} (x_4)
\end{split}
\end{equation}
Thus, we are able to systematically write down a Lagrangian contribution $\mathcal{L}_{\varphi \varphi g g} = \mathcal{O}_{\varphi} + \mathcal{O}_{\text{v}}$. A complete calculation of its associated four-point vertex rule (as it is a four-field operator, albeit non-local) $\mathcal{V}^{abcd}_{\mu \nu}$ which yields
\begin{equation}
\mathcal{V}^{abcd}_{\mu \nu} \epsilon^\mu \epsilon^\nu = \mathcal{A}_{\varphi\varphi g g }\,.
\end{equation}
In other words we can calculate the corresponding on-shell amplitude for two scalars and two gluons by dressing just the contact diagram: 
\begin{equation}
\contactFour{1}{2}{3}{4}
\end{equation}
with the Feynman rule $\mathcal{V}^{abcd}_{\mu \nu}$; this successfully reproduces $\mathcal{A}_{\varphi \varphi g g}$ as desired, with color-dual structure manifest and the non-localities in the operators $\mathcal{O}$ converted into the appropriate cubic propagators demanded by causality. We will denote this amplitude (and further amplitudes) schematically by drawing the graph contributions, with vertices labeled with the specific operator from which its dressing arises, 
\begin{equation}
\mathcal{A}_{\varphi \varphi g g} = \fourOp{1}{2}{3}{4}{\overline{\mathcal{L}}_{\varphi \varphi g g}} 
\end{equation}

We emphasize here that we are not done!  This is quite a distinct prescription from the traditional method of calculating an amplitude via Feynman diagrams. One must dress \textit{all} possible graphs at the specified multiplicity and order in quantum correction, not just the contact diagram.  If one does this with the operators we have introduced so far we will have a problem.  Remedying this problem involves removing redundancies by considering cuts as we now discuss.

\subsection{Writing the actual Lagrangian}
We have thus far given a prescription for generating a three-field operator, which, when turned into a Feynman vertex rule, reproduces the three-point amplitude; and a four-field operator, which, when used to dress a four-point contact diagram \textit{in isolation}, reproduces the four-point amplitude. By this, we mean that $\mathcal{A}_{\varphi \varphi g g}$ is computed solely by dressing a four-point contact diagram with the vertex rule arising from $\mathcal{L}_{\varphi \varphi g g}$. But of course, this is not how  Feynman rule calculations  transpire: one must also write down the cubic graphs at four points, and dress them with the rules arising from lower-multiplicity operators $\mathcal{L}_{\varphi \varphi g}$ and $\mathcal{L}_{\text{Yang-Mills}}$. 

With our operators as currently written, carrying out the standard procedure of writing both the cubic and quartic graphs and dressing with appropriate vertices specified by the candidate Lagrangian $ \mathcal{L}_{\text{Yang-Mills}} + \mathcal{L}_{\text{free-scalar}} + \mathcal{L}_{\varphi\varphi g}
+ \mathcal{L}_{\varphi\varphi g g}$ will not yield the correct amplitude, but rather, a gauge-dependent result containing redundant contributions. To construct a formally correct Lagrangian, then, we must remove this redundancy, which we will achieve by subtracting operators corresponding to the amplitude's cuts, so that the resulting four-field operator only contains the true local contact term. 

\subsubsection{Removing redundancy at four points}
For the two scalar, two gluon case, using the traditional form of the Lagrangian in equation~\ref{eqn:traditionalAction}, such a calculation is depicted schematically as follows, dressing the three cubic graphs in addition to the contact diagram: 
\begin{equation}
\begin{split}
\mathcal{A}_{\varphi\varphi g g}  &= \fourgraphScalarPropOp{4}{1}{2}{3}{\mathcal{L}_{\varphi \varphi g}}{\mathcal{L}_{\varphi \varphi g}}
+ \fourgraphScalarPropOp{3}{1}{2}{4}{\mathcal{L}_{\varphi \varphi g}}{\mathcal{L}_{\varphi \varphi g}} \\ 
&+ \fourgraphVectorPropOp{1}{2}{3}{4}{\mathcal{L}_{\varphi \varphi g}}{\mathcal{L}_{\text{YM}}} + \fourOp{1}{2}{3}{4}{\mathcal{L}_{\varphi \varphi g g}} 
\end{split}
\end{equation}
If we attempt to naively replicate this calculation but using the Feynman rules generated from our operators as constructed, we will arrive at an unphysical answer,
\begin{equation}
\begin{split}
\label{eqn:wrongAmp}
\mathcal{A}_{\varphi\varphi g g}  &\neq \fourgraphScalarPropOp{4}{1}{2}{3}{\overline{\mathcal{L}}_{\varphi \varphi g}}{\overline{\mathcal{L}}_{\varphi \varphi g}}
+ \fourgraphScalarPropOp{3}{1}{2}{4}{\overline{\mathcal{L}}_{\varphi \varphi g}}{\overline{\mathcal{L}}_{\varphi \varphi g}} \\ 
&+ \fourgraphVectorPropOp{1}{2}{3}{4}{\overline{\mathcal{L}}_{\varphi \varphi g}}{\overline{\mathcal{L}}_{\text{YM}}} + \fourOp{1}{2}{3}{4}{\overline{\mathcal{L}}_{\varphi \varphi g g}}  \, ,
\end{split}
\end{equation}
precisely because we have already encoded the content of the three cubic graph contributions within the four-point contact operator $\overline{\mathcal{L}}_{\varphi \varphi g g}$, so this calculation is redundant. While the resolution is simple --- only dress the contact diagram --- we now will aim to write down a full Lagrangian that yields the correct amplitude even using the traditional method of dressing all possible Feynman diagrams. This will necessitate the introduction of additional operators to compensate for this overcounting of the cubic graph contributions. Such information can be extracted from our bootstrap approach by careful consideration of the cuts of the desired amplitude. 

Schematically, we will write the Lagrangian in the following form: 
\begin{equation}
\mathcal{L} = \overline{\mathcal{L}}_{\text{Yang-Mills}} + \overline{\mathcal{L}}_{\text{free-scalar}} + \overline{\mathcal{L}}_{\varphi\varphi g}
+ \left(  \overline{\mathcal{L}}_{\varphi\varphi g g} -  \overline{\slashed{\mathcal{L}}}_{\varphi\varphi g g} \right) \, ,
\end{equation}
where we introduce the slashed operator $\overline{\slashed{\mathcal{L}}}_{\varphi\varphi g g}$ to remove this redundancy corresponding to the overcounting of cubic graph contributions. This will be a (non-local) four field contact operator, just like $\overline{\mathcal{L}}_{\varphi\varphi g g}$; hence, we will derive from it an additional Feynman rule for dressing the four particle vertex. The traditional application of this Lagrangian's Feynman rules to all possible graphs will then take the form: 
\begin{equation}
\begin{split}
\mathcal{A}_{\varphi\varphi g g}  = \fourgraphScalarPropOp{4}{1}{2}{3}{\overline{\mathcal{L}}_{\varphi \varphi g}}{\overline{\mathcal{L}}_{\varphi \varphi g}}
+ \fourgraphScalarPropOp{3}{1}{2}{4}{\overline{\mathcal{L}}_{\varphi \varphi g}}{\overline{\mathcal{L}}_{\varphi \varphi g}} \\
+ \fourgraphVectorPropOp{1}{2}{3}{4}{\overline{\mathcal{L}}_{\varphi \varphi g}}{\overline{\mathcal{L}}_{\text{YM}}} + \fourOp{1}{2}{3}{4}{\overline{\mathcal{L}}_{\varphi \varphi g g}} - \fourOp{1}{2}{3}{4}{\overline{\slashed{\mathcal{L}}}_{\varphi \varphi g g}}  \, .
\end{split}
\end{equation}
For this equality to hold, we must construct the new operator $\overline{\slashed{\mathcal{L}}}_{\varphi\varphi g g}$ to encode precisely the same information as the three cubic graph dressings, so that we can achieve a cancellation yielding the correct (and original form of) the amplitude:  
\begin{equation}
\mathcal{A}_{\varphi\varphi g g}  =  \fourOp{1}{2}{3}{4}{\overline{\mathcal{L}}_{\varphi \varphi g g}}  \, .
\end{equation}
We will achieve precisely this result by considering the cuts of the desired amplitude, and encoding said information in $\overline{\slashed{\mathcal{L}}}_{\varphi\varphi g g}$.

To understand the relationship between the cubic graph overcounting and the cuts of the amplitude, let's consider in closer detail precisely what goes wrong without the inclusion of this slashed operator $\overline{\slashed{\mathcal{L}}}_{\varphi\varphi g g}$. The schematic form of the calculation, as depicted in equation~\ref{eqn:wrongAmp}, does not initially raise any red flags: all diagrams are dressed with the rules from their corresponding operators. The issue lies in the fact that the four point contact diagram is dressed with non-local rule, rather than a truly local contact contribution, as would be the case when using the traditional Lagrangian,  when instead this graph is dressed with the rule arising from the familiar genuinely local four-field operator 
\begin{equation}
   \mathcal{L}_{\varphi \varphi g g} = \frac{g^2}{2} f^{abe}f^{ecd} \varphi^a A_{\mu}^b  \varphi^c A_{\nu}^d  \eta^{\mu \nu}
  \label{manifestPhiPhiGGcontact}
 \end{equation}
     To remedy this issue, we insist that the \textit{difference} of these four-field operators, $\left(  \overline{\mathcal{L}}_{\varphi\varphi g g} -  \overline{\slashed{\mathcal{L}}}_{\varphi\varphi g g} \right) $, be a truly local contact operator. Since the original operator $ \overline{\mathcal{L}}_{\varphi\varphi g g}$ was constructed from the full amplitude, it stands to reason that, to obtain just the local contact contribution, we must subtract the information living on the amplitude's cuts. 

To make this a bit more precise, let's write down the local contact piece of the amplitude in the following way, 
\begin{equation}
\mathcal{C}_{\varphi\varphi g g} = \sum_g \left(  \frac{c_g n_g}{d_g} - \frac{c_g \slashed{n}_g}{d_g} \right)
\end{equation}
where the slashed numerators $\slashed{n}_g$ are written down to encode that particular graph's unique cut contributions (this will be defined generically in the following subsection, but will be sufficiently intuitive at four points to construct using this explicit example). We then promote these slashed numerators to operators in precisely the same manner as standard numerators, so that the desired difference of operators can be written as:
\begin{equation}
\overline{\mathcal{L}}_{\varphi\varphi g g} -  \overline{\slashed{\mathcal{L}}}_{\varphi\varphi g g} =
\mD{4} \left( \sum_{\uptau \in  \mathcal{T}^4_3{}_{\varphi\varphi g g}} s_{\uptau} \frac{c_{\uptau}  \left(  \oop{n}_{\uptau}-  \oop{\slashed{n}}_{\uptau} \right)}{\oop{d}_{\uptau}} \right)   \varphi^{a} \varphi^{b} A^{c}_\mu A^{d}_\nu\,.
\end{equation}
The distinct cubic topologies are our familiar scalar propagator and vector propagator graphs, so the sum expands as follows:
\begin{equation}
\sum_{\uptau \in  \mathcal{T}^4_3{}_{\varphi\varphi g g}}  \left( s_{\uptau} \frac{c_{\uptau} \left( \oop{n}_{\uptau}-  \oop{\slashed{n}}_{\uptau} \right)}{\oop{d}_{\uptau}} \right) = \left( s_{\varphi} \frac{c_{\varphi} \left( \oop{n}_{\varphi} - \slashed{\oop{n}}_{\varphi}\right)}{\oop{d}_{\varphi}} + 
s_{\text{v}} \frac{c_{\text{v}} \left( \oop{n}_{\text{v}} - \slashed{\oop{n}}_{\text{v}} \right)}{\oop{d}_{\text{v}}} \right) \,.
\end{equation}

Schematically, these slashed numerators correspond to how the graph topology can contribute to any possible cuts of the amplitude.  We note that in this form the color-dual nature of the amplitudes themselves is manifest.  The scalar propagator topology contains information that survives the $s_{14}$ cut of the amplitude, depicted schematically as: 
\begin{equation}
\slashed{\oop{n}}_{\varphi} = n \circ \fourgraphScalarPropCut{4}{1}{2}{3}{\mathcal{L}_{\varphi \varphi g}}{\mathcal{L}_{\varphi \varphi g}}
\end{equation}
This can be calculated by imposing the cut conditions to arrive at: 
\begin{equation}
\slashed{n}_{\varphi}(1234) = - 4 g^2 (k_2\cdot\varepsilon_3)(k_1 \cdot\varepsilon_4) 
\end{equation}
Similarly, the vector propagator topology dressing has a contribution that survives the $s_{12}$ channel cut, 
\begin{equation}
\slashed{n}_{\text{v}} = n \circ \fourgraphVectorPropCut{1}{2}{3}{4}{\mathcal{L}_{\varphi \varphi g}}{\mathcal{L}_{\varphi \varphi g}}
\end{equation}
\begin{equation}
\begin{split}
\slashed{n}_{\text{v}}(1234) &= \, 4 g^2 (k_1\cdot \varepsilon_3) (k_2\cdot \varepsilon_4) - 
 4 g^2 (k_1\cdot \varepsilon_4) (k_2\cdot \varepsilon_3) \\ &\, \,- 
 4 g^2 (k_2\cdot k_3) (\varepsilon_3\cdot \varepsilon_4) 
\end{split}
\end{equation}
One should not be concerned by the absence of $s_{13}$ cut information, as our procedure, in summing over all possible permutations of particle labels, will encode this content in the Feynman rule.

These cut numerators $\slashed{n}$ are promoted to operators just like the standard numerators, 
\begin{equation}
\slashed{\mathcal{O}}_{\varphi} =  - g^2 f^{dae}f^{ebc} \mD{4} \, \frac{\big( 2\, \partial_2^{\mu} \,\partial_1^{\nu})}{2 \left( \partial_2 \cdot \partial_3 \right)} \varphi^a(x_1) \varphi^b(x_2) A^{c}_{\mu} (x_3) A^{d}_{\mu} (x_4) \,,
\end{equation}
\begin{equation}
\begin{split}
\slashed{\mathcal{O}}_{\text{v}} = - \frac{1}{2} g^2 f^{abe}f^{ecd} \mD{4} \, & \frac{\big( 2\, \partial_2^{[\mu ,} \partial_1^{\nu]} + 2\, \eta^{\mu\nu} \, \partial_2^{\rho} \, \partial_{3,\rho} \big)}{2 \left( \partial_1 \cdot \partial_2 \right)}   \varphi^a(x_1) \varphi^b(x_2) A^{c}_{\mu} (x_3) A^{d}_{\mu} (x_4)\,.
\end{split}
\end{equation}

Using conservation of momentum and the results in equations~\ref{eqn:Os} and~\ref{eqn:Ov} we therefore find,
\begin{equation}
\begin{split}
\mathcal{L}_{\varphi\varphi g g} - \slashed{\mathcal{L}}_{\varphi\varphi g g} = &- g^2 f^{dae}f^{ebc} \mD{4} \, \frac{\big(\eta^{\mu\nu} \, \partial_2 \cdot \partial_3 \big)}{2 \left( \partial_2 \cdot \partial_3 \right)} \varphi^a(x_1) \varphi^b(x_2) A^{c}_{\mu} (x_3) A^{d}_{\mu} (x_4)   \\
&- \frac{1}{2} g^2 f^{abe}f^{ecd} \mD{4} \, \frac{\big( \eta^{\mu\nu} \, \partial_1 \cdot \partial_2 \big)}{2 \left( \partial_1 \cdot \partial_2 \right)}   \varphi^a(x_1) \varphi^b(x_2) A^{c}_{\mu} (x_3) A^{d}_{\mu} (x_4)
\end{split}
\end{equation}
Noting the desired explicit cancellation of the propagators, we can simplify this to recover the familiar manifestly local four-field contact term of \cref{manifestPhiPhiGGcontact}, 
\begin{align}
\mathcal{L}_{\varphi\varphi g g} - \slashed{\mathcal{L}}_{\varphi\varphi g g} &= - \frac{1}{4} g^2  \mD{4}  \, \left(2 f^{dae}f^{ebc} + f^{abe}f^{ecd} \right) \varphi^a(x_1) \varphi^b(x_2) A^{c \, \mu} (x_3) A^{d}_{\mu} (x_4)  \\
 & = - \frac{1}{4} g^2  \left(2 f^{dae}f^{ebc} + f^{abe}f^{ecd}  \right) \varphi^a(x)A^{d}_{\mu} (x) \varphi^b(x)  A^{c \, \mu} (x)  \\
 & = - \frac{1}{4} g^2  \left(2 f^{dae}f^{ebc} \right) \varphi^a(x)A^{d}_{\mu} (x) \varphi^b(x)  A^{c \, \mu} (x)  \\
 & = - \frac{1}{2} g^2  \left(   f^{bae}f^{ecd}  \right) \varphi^a(x)A^{b}_{\mu} (x) \varphi^c(x)  A^{d \, \mu} (x)  \\      
 & =  \frac{1}{2} g^2  \left( f^{abe}f^{ecd}  \right) \varphi^a(x)A^{b}_{\mu} (x) \varphi^c(x)  A^{d \, \mu} (x)  \\      
 &=   \mathcal{C}_{\varphi\varphi g g}\,.
\label{fourPointContactFromAmp}
\end{align}
In going from the second to third line we note that the contraction of the antisymmetric structure constants with symmetric field labels (both in scalars and gluons)  vanishes after the fields are all brought locally, and in going from the third to fourth line simply relabel indices to line up field labeling with the usual form of the contact written in the action \cref{manifestPhiPhiGGcontact}.

Of course, there is a generalized gauge freedom in how one writes zero on the
cuts.  Different choices yield the same on-shell contribution at a given
multiplicity, but can propagate differently into higher-multiplicity
operators.  Any operator ambiguity proportional to the equations of motion is
invisible to on-shell amplitudes and corresponds to a field redefinition; such
terms may affect intermediate off-shell representations but cannot modify
physical S-matrix elements.  There are also on-shell identities that are
genuinely ambiguous, such as writing $s_{12}=2(k_1\!\cdot k_2)$ versus
$k_1^2+k_2^2+2(k_1\!\cdot k_2)$.

To illustrate this point, one may add terms that vanish on-shell at four points.
Such terms, if they do not vanish by the equations of motion, can be equivalent,
up to field redefinitions, to higher-multiplicity operators that only become
visible once sufficiently many legs are taken off shell.  These typically arise
from structures that vanish due to gauge invariance at fixed multiplicity
(e.g.\ $k_1\!\cdot\varepsilon_1$) or from imposing explicit on-shell conditions
(e.g.\ $k_i^2=0$).  For this reason, it is natural and recommended to work in an
on-shell minimal basis using momentum conservation, though one is of course
free to write zero in any convenient form.

If higher-point contributions are implicitly introduced at lower multiplicity,
they must be subtracted at the appropriate stage to avoid double counting.
This bookkeeping is made explicit when comparing higher-point input amplitudes
with amplitudes generated from the action constructed so far, as discussed in
\cref{sec:opPromotionCuts}.

As a contrasting example, we will carry out the same procedure for the four scalar amplitude (calculated in equation~\ref{eqn:fourScalarAmp}) and its associated operators. We again note that there is no apparent contact contribution: 
\begin{equation}
\mathcal{A}_{\varphi\varphi\varphi\varphi} = 2 g^2 \left( \frac{c_s (u - t)}{s} + \frac{c_t (u-s)}{t} + \frac{c_u (s-t)}{u}  \right) \, . 
\end{equation}
The operator in can be written down in the same manner as elaborated in the previous subsection; we write down $\mathcal{L}_{\varphi\varphi\varphi\varphi} = \mD{4} s_4 \left( c_4 \oop{n}_4 / \oop{d}_4 \right)\varphi^4$ from the constructed numerator $n_4$: 
\begin{equation}
n_4 (1234) = 2 g^2 \left( s_{13} - s_{23} \right)
\end{equation}
\begin{equation}
\mathcal{L}_{\varphi\varphi\varphi\varphi} = \frac{1}{8} \left( 2 g^2 \right) f^{abe}f^{ecd} \mD{4} \, \frac{ \left( \partial_1 \cdot \partial_3 \right) -  \left( \partial_2 \cdot \partial_3 \right) }{ \left( \partial_1 \cdot \partial_2 \right)} \varphi^a(x_1) \varphi^b(x_2) \varphi^c(x_3) \varphi^d(x_4)
\end{equation}
This operator, when used to dress the four-scalar contact graph, yields the desired four-scalar amplitude. To remove redundancy associated with dressing both the contact and the relevant cubic graphs, we must write: 
\begin{equation}
\mathcal{L}_{\varphi\varphi\varphi\varphi} - \slashed{\mathcal{L}}_{\varphi\varphi\varphi\varphi} = \mD{4} \left(s_4 \frac{c_4 \oop{n}_4}{\oop{d}_4}  - s_4 \frac{c_4\slashed{\oop{n}}_4}{\oop{d}_4}\right)\varphi^4
\end{equation}
The cut contribution $\slashed{\oop{n}}_4$ is simply the operator promotion of the numerator's unique cut contributions, $\slashed{n}_4$. In this case, as no contact is required for the duality between color and kinematics, this leaves the numerator unchanged, 
\begin{equation}
\slashed{n}_4 (1234) = 2 g^2 \left( s_{13} - s_{23} \right) = n_4 (1234) \, , 
\end{equation}
so the slashed operator will give the same contribution as the original operator. Hence, the overall contribution to the Lagrangian vanishes, 
\begin{equation}
\mathcal{L}_{\varphi\varphi\varphi\varphi} - \slashed{\mathcal{L}}_{\varphi\varphi\varphi\varphi} = 0 \, ,
\end{equation}
consistent with our finding that there is no contact contribution to the full four-scalar amplitude $\mathcal{A}_{\varphi\varphi\varphi\varphi}$.

\subsection{Summary}

Let us take the opportunity to summarize what we accomplished this invitation. We bootstrapped the amplitudes associated with the theory of the minimally coupled real adjoint scalar by imposing the duality between color and kinematics, and consistent factorization.  We chose this theory for the invitation because the presence of external scalars allows for color-dual scattering amplitudes with non-trivial kinematic factors, but without necessarily drowning in polarization vectors.  We then took that perturbative interacting data and constructively wrote down the corresponding interaction operators required to generate that data.  In this section we restricted ourselves to operators involving scalar fields, and will demonstrate the promotion of gluon operators after we discuss our procedure in full

The familiar action for the covariantized free scalar theory is given by the Lagrangian,
\begin{align} 
\mathcal{L} &= -\frac{1}{4}\Tr \left( F^2 \right) + \frac{1}{2}\left(D_{\mu} \varphi \right)^{a}\!\left(D^{\mu}\varphi \right)^{a} \\ 
  &= -\frac{1}{4}\Tr \left( F^2 \right) + \frac{1}{2} (\partial \varphi)^2 + g f^{abc} (\partial^\mu \varphi^a)  A^b_\mu  \varphi^{c}   
  + \frac{g^2}{2} f^{abe}f^{ecd} \varphi^a A_{\mu}^b  \varphi^c A_{\nu}^d  \eta^{\mu \nu}
\end{align}

We reproduced the cubic scalar interaction  $g f^{abc} \phi^{a}  A^b_\mu \partial^\mu \phi^c $ by considering  the three-point bootstrapped amplitude \cref{eqn:a3pt}, $\mathcal{A}_3= 2 g f^{abc} \left( k_2 \cdot \varepsilon_3 \right)$.  Applying \cref{eqn:Lform} to the 3-point ampllitudes with graph symmetry factor $1/2$ associated with the automorphic exchange of scalar legs $1\leftrightarrow 2$, we simply found, \cref{3ptAmpAsOp},
\begin{align}
\mathcal{L}_{\varphi\varphi g} &=  \mD{3} (s_3 c_3 \oop{n}_{\varphi\varphi g}) \varphi^2 A\\
&= g f^{abc} \mD{3} \, \partial_{2}^{\mu} \left[ \varphi^a(x_1) \varphi^b(x_2) A^{c}_{\mu} (x_3) \right] \, \\
&=  g f^{abc} \varphi^a (\partial^\mu \varphi^b) A^c_\mu 
\end{align}

We reproduced the four point contact interaction  $\frac{g^2}{2} f^{bce}f^{eda} \varphi^b \varphi^d \eta^{\mu \nu} A_{\mu}^c A_{\nu}^a$  by isolating the contact in the four-point amplitude and promoting it to an operator.  How did we isolate the contact? We exploited the fact that Feynman rules encode permutation symmetry.  We bootstrapped the necessary topologies to construct the full 4-point amplitude and promoted them to operators --- which of course would result in overcounting, so then we subtracted the operators associated with the cut of the amplitude, yielding (\cref{fourPointContactFromAmp}) precisely 
\begin{align}
\mathcal{L}_{\varphi\varphi g g} - \slashed{\mathcal{L}}_{\varphi\varphi g g} &=
\mD{4} \left( s_{\varphi} \frac{c_{\varphi} \left( \oop{n}_{\varphi} - \slashed{\oop{n}}_{\varphi}\right)}{\oop{d}_{\varphi}} + 
s_{\text{v}} \frac{c_{\text{v}} \left( \oop{n}_{\text{v}} - \slashed{\oop{n}}_{\text{v}} \right)}{\oop{d}_{\text{v}}} \right)\varphi^{a} \varphi^{b} A^{c}_\mu A^{d}_\nu\\
&= - \frac{1}{4} g^2  \mD{4}  \, \left(2 f^{dae}f^{ebc} + f^{abe}f^{ecd} \right) \varphi^a(x_1) \varphi^b(x_2) A^{c \, \mu} (x_3) A^{d}_{\mu} (x_4)  \\
& = - \frac{1}{4} g^2  \left(2 f^{dae}f^{ebc} + f^{abe}f^{ecd} \right) \varphi^a(x) \varphi^b(x) A^{c \, \mu} (x) A^{d}_{\mu} (x) \\
&=  \frac{g^2}{2} f^{bce}f^{eda} \varphi^b \varphi^d \eta^{\mu \nu} A_{\mu}^c A_{\nu}^a\,.
\end{align}
\label{4otAmpMinusCutAsOp}

We also demonstrated that the approach of building operators and subtracting cuts 
suggests no four-scalar contact associated with this theory even though there is a 
non-trivial color-dual scattering amplitude. 

While the examples considered in this invitation involve massless fields, the
approach applies verbatim to massive fields as well.  Massive propagators modify
the kinematic basis but not the underlying algorithm.  Concrete examples include
massive amplitudes in the fundamental representation, bootstrapped directly from
on-shell data without reference to an underlying action, as constructed in
\cite{Carrasco:2021ptp,Carrasco:2023vjg}.  These amplitudes were constructed
entirely via the method of maximal cuts, with contact terms specified when
necessary by principles such as gauge invariance and the duality between color
and kinematics.

\section{Framework Summary: From Amplitudes to Operators}
\label{sec:algorithm}

Before turning to explicit constructions, we summarize the logic of the
amplitude-to-operator map introduced in this work.
This section is intended as a practical recipe rather than a derivation, and
serves to clarify how the various ingredients fit together.
All technical details are developed in subsequent sections.

\medskip

\noindent\textbf{Input.}
We begin with on-shell, gauge-invariant tree-level amplitudes written in a
color-dual representation.  At fixed multiplicity $m$, the amplitude is assumed
to be expressed as a sum over cubic graphs,
\begin{equation}
\mathcal{A}_m = \sum_{g\in\Gamma_3^m} \frac{c_g\, n_g}{d_g}\,,
\end{equation}
with kinematic numerators $n_g$ satisfying the same algebraic relations
(Jacobi identities and antisymmetries) as the color factors $c_g$.

\medskip

\noindent\textbf{Step 1: Separate novel contact information.}
Using generalized unitarity, we identify the part of the amplitude that is
fully reconstructible from lower-multiplicity data,
denoted $\slashed{\mathcal{A}}_m$.
The difference
\begin{equation}
\label{eq:contact_isolation}
\mathcal{C}_m \equiv \mathcal{A}_m - \slashed{\mathcal{A}}_m
\end{equation}
isolates the genuinely new local information required at $m$ points.
This step, described in detail in \cref{sec:contactExtraction}, is equivalent
to the identification of contact terms in the method of maximal cuts.
Because there are many ways of writing zero, the choice of cut data reflects
what is understood in amplitudes as a generalized gauge freedom: a choice of
how much information to encode in contact terms.
Any choice satisfying \cref{eq:contact_isolation} reproduces the same physical
amplitudes, but can affect the explicit form of higher-multiplicity operators,
as emphasized in \cref{sec:opPromotionCuts}.

\medskip

\noindent\textbf{Step 2: Promote amplitudes to operators.}
(Operator promotion is described in detail in \cref{sec:promotion}.)
Each ingredient of the amplitude is promoted to position space:
momenta are replaced by derivatives acting on the appropriate fields,
polarizations are replaced by explicit field insertions,
and propagator factors are represented by inverse differential operators.
This defines an operator-valued map
\begin{equation}
\OpPromotion(\,\cdot\,): \quad \mathcal{A}_m \;\longrightarrow\; \overline{\mathcal{L}}_m
\equiv \OpPromotion(\mathcal{A}_m)\,,
\end{equation}
where $\overline{\mathcal{L}}_m$ is a non-local operator that reproduces the full
$m$-point amplitude when used as a single contact vertex, and serves only as an
intermediate bookkeeping device.

\medskip

\noindent\textbf{Step 3: Subtract cut-constructible contributions.}
Applying the same promotion to $\slashed{\mathcal{A}}_m$ yields
$\OpPromotion(\slashed{\mathcal{A}}_m)$, which represents the contribution
already generated by lower-point interactions and propagators in the theory.
The physical local interaction at multiplicity $m$ is then defined by
\begin{equation}
\mathcal{L}_m \equiv
\OpPromotion(\mathcal{A}_m) - \OpPromotion(\slashed{\mathcal{A}}_m)\,.
\end{equation}
Crucially, by construction, all inverse-derivative structures cancel in this
difference, leaving a manifestly local operator.  This cancellation is
exemplified explicitly in \cref{sect:YMops}, where we show that through five
points no inverse-derivative operators survive.  From an amplitudes
perspective this is natural: once a full amplitude is stripped of all
cut-constructible contributions, the remainder can only consist of purely
local contact terms.

\medskip

\noindent\textbf{Step 4: Consistency with previous choices.}
Because generalized gauge choices and field redefinitions affect the precise
form of $\slashed{\mathcal{A}}_m$, the resulting $\mathcal{L}_m$ must be
consistent with the conventions adopted at lower multiplicity.
This consistency check is automatic when the same color-dual numerator
representation is used throughout, but it can be verified explicitly by
reconstructing amplitudes from the action as specifically discussed in \cref{sec:opPromotionCuts}.

\medskip

\noindent\textbf{Output.}
Iterating this procedure over all multiplicities defines an effective action
\begin{equation}
\mathcal{S} = \sum_{m\geq 3} \int d^D x\, \mathcal{L}_m
\end{equation}
whose Feynman rules reproduce the original color-dual amplitudes.
When the input amplitudes are related by double copy, the resulting operators
inherit a manifest double-copy structure at the level of the action.

\medskip

This algorithm provides a constructive inverse to the usual amplitude
computation: rather than deriving amplitudes from a given Lagrangian, we build
a local action directly from on-shell data while preserving the algebraic
structure responsible for color-kinematics duality and the double copy.

\section{Extracting Novel Contact Terms From Scattering Amplitudes}
\label{sec:contactExtraction}

The core challenge in constructing higher-derivative operators from scattering amplitudes lies in identifying what information is genuinely new at each multiplicity. That is, which terms in an $m$-point amplitude require the introduction of new $m$-field operators --- and which simply arise from lower-multiplicity physics via unitarity. This section presents a sharp and general method for isolating these novel contributions using graph-local data in color-dual or double-copy 
representations inspired by the systematic   application of generalized unitarity~\cite{Bern:2007ct, Bern:1994zx, Bern:1994cg, Bern:1995db,  Bern:1997sc, Britto:2004nc}  known as the method of maximal cuts~\cite{Bern:2007ct, Bern:2008pv, Bern:2010tq, Carrasco:2011hw,Carrasco:2015iwa}.  

\subsection{General Procedure}

An $m$-point amplitude contains both \emph{old} and \emph{new} information. The old content arises from lower-point interactions --- it propagates through factorization channels and corresponds to residues on physical poles. The new content, by contrast, originates from local $m$-field contact operators that contribute only starting at $m$-points.

To determine which new operators are needed at multiplicity $m$, we must isolate these novel contact contributions. This is possible because all cut-constructible terms --- those with physical poles --- are already determined by lower-multiplicity data. The remaining contact terms represent genuine new operator content.

Schematically we can simply say each amplitude is the sum over any new contacts, $\mathcal{C}$, plus residues of cuts over their uncut propagators, $\slashed{\mathcal{A}}$,
\begin{equation}
  \mathcal{A}_m =  \mathcal{C}_m + \slashed{\mathcal{A}}_m\,.
\end{equation}
This can be used  as a functional definition:
\begin{equation}
\label{fancyContactDef}
\boxed{
  \mathcal{C}_m \equiv   \mathcal{A}_m -  \slashed{\mathcal{A}}_m \,.}
\end{equation}
This functional identity defines the $m$-point contact term as the difference between the full amplitude and the sum of all contributions reconstructible from lower multiplicity via cuts.

In \sect{sec:promotion} we will define a prescription,  $\OpPromotion$, for promoting $m$-point amplitude expressions, $X_m$, to $m$-field operators, $\mathcal{O}_m$.  Specifically  the promotion $\OpPromotion$ will be defined such that the Feynman rule associated with the operator,
$\mathcal{O}_m \equiv \oop{\mathcal{O}} \circ ( X_m )$,
dresses the $m$-point contact graph to contribute precisely the expression $X_m$.
As such, we can write the unique $m$-field interaction, $\mathcal{L}_m$, from the novel contact information within the $m$-point amplitude,
\begin{align}
\mathcal{L}_m &= \oop{\mathcal{O}}\circ\mathcal{C}_m \\
&= \oop{\mathcal{O}}\circ \left(\mathcal{A}_m -  \slashed{\mathcal{A}}_m \right)\\
&= \oop{\mathcal{O}}(\mathcal{A}_m) -  \oop{\mathcal{O}}(\slashed{\mathcal{A}}_m)
\end{align}
This is morally how we will promote the color-dual representations of our amplitudes to  operator form in our Lagrangian density without overcounting the redundant information that lives on the cuts,
\begin{align}
\mathcal{S} &= \int d^d x \left[ \mathcal{L}_\text{free} + \sum_m \mathcal{L}_m \right] \\
&=   \int d^d x \left[ \mathcal{L}_\text{free} + \sum_m  \oop{\mathcal{O}}(\mathcal{A}_m) -   \sum_m \oop{\mathcal{O}}(\slashed{\mathcal{A}}_m) \right]\,.
\end{align} 
Of course the alignment of cut construction with Feynman rules in a particular gauge only has to hold on-shell after sum over all diagrams contributing to cuts, and generalized gauge choice consistency between contact terms is a matter of book-keeping given particular gauge choices. This means that practically one should ensure that one's generalized gauge choices are consistent with the desired amplitude form of the representation of cut data, and adjust the presentation of  $\oop{\mathcal{O}}(\slashed{\mathcal{A}}_m)$ consistent with the gauge choices being made in the action. Such representation accounting will be discussed in the operator promotion ~\cref{sec:opPromotionCuts}.  Being careful about the compatibility between operators at different multiplicity will not be new for EFT.  For amplitudes practitioners this is nothing more than the requirement that one's representation of the $k$-collapsed propagator contact-graph contributions depend crucially on decisions made in assigning cut data to fewer collapsed propagator contributions.

Now we will introduce a simple method to entirely isolate contact terms, multiplicity by multiplicity, via a systematic generalization of the method of maximal cuts.  The method of maximal cuts can always be applied to isolate novel contact contributions directly to the full amplitude. In general this involves a factorial number of operations due to the $(2n-5)!!$ graphs contributing.  If an amplitude can be decomposed into smaller gauge-invariant blocks such as color-ordered (or stripped) amplitudes all the better.  Each ordered amplitude of multiplicity $n$ has only $n(n-3)/2$ channels contributing to it, which means a Catalan number of cuts are required to isolate any novel contact.   If however we are in the privileged position of having a color-dual representation we can simply look at the half-ladder graph that has $(n-3)$ propagators, so only requires at most $2^{(n-3)}-1$ cuts to entirely isolate its contact.  In more general local representations, the same recursive subtraction can be applied graph-by-graph to each cubic topology---though the cut isolation can be performed on each graph independently. 

To do this, we make manifest the contribution of each cubic graph $g \in \mathcal{T}^{m}_3$ to the $m$-point contact term in the full amplitude. This is achieved by dressing the numerator $n(g)$ in terms of a minimal independent basis of momentum invariants chosen such that all inverse propagators of the graph $g$ appear as basis elements; this will allow us to identify the contact contribution as arising from terms proportional to the product of all inverse propagators of the graph. Take the set of $(m-3)$ inverse-propagators $\Delta(g)$ of a cubic $m$-point graph, $g$, such that $d(g)= \prod_{i=1}^{|\Delta|} \Delta_i$. We decompose the numerator such that each term is classified by the specific subset $P \subset \Delta(g)$ of inverse propagators to which it is proportional, 
\begin{align}
n(g) &= \sum_{P \subset \Delta(g)} \left(\prodList{P}\equiv P_1 \times \cdots \times P_{|P|} \right) n^{|P|}_P(g)\nonumber \\
 &= n^{(0)}(g) + \sum_{p\in \Delta(g)} p\, n^{(1)}_p(g) + \sum_{\{p,q\} \subset \Delta(g)} \,p \times q \times n^{(2)}_{\{p,q\}}(g) + \cdots + d(g)\, n^{|\Delta|}_\Delta(g)\,.
\label{eq:decompNumIntoProps}
\end{align}

For generic kinematics in $D$ dimensions (i.e., avoiding specific kinematic configurations leading to Gram determinant constraints among external momenta), the $m-3$ inverse propagators $\Delta(g) = \{p_1, \dots, p_{m-3}\}$ associated with a specific $m$-point tree-level cubic graph topology $g$ form a set of linearly independent kinematic variables. The presence of non-zero external masses $m_i^2$ does not diminish, and generally enhances, the space of available independent kinematic invariants, further ensuring the independence of these $m-3$ internal propagator variables for generic external momenta. This linear independence guarantees that any polynomial numerator $n(g)$ can be uniquely decomposed into terms classified by their explicit dependence on products of these inverse propagators, as in Eq.~\eqref{eq:decompNumIntoProps}. The coefficients $n^{(|P|)}_P(g)$ in this expansion are, by construction, polynomial functions of the kinematic invariants but are free of any further explicit factors from $\Delta(g) \setminus P$.

We therefore uniquely identify $n^{(|P|)}_P(g)$ such that it contains no factors of any inverse propagators present in the complement of $P$ in $\Delta$; it can, however, contain additional powers of $P_i \in P$. This means that $n^{(0)}(g)$ has no dependence on any inverse propagators $\Delta_i$ relevant to the graph $g$, and in general, for higher-derivative contributions, $n^{|\Delta|}_\Delta(g)$ could contain any number of additional powers of inverse propagators. For an $m$-point amplitude organized in terms of cubic graphs, it should be clear that the $m$-point contact contribution to the full amplitude is given simply by summing over the $n^{|\Delta(g)|}_{\Delta(g)}(g)$ contribution from each graph.
\begin{align}
   \mathcal{C}_m &= \sum_{g\in \Gamma^3_m} \frac{\left(d(g) n^{|\Delta(g)|}_{\Delta(g)}(g) \right)}{d(g)}\\
   &= \sum_{g\in \Gamma^3_m } n^{|\Delta(g)|}_{\Delta(g)}(g)\,.
\end{align}

This decomposition of the kinematic numerators into contributions $n^{(|P|)}_P(g)$ identified by their inverse-propagator dependence has a recursive definition in terms of applying cut conditions (taking inverse-propagators to vanish for massless particles).  The following subtraction recursively removes all subleading cut contributions from $n(g)$, isolating the term proportional to the full set of inverse propagators in $P$.\\
\noindent\textbf{Recursive extraction of contact numerators:}
\begin{equation}
\left(\prodList{P}\right) n^{|P|}_P(g)
= n(g)\big|_{(\Delta \setminus P) \to 0}
- \sum_{Q \subsetneqq P} \left(\prodList{Q}\right) n^{|Q|}_Q(g)
\label{eqn:contactRecursion}
\end{equation}
The first term corresponds to the graph numerator under cut conditions corresponding to all inverse propagators in the complement of $P$ in $\Delta$, written as $(\Delta \setminus P)$, taken on-shell. The second term is a sum over numerator contributions from $n^{(|Q|)}_Q(g)$ with inverse-propagator dependence $Q$ that is a subset of $P$, but excluding $P$ itself.
We can then define the unique cut contributions to an amplitude, graph by graph, as:
\begin{equation}
\label{slashedDef}
     \slashed{n}(g) = n(g) - d(g) n^{|\Delta|}_{\Delta}(g)\,.
\end{equation}

Now it is important to realize that what we mean here is the full numerator dressing over the propagators, so in gravitational theories, or double-copy theories where both copies contain kinematics, the situation is more subtle.
 Contact extraction must act on the full double-copy numerator $n \tilde{n}$, since cut conditions act on kinematics but not on color. While in Yang–Mills $\slashed{n}_{\text{YM}} = c \cdot \slashed{n}$, in gravity one must evaluate
\[
\slashed{n}_{\text{GR}} = \textbf{cut}(n \tilde{n})\,.
\]
This ensures that \(\mathcal{C}_m = \mathcal{A}_m - \slashed{\mathcal{A}}_m\) continues to isolate purely contact contributions even in the double-copy theories with two kinematic numerator weights. 

We summarize the algorithm as follows:

\begin{enumerate}
  \item Choose a graph $g$ in the amplitude.
  \item Identify the set of inverse propagators $\Delta(g)$.
  \item Write $n(g)$ as a sum over terms proportional to subsets $P \subset \Delta(g)$.
  \item Recursively extract $n^{(|P|)}_P(g)$ using cut conditions on $\Delta \setminus P$.
  \item The contact numerator is $n^{(|\Delta|)}_{\Delta}(g)$.
  \item Sum over all graphs to get $\mathcal{C}_m$.
\end{enumerate}

\subsection{Five point example}

As a schematic example, we consider a generic theory of a single particle type at five point tree level. There is a single cubic graph topology (the half ladder); we parameterize its numerator dressing in the following form: 
\begin{equation}
n(12345) = n \circ \fivegraphAct{1}{2}{3}{4}{5} = n^{(0)} + s_{12} \, n^{(1)}_{12} + s_{45} \, n^{(1)}_{45} + s_{12} s_{45} \, n^{(2)}_{12,45}
\end{equation}
The generic kinematic functions $n^{(0)}$, $n^{(1)}_{12}$, $ n^{(1)}_{45}$, and $n^{(2)}_{12,45}$ are defined such that dependence on the graph's inverse-propagators is explicit. Specifically, $n^{(0)}$ does not depend on $s_{12}$ or $s_{45}$; $n^{(1)}_{12}$ does not depend on $s_{45}$; and $n^{(1)}_{45}$ does not depend on $s_{12}$. 

This form is highly suggestive of the existence of contact operators in the theory. As the denominator for this graph is simply $d_g = s_{12}s_{45}$, the contributions arising from five-point contact operators are encoded entirely and exclusively by the function $n^{(2)}_{12,45}$. The other functions ($n^{(0)}$, $n^{(1)}_{12}$ and $ n^{(1)}_{45}$) are associated with the contributions of this graph to particular cuts of the five point amplitude; in particular, the existence of non-vanishing $n^{(1)}_{12}$ and $ n^{(1)}_{45}$ indicate the presence of a four-point contact. In particular theories, some of these functions can vanish entirely, corresponding to the absence of such contact operators; conversely, for theories with higher-derivative corrections of sufficient order, these functions can contain additional powers of their corresponding off-shell propagators. 

There are three possible cuts of the corresponding amplitude $\mathcal{A}_5$ to which this particular graph dressing can contribute: the maximal cut $\mathcal{A}_3(1,2,i)\mathcal{A}_3(-i,3,j)\mathcal{A}_3(-j,4,5)$ and the two next-to-maximal cuts $\mathcal{A}_3(1,2,i)\mathcal{A}_4(-i,3,4,5)$ and $\mathcal{A}_4(1,2,3,i)\mathcal{A}_3(-i,4,5)$.  Now, in some of the cuts below, the factorized objects correspond to higher-point
amplitudes with multiple channels. Here, however, we display only the contribution of
the specific cubic graph under consideration to each cut. 
Expressions are found by applying relevant cut conditions to the original form of the numerator: 
\begin{align}
\mathcal{A}_3(1,2,i)\mathcal{A}_3(-i,3,j)\mathcal{A}_3(-j,4,5):&& n \circ \fivegraphMcut{1}{2}{3}{4}{5} &=   n(12345)|_{ \left\{s_{12} ,s_{45} \right\} \to 0} \\
&&&=n^{(0)}\\
\mathcal{A}_3(1,2,i)\mathcal{A}_4(-i,3,4,5):&& n \circ \fivegraphNMcutLeft{1}{2}{3}{4}{5} &= n(12345)|_{s_{12} \to 0} \\
&&&=n^{(0)} + s_{45} \, n^{(1)}_{45} \\
\mathcal{A}_4(1,2,3,i)\mathcal{A}_3(-i,4,5):&& n \circ \fivegraphNMcutRight{1}{2}{3}{4}{5} &= n(12345)|_{s_{45} \to 0}\\
&&&=  n^{(0)} + s_{12} \, n^{(1)}_{12}
\end{align}
where we  indicate which cuts are responsible for the relevant kinematic limits. We see that all the kinematic information in the numerator except for the contact function $n^{(2)}_{12,45}$ contribute to these cuts! This is consistent with the removal of non-contact terms $\slashed{n}$ from the numerator $n$ by carefully constructing combinations of cut conditions.  

We can see clearly that simply adding all kinematic limits will result in overcounting of certain cut contributions. For example, each next-to-maximal cut contains data already specified by the maximal cut (namely, what we have parameterized as $n^{(0)}$). But as outlined in \eqn{eqn:contactRecursion} the approach of identifying each individual $n^{(|P|)}_P$ can occur recursively.

Starting with the $s_{12}$ cut condition,
\begin{equation}
n \circ \fivegraphNMcutLeft{1}{2}{3}{4}{5} =  n^{(0)} + s_{45}\, n^{(1)}_{45}\,
\end{equation} and subtracting the information from the maximal cut isolates the  $n^{(1)}_{45}$ term:
\begin{equation}
n \circ \fivegraphNMcutLeft{1}{2}{3}{4}{5} -n \circ \fivegraphMcut{1}{2}{3}{4}{5} = s_{45} \, n^{(1)}_{45} 
\end{equation}
We will denote this unique contribution by marking on the diagram the inverse propagator to which it is proportional, allowing us the diagramatic illustration of the recursive definition of
$s_{45}  \, n^{(1)}_{45}$,
\begin{equation}
n \circ \fivegraphNMcutLeftUnique{1}{2}{3}{4}{5} \equiv  n \circ \fivegraphNMcutLeft{1}{2}{3}{4}{5} - n \circ \fivegraphMcut{1}{2}{3}{4}{5}\,.
\end{equation}
Similarly,
\begin{equation}
n \circ \fivegraphNMcutRightUnique{1}{2}{3}{4}{5} \equiv  n \circ \fivegraphNMcutRight{1}{2}{3}{4}{5} - n \circ \fivegraphMcut{1}{2}{3}{4}{5}\,.
\end{equation}

And at last we come to the contact term $ s_{12} s_{45} n^{(1)}_{12,45}$ whose recursive definition, following \eqn{eqn:contactRecursion}, is simply:
\begin{align}
n \circ \fivegraphContactLeftRightUnique{1}{2}{3}{4}{5} &= n \circ \left(\fivegraphAct{1}{2}{3}{4}{5} -  \fivegraphNMcutRightUnique{1}{2}{3}{4}{5} -
  \fivegraphNMcutLeftUnique{1}{2}{3}{4}{5}  \right. \nn \\
  &~~~~~~~~~~ \left.-  \fivegraphMcut{1}{2}{3}{4}{5}\right)\,.
\end{align}
\begin{align}
n \circ \fivegraphContactLeftRightUnique{1}{2}{3}{4}{5} &= n \circ \left(\fivegraphAct{1}{2}{3}{4}{5} -  \fivegraphNMcutRightUnique{1}{2}{3}{4}{5} \right. \nn \\
&\qquad \left. -
  \fivegraphNMcutLeftUnique{1}{2}{3}{4}{5}   -  \fivegraphMcut{1}{2}{3}{4}{5}\right)\,. \nn
\end{align}
  
\subsection{Six point example}

We continue now to six-point with: 
\begin{equation}
n(123456) = n \circ \sixgraph{1}{2}{3}{4}{5}{6} 
\end{equation}

We again begin by parameterizing the numerator so that dependence on inverse propagators is made manifest: 
\begin{equation}
\begin{split}
n(123456) = & \, n^{(0)} + s_{12} \, n^{(1)}_{12} + s_{123} \,n^{(1)}_{123} + s_{56} \, n^{(1)}_{56} + s_{12}\, s_{123} \, n^{(2)}_{12,123}  \\ &+ s_{12}\, s_{56} \,  n^{(2)}_{12,56}  + s_{123}\, s_{56} \,  n^{(2)}_{123,56}  + s_{12}\,s_{123}\,s_{56} \,  n^{(3)}_{12,123,56} \,.
\end{split}
\end{equation}

Using the graphical conventions explained in the five point example, this can be written,
\begin{multline}
n \circ \sixgraph{1}{2}{3}{4}{5}{6}  = n\circ \left(  \sixGraphMC{1}{2}{3}{4}{5}{6}  \right. \\
\left. + \sixGraphNMCleftU{1}{2}{3}{4}{5}{6} + \sixGraphNMCmiddleU{1}{2}{3}{4}{5}{6} +\sixGraphNMCrightU{1}{2}{3}{4}{5}{6}  \right.  \\
 \left. + \sixGraphNNMCleftU{1}{2}{3}{4}{5}{6} + \sixGraphNNMCmiddleU{1}{2}{3}{4}{5}{6} +  \sixGraphNNMCrightU{1}{2}{3}{4}{5}{6} \right.  \\
\left.  + \sixGraphUniqueContact{1}{2}{3}{4}{5}{6}  \right)
\end{multline}

The recursive definition of the contact is then simply given by:
\begin{multline}
n \circ \sixGraphUniqueContact{1}{2}{3}{4}{5}{6} = n\circ \left(\sixgraph{1}{2}{3}{4}{5}{6} - \right. \\
 \left. - \sixGraphNNMCleftU{1}{2}{3}{4}{5}{6} - \sixGraphNNMCmiddleU{1}{2}{3}{4}{5}{6} -  \sixGraphNNMCrightU{1}{2}{3}{4}{5}{6} \right.  \\
\left. -\sixGraphNMCleftU{1}{2}{3}{4}{5}{6} - \sixGraphNMCmiddleU{1}{2}{3}{4}{5}{6} -\sixGraphNMCrightU{1}{2}{3}{4}{5}{6}  \right.  \\
\left. -\sixGraphMC{1}{2}{3}{4}{5}{6}  \right)
\end{multline}\,
with e.g.
\begin{align}
\sixGraphNNMCrightU{1}{2}{3}{4}{5}{6} &=  s_{12} \, s_{123} \, n^{(2)}_{12,123} \nn \\
&= n\circ \left(  \sixGraphNNMCright{1}{2}{3}{4}{5}{6}  -\sixGraphNMCleftU{1}{2}{3}{4}{5}{6} \right.\nn \\ 
     & \left.  - \sixGraphNMCmiddleU{1}{2}{3}{4}{5}{6}  - \sixGraphMC{1}{2}{3}{4}{5}{6}  \right) \nn\\ 
   &= n(123456)|_{s_{56}\to0} - \left( s_{12} n^{(1)}_{12} \equiv  n(123456)|_{\left\{s_{123},s_{56}\right\}\to0} - n^{(0)} \right) \nn  \\
   &~~~-  \left(  s_{123} n^{(1)}_{123}  \equiv  n(123456)|_{\left\{s_{12},s_{56}\right\}\to0}-  n^{(0)} \right) - n^{(0)}  \,.
 \end{align}

\subsection{More on gravitational contact isolation}
\label{sect:gravContact}
We take this as an opportunity to detail how all-multiplicity gravitational ($\sqrt{-g}R$) contact terms can appear via double copy. The basic mechanism is conceptually straightforward, but it is instructive to see it illustrated explicitly in the context of local color-dual representations of Yang–Mills theory.

We will depict the emergence of the gravitational contacts required for linearized diffeomorphism below, first noting how $n_{\text{YM}}(12345)$ behaves for the half-ladder with at most a $4$-point contact contributing singularly per channel,
\begin{align}
n_{\text{YM}}(12345) 
&= n_{\text{YM}}^{(0)} + s_{12} \, n_{\text{YM}}^{(1)}{}_{12} + s_{45} \, n_{\text{YM}}^{(1)}{}_{45} \\
&=  \left( {\fivegraphMcut{1}{2}{3}{4}{5}} +\fivegraphNMcutLeftUnique{1}{2}{3}{4}{5}  +
 \fivegraphNMcutRightUnique{1}{2}{3}{4}{5}  \right)\,. 
\end{align}
The double-copy numerator for gravity then is quite simply of the following form,
\begin{align}
n_{\text{GR}}&= n_{\text{YM}} \times \widetilde{n_{\text{YM}}}\\
&=  \left( {\fivegraphMcut{1}{2}{3}{4}{5}} +\fivegraphNMcutLeftUnique{1}{2}{3}{4}{5}  +
 \fivegraphNMcutRightUnique{1}{2}{3}{4}{5}  \right)\times \\
&\qquad  \left( 
\widetilde{ \fivegraphMcut{1}{2}{3}{4}{5}  } +
 \widetilde{ \fivegraphNMcutLeftUnique{1}{2}{3}{4}{5} } +
 \widetilde{ \fivegraphNMcutRightUnique{1}{2}{3}{4}{5} } \right)\,.
\end{align}
The novel five-point gravitational contact arises from the cross terms that in concert from both sides cancel all propagators,
\begin{align}
n_\text{GR} \circ \fivegraphContactLeftRightUnique{1}{2}{3}{4}{5} &=  n_{\text{GR}} - \slashed{n}_{\text{GR}}\\
 &= \fivegraphNMcutLeftUnique{1}{2}{3}{4}{5}  \widetilde{ \fivegraphNMcutRightUnique{1}{2}{3}{4}{5} }  \nonumber \\
&\qquad   +  \widetilde{ \fivegraphNMcutLeftUnique{1}{2}{3}{4}{5} }   \fivegraphNMcutRightUnique{1}{2}{3}{4}{5} \\
 &= s_{12} s_{45}  \left( n^{(1)}_{12}  \tilde{n}^{(1)}_{45}+
 n^{(1)}_{45}  \tilde{n}^{(1)}_{12} \right) \,.
\end{align}

Similarly at six points, a color-dual Yang-Mills numerator can have the following form with at most two collapsed propagators at a time,
\begin{multline}
n^{\text{YM}} \circ \sixgraph{1}{2}{3}{4}{5}{6}  = \left(  \sixGraphMC{1}{2}{3}{4}{5}{6}  \right. \\
\left. + \sixGraphNMCleftU{1}{2}{3}{4}{5}{6} + \sixGraphNMCmiddleU{1}{2}{3}{4}{5}{6} +\sixGraphNMCrightU{1}{2}{3}{4}{5}{6}  \right.  \\
 \left. + \sixGraphNNMCleftU{1}{2}{3}{4}{5}{6} + \sixGraphNNMCmiddleU{1}{2}{3}{4}{5}{6} +  \sixGraphNNMCrightU{1}{2}{3}{4}{5}{6} \right)
\end{multline}
Notice for example the appearance of what appear to be five-point contacts in say the last contact graph which collapses propagators for $(k_1+k_2)^2$ and $(k_1+k_2+k_3)^2$.  These are of course spurious and cancel against similar terms in the sum over graphs contributing to gauge-invariant objects such as the full-amplitude or even ordered cuts like $A(1234l) A(-l 56)$. Such terms can contribute in cross-terms to the gravitational contact depending on the generalized gauge and so must in general still be considered in all topologies that contribute to the full amplitude.
 
 \subsection{Spurious non-locality}
 It is possible to formally dress kinematic numerators of cubic graphs with functions that may contain poles.   A fine example would be the virtuous numerators of ref.~\cite{Broedel:2011pd}.  Introducing spurious non-locality in numerator dressings is allowed as long as any gauge invariant construction involving the numerators, such as ordered amplitudes, has poles only where physical.  Virtuous representations where external states are encoded in ordered amplitudes make this required property manifest.  In such cases, however, the above individual numerator contact isolation procedure is not guaranteed to work. One can resort to method of maximal cuts as applied to local gauge-invariant blocks --- ordered amplitudes for gauge theories, and full amplitudes otherwise. We hope it does not escape the reader that this is literally the equivalent of building up a local representation.

\section{The Operator Promotion}
\label{sec:promotion}

We will now to establish a mapping from momentum space amplitudes to position space operators. More sharply, for some term $n$ contributing to an amplitude $\mathcal{A}_m$, we will prescribe an operator promotion $n \rightarrow \oop{\mathcal{O}}(n) \equiv \oop{n}$ to an $m$-field operator $\oop{n}$. This operator will by construction yield a Feynman rule that dresses the $m$-point vertex, and thus first shows up as a contribution to the $m$-point contact diagram. We must now establish which contributions $n$ to the amplitude we need to specify in order to describe the entire theory at said multiplicity. 

Graph-organized double-copy amplitudes, as established, are written as sum over contributions from all relevant distinct cubic graphs $\Gamma$. We will write this contribution from each graph $g \in \Gamma$ as $Q(g)$, so $\mathcal{A} = \sum_{g \in \Gamma} Q(g)$. Such a contribution $Q(g)$ is generically a function of particle momenta $k_i$, external state wavefunctions\footnote{\textit{Exempli gratia}, polarizations $\varepsilon_i(k_i)$; spinors $u_i(k_i)$, $v_i(k_i)$; 1; etcetera}, and gauge group structure constants $f^{abc}$ and generators $T^a_{ij}$. We can map this back to position space via Fourier transform, replacing momenta $k^\mu$ with derivatives $\partial^\mu$ and replacing wavefunctions with appropriately Lorentz indexed external fields, e.g. $\varepsilon_\mu \to A_\mu$.

While generically there are many distinct cubic graphs  relevant to a given set of external states, we only need to consider the contribution of one graph per basis topology when writing down the operator. Schematically, we can understand this as follows. 

Let us initially restrict ourselves to an amplitude with only a single cubic graph topology; for example, Yang Mills at four points. All three graphs $\Gamma = \{g_s, g_t, g_u\}$ can be considered permutations of the external leg labels on the first graph (chosen arbitrarily to be the $s$-channel). If we label $g_s$ as $g(1234)$, we can identify $g_t$ as $g(4123)$ and $g_u$ as $g(4231)$. These are only 3 of 24 permutations of the labels $(1234)$; all remaining permutations are isomorphic to one of the three channels. We can instead choose to write the full amplitude as a sum over all such permutations, modded out by the appropriate symmetry factor:

\begin{equation}
\mathcal{A}_4 = \frac{1}{8} \sum_{\sigma \in S_4} Q(g(\sigma))
\end{equation}

Consider the operator obtained from mapping just the $s$-channel contribution $Q(g(1234))$ to position space as specified above. This four-field operator will give rise to a four-point vertex Feynman rule $\mathcal{V}_4$. Feynman rules, at their core, are obtained by performing functional derivatives on momentum-space operators, which has the effect of summing over all permutations of labels on identical fields: 

\begin{equation}
\mathcal{V}_4 \ce_1 \ce_2 \ce_3 \ce_4  = \sum_{\sigma \in S_4} Q(g(\sigma))
\end{equation}

Feynman rules inherently sum over all such permutations, which we can identify as precisely identical to the sum over all graphs in the full amplitude $\mathcal{A}_4$. Thus, we need only consider the contribution from \textit{a single graph for each topology} in order to write an operator that encapsulates the contributions from \textit{all graphs.}

Color-kinematics duality simplifies this description further. Color-kinematics Jacobi identities relate the numerator dressings of distinct cubic graph topologies; all such dressings can be written in terms of linear relations upon relabelings of those of the basis graphs. As a result, from a single graph dressing for each \textit{basis} topology, we can write down the full operator that gives rise to the complete full amplitude $\mathcal{A}$.

In this approach, each multiplicity/set of external states is considered separately and gives rise to a (generically non-local) contact operator. This contrasts significantly with the typical approach, where there can be a limited number of low-point operators, and higher multiplicity amplitudes are generated by sewing together such vertices with internal propagators. By focusing on each multiplicity separately, we are able to preserve a manifestly color-dual structure for the operators. This allows us to see clearly that the operators can be double copied to write down operators for different theories. 

\subsection{Double-Copy Structure}

How do we represent double-copy structure at the level of fields? We want both sides of the double copy to be on equal footing, so each conceptual copy is assigned field content. This decomposition is most easily understood through some examples for common double-copy theories, summarized in Table~\ref{tab:dc_fields}. In this notation, $\otimes$ signifies the conceptual double-copy product, and the fields on the right-hand side represent the constituent `single-copy' or root structures.

\begin{table}[htbp]
\centering
\caption{Color-dual double-copy field decompositions for various theories. $A_\mu$ represents a generic vector (kinematic) structure, $\varphi^a$ an adjoint scalar (color) structure, and $\pi$ a generic scalar (kinematic, e.g., NLSM pion) structure. Tilded objects denote distinct copies if necessary.}
\label{tab:dc_fields}
\begin{tabular}{l l c c l}
\toprule
Double-Copied Field & Symbol & Root 1 & Root 2 & Interpretation \\
\midrule
Graviton & $h_{\mu\nu}(x)$ & $A_\mu(x)$ & $A_\nu(x)$ & Vector $\otimes$ Vector \\
Gauge Boson (YM) & $A_\mu^a(x)$ & $A_\mu(x)$ & $\varphi^a(x)$ & Vector$\otimes$ Scalar (color) \\
Bi-adjoint Scalar & $\phi^{a\tilde{a}}(x)$ & $\varphi^a(x)$ & $\tilde{\varphi}^{\tilde{a}}(x)$ & Scalar$_1$ $\otimes$ Scalar$_2$ \\
Pion (NLSM-like) & $\pi^a(x)$ & $\varphi^a(x)$ & $\pi(x)$ &  Scalar ($\partial^2$) $\otimes$ Scalar (color) \\
Born-Infeld Photon & $A_\mu^{\text{BI}}(x)$ & $A_\mu(x)$ & $\pi(x)$ & Vector$\otimes$ Scalar ($\partial^2$) \\
Z-theory & $Z^{aA}(x)$ & $\varphi^{a}(x)$ & $z^{A}(x)$ & Scalar (color) $\otimes$  $\alpha'^\infty$-Scalar (color) \\
Open Superstring (vector)& $\text{OSS}^{A}_\mu(x)$ & $A_\mu(x)$ & $z^{A}(x)$ & Vector $\otimes$  $\alpha'^\infty$-Scalar (color)\\
$($DF$){}^2$+YM &               $ B^a_{\mu} (x)$ & $ B_{\mu}(x)$ & $ \varphi^a(x) $ & Vector $\otimes $ Scalar (color) \\
Open Bosonic String (vector)& $\text{OBS}^{A}_\mu(x)$ & $B_{\mu}(x)$ & $z^{A}(x)$ & Vector $\otimes$ $\alpha'^\infty$-scalar (color)  \\
Closed Superstring (graviton)& $\text{CSS}_{\mu\nu}$ & $A_{\mu}(x)$ & $(A)^{\text{sv}}_{\mu}(x)$ & Vector $\otimes$  $\alpha'^\infty$-Vector \\
Heterotic String (graviton)& $\text{HS}_{\mu\nu}$ & $B_{\mu}(x)$ & $(A)^{\text{sv}}_{\mu}(x)$ & Vector $\otimes$  $\alpha'^\infty$-Vector \\
\bottomrule
\end{tabular}
\end{table}

For the graviton, this decomposition is of course very familiar. For something like the Yang-Mills gauge field, $A_\mu^a(x)$, our decomposition implies that the color information (carried by $\varphi^a$) is conceptually separated from the vector kinematic structure (carried by $A_\mu$). The scalar fields $\varphi^a$ will appear in the operator promotions of color factors $c_g$, while the colorless vector structures $A_\mu$ will correspond to kinematic numerators $n_g$. Similarly, the bi-adjoint scalar field $\phi^{a\tilde{a}}(x)$ is decomposed into two distinct color-carrying scalar structures. This 'factorized' field representation is advantageous as it allows us to systematically construct operators for a plethora of theories related by double copy by mixing and matching these constituent operator types, as long as one is diligent about the bookkeeping of symmetries and physical state projections.  It would be an error to assign dilaton or antisymmetric 2-form operators to gravity --- as it would be a mistake at the level of amplitudes. This places the burden of correct book-keeping on physical observables --- where projection to desired physical states is part and parcel to unitarity methods.  Some of the more exotic fields ($\alpha'^\infty$) involve non-local interactions (at least, infinite towers of higher-derivative operators), and yet we can still understand their tree-level amplitudes at least as field-theory double-copies, and can use the approach of this current paper to write operators to reproduce those scattering amplitudes.  In the spirit of sheer pragmatism of writing down operators which generate the amplitudes at hand, we will defer addressing almost all subtleties of this construction to former work (validity of the theories as quantum field theories, unitarity, etc), and simply discuss examples of such constructions in \cref{sec:hdopAndStrings}.

Let us address a critical point right here.  It is perfectly acceptable to regard the above mapping as merely a creative way of labeling well known and familiar fields.  That would be the most conservative stance and entirely appropriate if all one wished to do was have an elegant way of classifying and constructing operators.  There is a path forward to a state-level understanding and appreciation of double-copy that is however now available to us, and one that we will address directly in \cref{sec:doublecopyGrav}.

For Yang Mills theory, this means we will write down separate operators for the color factor $c_g$, the kinematic numerator dressing $n_g$, and the inverse propagator $1/d_g$, and then multiply these together to constitute the overall graph contribution $Q(g)$. For $n_g$, the steps to arriving at an operator are fairly familiar: convert polarizations to vector fields, and convert momenta to derivatives acting on the appropriate vector fields.

The color factor $c_g$ is simply a contraction of structure constants for the relevant gauge group, so it can be tempting to identify the corresponding operator as simply this contraction, as it needs no conversion from momentum to position space. However, we would like our operator promotions to be as theory-agnostic as possible: each numerator factor, no matter its nature, should have an associated set of fields. We will choose to write down the $c_g$ operator promotion by inserting scalar fields carrying adjoint color indices. For example, at four points,

\begin{equation}
\oop{c}_g = f^{abe}f^{ecd} \varphi^a(x) \varphi^b(x) \varphi^c(x) \varphi^d(x)
\end{equation}  

Finally, we promote the inverse propagator by generically allowing the Lagrangian to be non-local. We want to be able to introduce an operator that targets propagator structure associated with a Feynman graph.  For example, consider an $s$-channel, $(k_1+k_2)^2$, four point graph.  We will want a propagator operator $\frac{1}{\oop{s}}$ that yields a Feynman rule that generates propagator kinematics, schematically,
\begin{equation}
\frac{1}{8} \frac{\oop{c}_s \oop{n}_s }{\oop{s}} \to  \frac{c_s n_s}{s} + \frac{c_t n_t}{t} + \frac{c_u n_u}{u}\,.
\end{equation} 
All such non-locality will be canceled upon subtraction of cut-terms, but it will be key to our organization to be able to target specific graphs. It turns out the writing down of such operators is greatly facilitated by introducing a redundancy in spacetime variables as we describe in the next section.

\subsection{Operators can look like Amplitudes} 

As theories increase in complexity (especially as higher-derivative corrections are included), the process of mapping momenta in numerators to derivatives in operators becomes quite verbose. A key issue is that $k_i^\mu$ must be translated to $\partial^\mu$ acting specifically upon field $\phi_i$. This can obfuscate the structure of the operator, as one must match up Lorentz indices of derivatives acting upon different fields to realize the corresponding Lorentz invariant dot product that will arise in the amplitude.

We introduce the following measure: 
\begin{equation}
\boxed{
\mD {m} \equiv \int \prod_i^m \,d^D x_i\,  d^D \, \tilde{x}_i \, \delta^{D}(x - x_i)\,\delta^{D}(x - \tilde{x}_i)\,
}
\end{equation}
where we include both $x_i$ and $\tilde{x}_i$ as auxiliary coordinates for the left and right sides of the double-copy inspired field decomposition. 

We will also write: 
\begin{equation}
\boxed{
\fieldsBold{\phi}{a}{m}{q} \equiv \phi^{a_1}(q_1) \cdots \phi^{a_m}(q_m)\,.
}\end{equation}

As an example, let us consider the four point amplitude for the non-linear sigma model.  The pion fields cary color, so in our double-copy framework we will annotate this as:
\begin{equation}
 \pi^{a}(x)\equiv(\phi^{a} \otimes \pi)(x) =\mD {1} \phi^{a}(\tilde{x}_1) \otimes \pi(x_1)
\end{equation}
The amplitude is given in terms of cubic graphs via some color-factor, kinematic numerator, and propagator as:
\begin{equation}
\mathcal{A}_4 = \frac{ c_s n_s}{s}+\text{perms}\,.
\end{equation}
We will see that we can write the position space operator that generates this amplitude as follows:
\begin{equation}
\mathcal{O}_4 =\mD{4} \,\oop{\mathcal{A}}_4\, \fieldsBold{\pi}{a}{4}{x} \,.
\end{equation}
with
\begin{equation}
\oop{\mathcal{A}}_4\equiv \frac{1}{8} \frac{ \oop{c}_s \oop{n}_s}{\oop{s}}\,.
\end{equation}

To see what the hatted objects will mean let's first be very clear about the building blocks of the amplitude written in terms of cubic graphs.
We have a color factor for the $s$-channel graph given by:
\begin{equation}
c_s\equiv f^{a_1 a_2 e} f^{e a_3 a_4}\,.
\end{equation}
a propagator $s\equiv (k_1+k_2)^2$, and 
a numerator factor
\begin{equation}
n_s = \frac{1}{3} s(u-t)
\end{equation}
We note the presence of the inverse propagator $s$ in the kinematic numerator means that we are really describing a four-point contact amplitude.  Indeed the entire amplitude can be written as:
\begin{align}
\mathcal{A}_4 &= c_s (u-t) + c_t (u-s) + c_u (s -t) \\
    &=  c_s (2 u -s -t) + c_u (-u-t)\\
    &= c_s u + ( s\leftrightarrow u )\,.
\end{align}
We will stick with our graph organization at the moments as it allows the duality between color and kinematics to be manifest, and will make it straightforward to double-copy to generate Born-Infeld shortly.

One could imagine writing a position space operator that generates something like $n_s$ at four-points (recalling that $s_{ij}=(k_i+k_j)^2=2 k_i\cdot k_j$ for on-shell $k_i^2=0$) as follows,
\begin{multline}
\mathcal{O} = 4\times \frac{1}{3}   \left[ \left(  \partial_\mu \partial_\nu \pi^{a_1}(x) \right) 
\left(\partial^\mu \pi^{a_2} (x) \right) \left(\partial^\nu \pi^{a_3}(x) \right) \left( \pi^{a_4}(x) \right)  \right. \\
\left. -  \left(\partial_\mu \partial_\nu \pi^{a_1}(x)\right) \left(\partial^\mu \pi^{a_2} (x) \right) \left( \pi^{a_3}(x) \right) \left(\partial^\nu \pi^{a_4}(x) \right)\right]\,.
\end{multline}
Arguably this is far less pleasant than the amplitude.  We can make the structure more clear by borrowing a trick of ref.~\cite{deRoo:2003xv} and assigning each field its own auxiliary position variable:
\begin{equation}
\phi_i(x) = \int d^d x_i \delta^{(d)}\left(x - x_i \right) \phi_i(x_i)\,.
\end{equation}
As such, $k_{i,\mu}$ can be mapped to: 
\begin{equation}
\partial_\mu\phi_i(x) =   \int d^d x_i \delta^{(d)}\left(x - x_i \right) \left(\frac{\partial}{\partial x_i^\mu} \right) \phi_i(x_i)\,.
\end{equation}
We will define $\partial_{i,\mu} \equiv \partial / \partial x_i^\mu$ for notational convenience. So far, this is, of course, a trivial statement. However, the benefits of this become clear when we consider the full product of fields appearing in an operator. For the NLSM numerator, we can rewrite the operator promotion using these auxiliary spacetime coordinates as follows: 
\begin{align}
\mathcal{O} &= \frac{4}{3}  \int \prod_{i = 1}^{4} \left( d^d x_i \delta^{(d)} \left(x - x_i \right) \right)  [  \\ 
&  \left(\partial_{1,\mu} \partial_{1,\nu} \pi^{a_1}_1(x_1) \right) \left(\partial_2^\mu \pi^{a_2}_2 (x_2) \right) \left(\partial_3^\nu \pi^{a_3}_3(x_3) \right) \left( \pi^{a_4}_4(x_4) \right) \nonumber \\ &-  \left(\partial_{1,\mu} \partial_{1,\nu} \pi^{a_1}_1(x_1) \right) \left(\partial_2^\mu \pi^{a_2}_2 (x_2) \right) \left( \pi^{a_3}_3(x_3) \right) \left(\partial_4^\nu \pi^{a_4}_4(x_4) \right) ] \nonumber
\end{align}
 Since $\partial_{i,\mu}$ only acts on the $i^{th}$ field, we can factor all the partial derivatives out from the product of fields (and dot them together into Lorentz invariants):
\begin{align}
\int\mathcal{D}_4 \oop{n}  \fieldsBold{\pi}{a}{4}{x}   &=\frac{4}{3} \int \prod_{i = 1}^{4} \left( d^d x_i \delta^{(d)} \left(x - x_i \right) \right)  \left(\partial_1 \cdot \partial_2\right) \left( \left(\partial_1 \cdot \partial_3\right) -  \left(\partial_1 \cdot \partial_3\right) \right)     \prod \pi^{a_i}_i(x_i)
\end{align}

Here, we can identify the $s(u-t)$ structure immediately by inspection!   Indeed let's sharpen the game by introducing 
\begin{equation}
\boxed{
\oop{s}_{ij}\equiv(\partial_i+\partial_j)^2\,,
}
\end{equation}
with then $\oop{s}\equiv\oop{s}_{12}$, $\oop{t}\equiv\oop{s}_{23}$, $\oop{u}\equiv\oop{s}_{13}$, so that:
\begin{equation}
\oop{n} = \oop{s} \left(\oop{t}-\oop{u}\right) \,.
\end{equation}

With $\oop{c}^{\mathbf{a}}_s \equiv c^{\mathbf{a}}_s$,  our amplitude is generated by an operator written:
\begin{align}
\mathcal{O}_4 &= \int \mathcal{D}_4\,  \frac{1}{8}  \frac{ \oop{c}^{\mathbf{a}}_s \oop{n}_s}{\oop{s}} \, \fieldsBold{\pi}{a}{4}{x}\\
    &= \int \mathcal{D}_4\,  \frac{1}{8} \oop{c}^{\mathbf{a}}_s( \oop{t}-\oop{u} ) \, \fieldsBold{\pi}{a}{4}{x}
\end{align}

How should we maximize the ease of translating vector kinematic numerators to operators?  Let vector fields $A_\mu$ carry their own Lorentz indices, so our $m$-field vector product, $\fieldsBold{A}{a}{\mu,m}{q}$, will be defined as:
\begin{equation}
\boxed{
\fieldsBold{A}{a}{\mu, m}{q} \equiv A^{a_1}_{\mu_1}(q_1) \cdots A^{a_m}_{\mu_m}(q_m)\,.
}\end{equation}

Polarizations $\varepsilon_{\mu_i}(k_i)$ appearing in Lorentz invariants in scattering amplitude expressions are then simply removed to allow the metric contractions to indicate which field is being contracted,
\begin{equation}
\eta^{\mu_i \nu} \varepsilon_{\mu_i} \to \eta^{\mu_i \nu} \oop{\varepsilon}_{\mu_i} \equiv \eta^{\mu_i \nu} \,.
\end{equation}
We often annotate the polarization associated with the $j$th external leg as $\varepsilon_j$ when its Lorentz indices are suppressed in expressions, as per:
\begin{equation}
n_3 =\left[ \left( \varepsilon_1 \cdot \varepsilon_2 \right) \left(\varepsilon_3 \cdot (k_1 -k_2) \right) + \text{cyclic}  \right]\,.
\end{equation}
As such we would write the three-poing Yang-Mills amplitude as
\begin{equation}
\mathcal{A}^{\text{YM}}_3=g f^{a_1 a_2 a_3} n_3 \,,
\end{equation}
and the three-point operator associated with this cubic vertex interaction is simply given as:
\begin{equation}
\mathcal{O}^{\text{YM}}_3 = g \int \mathcal{D}_3 \frac{1}{6} f^{a_1 a_2 a_3} \, \oop{n}^{\boldsymbol{\mu}}_3 \,\fieldsBold{A}{a}{\mu,3}{x} \,,
\end{equation}
with the factor of 6 from the automorphic symmetry of the 3 vertex, $\mathcal{D}_3$ and 
\begin{align}
\oop{n}^{\boldsymbol{\mu}}_3(x) &=\left[ \left( \oop{\varepsilon}_1 \cdot \oop{\varepsilon}_2 \right) \left(\oop{\varepsilon}_3 \cdot (\oop{k}_1 -\oop{k}_2) \right) + \text{cyclic}  \right] \, \\
&=\left[ \left( \eta^{\mu_1 \mu_2} \right) \left(  \eta^{\mu_3 \rho} ( \partial_\rho{}_{x_1} - \partial_\rho{}_{x_2}) \right) + \text{cyclic}  \right]\\
&=-2 \left[ \left( \eta^{\mu_1 \mu_2} \right) \left(  \eta^{\mu_3 \rho} \partial_\rho{}_{x_2}) \right) + \text{cyclic}  \right]\,.
\end{align}
It is not hard to see that this explicitly reproduces the cubic term from $\mathcal{L}^{\text YM}=-\frac{1}{4} F^{a}_{\mu \nu} F^{a\,\mu\nu}$,
\begin{align}
\Lag^{\text{YM}} _3 &=  \frac{1}{6} \int \mathcal{D}_3 \, g \, f^{a_1 a_2 a_3} \, \oop{n}^{\boldsymbol{\mu}}_3 \,\fieldsBold{A}{a}{\mu,3}{x}  \\
&= f^{a_1 a_2 a_3}  \int \prod_i^3 \,d^d x_i\, \delta^{(d)}(x - x_i) \oop{n}^{\boldsymbol{\mu}}_3\fieldsBold{A}{a}{\mu,3}{x} \\ 
  &=-2\frac{g}{6}  f^{a_1 a_2 a_3} \left[ 
   A^{a_1\, \mu} \partial^\nu A^{a_2}_{\mu} A^{a_3}_{\nu} + \text{cyclic} \right]\, \\
  &=- g f^{a_1 a_2 a_3} A^{a_1\, \mu} \partial^\nu A^{a_2}_{\mu} A^{a_3}_{\nu} \, .
  \label{3ptLymOp}
 \end{align}
 
\newcommand{\optildeps}[1]{ { \oop {\tilde{ \varepsilon}}}_{#1}}

Given the fact that graviton polarizations factorize, and that graviton graph numerators at tree-level can be written as the double-copy of color-dual Yang-Mills numerators we can easily write the three-graviton interaction operator as simply:
\begin{equation}
\mathcal{O}^{\text{GR}}_3 = \frac{\kappa}{2} \frac{1}{6} \int \mathcal{D}_3 \, (\oop{n}^\mu_3 \,\oop{\tilde{n}}^\nu_3)_{\text{ST}} \, \fieldsBold{A}{}{\mu}{x} \otimes_{\text{ST}} \fieldsBold{\tilde{A}}{}{\nu}{\tilde{x}}  \,.
\end{equation}
Now of course, written in our new form this it's a delight in contrast to manifestly position space operators.  This is a feature shared in general with EFT operators, especially as number of derivatives increase. To see how to begin expose a more verbose form we begin to expand out:
\begin{align}
(\oop{n}^\mu_3\oop{\tilde{n}}^\nu_3)_{\text{ST}} &= 4  \left . \Big( ( \oop{\varepsilon}_1 \cdot \oop{\varepsilon}_2)\left( \oop{\varepsilon}_3 \! \cdot \! \oop{k}_2 \right)  + \text{cyclic} \Big) \Big( ( \optildeps{1} \cdot \optildeps{2}) \left( \optildeps{3}  \! \cdot \! \oop{\tilde{ k}}_2 \right)  + \text{cyclic} \Big) ~ \right |_{~\text{ sym}} \\
&= 4 \left . \Big( ( \eta^{\mu_1 \mu_2}) \left(\eta^{\mu_3 \rho} \partial_2{}_\rho \right)  + \text{cyclic} \Big) \Big( (
 \eta^{\nu_1 \nu_2}) \left(  \eta^{\nu_3 \sigma} \partial_{\tilde{2}}{}_\sigma \right)  + \text{cyclic}  \Big) \right |_{~\text{sym}} \,.
\end{align}
A brute writing out of a more familiar form of the operator proceeds as the other cases
\begin{align}
\mathcal{O}^{\text{GR}}_3 &= \frac{\kappa}{2} \frac{1}{6} \int \mathcal{D}_3 \, (\oop{n}^\mu_3 \,\oop{\tilde{n}}^\nu_3)_{\text{ST}} \, ( \fieldsBold{h}{}{\mu\nu,3}{x}  \equiv \fieldsBold{A}{}{\mu}{x} \otimes \fieldsBold{\tilde{A}}{}{\nu}{\tilde{x}} )\,. \\
&= \frac{\kappa}{2} \frac{4}{6} \int \mathcal{D}_3 \,   \Big( ( \eta^{\mu_1 \mu_2}) \left(\eta^{\mu_3 \rho} 
\partial_2{}_\rho \right)  + \text{cyclic} \Big) \Big( (
 \eta^{\nu_1 \nu_2}) \left(  \eta^{\nu_3 \sigma}\partial_{\tilde{2}}{}_\sigma \right)  + \text{cyclic}  \Big)  \, \fieldsBold{h}{}{\mu\nu,3}{x} \, \nonumber \\
&=\frac{\kappa}{4} \left(
 \eta^{\mu_1\mu_2}\eta^{\mu_3 \rho } \eta^{\nu_1\nu_3} \eta^{ \nu_2\sigma} \partial \text{}_{\sigma
}\text{}h_{\mu_1\nu_1}\text{} \partial \text{}_{\rho }\text{}h_{\mu_2\nu_2} h_{\mu_3\nu_3} +
 \eta^{\mu_1\mu_2}\eta^{ \mu_3 \rho}  \eta^{\nu_1\nu_2} \eta^{ \nu_3\sigma} h_{\mu_1\nu_1}
\partial \text{}_{\rho ,\sigma }\text{}h_{\mu_2\nu_2}h_{\mu_3\nu_3} \text{} \right. \nonumber \\
&\left.+
\eta^{\mu_1\mu_2} \eta^{ \mu_3\rho} \eta^{ \nu_1\sigma} \eta^{\nu_2\nu_3} h_{\mu_1\nu_1} \partial \text{}_{\rho}\text{}h_{\mu_2\nu_2}\text{} \partial \text{}_{\sigma }\text{}h_{\mu_3\nu_3}\text{}+ 
\eta^{\mu_1\mu_3}\eta^{ \mu_2\rho}  \eta^{\nu_1\nu_3}\eta^{ \nu_2\sigma}  h_{\mu_2\nu_2} h_{\mu_3\nu_3} \partial \text{}_{\rho ,\sigma }\text{}h_{\mu_1\nu_1}\text{} \right. \nonumber \\
&\left.+
\eta^{\mu_1\mu_3} \eta^{\mu_2\rho }  \partial \text{}_{\rho }\text{}h_{\mu_1\nu_1}\text{} \left(\eta^{\nu_3\sigma } \eta^{\nu_1\nu_2} h_{\mu_3\nu_3} \partial \text{}_{\sigma }\text{}h_{\mu_2\nu_2}\text{}+\eta^{ \nu_1 \sigma} \eta^{\nu_2\nu_3} h_{\mu_2\nu_2} \partial \text{}_{\sigma }\text{}h_{\mu_3\nu_3}\text{}\right) \right. \nonumber\\
&\left.+
\eta^{\mu_2\mu_3} \eta^{ \mu_1\rho}  \left( \eta^{\nu_1\nu_3} \eta^{ \nu_2 \sigma} h_{\mu_2\nu_2} \partial \text{}_{\sigma }\text{}h_{\mu
_1\nu_1}\text{}+\eta^{ \nu_3\sigma} \eta^{\nu_1\nu_2} h_{\mu_1\nu_1} \partial \text{}_{\sigma }\text{}h_{\mu_2\nu_2}\text{}\right) \partial
\text{}_{\rho }\text{}h_{\mu_3\nu_3}\text{}\right. \nonumber \\
&\left . +
\eta^{\rho \mu_1} \eta^{\mu_2\mu_3} \eta^{\sigma \nu_1} \eta^{\nu_2\nu_3} h_{\mu_1\nu_1} h_{\mu_2\nu_2} \partial \text{}_{\rho,\sigma }\text{}h_{\mu_3\nu_3}\text{}\right) \nonumber \\
&=\frac{\kappa}{4}h^{\mu \nu } 
\left(\partial \text{}_{\nu'}\text{}h_{\mu'\nu }\text{} \partial \text{}_{\mu }\text{}h^{\mu'\nu'}\text{}+
\left(\partial \text{}_{\mu'}\text{}h_{\mu \nu'}\text{}-\partial \text{}_{\mu }\text{}h_{\mu'\nu'}\text{}\right) \partial \text{}_{\nu }\text{}h^{\mu'\nu'}\text{}\right)
\end{align}
Arriving at the final expression simply applied naive integration by parts and relabeling. 
It is not hard to difficult to see this correctly reproduces the scattering amplitude. Even the final line one can write back in terms of $A$ and $\tilde{A}$, via $h^{\mu\nu}\equiv A\otimes \tilde{A}$, appreciating that $\mu$'s are only contracted with $\mu's$ and $\nu$'s are only contracted with $\nu$'s.

\subsection{Generalized gauge accounting in $\oop{\mathcal{O}}(\slashed{\mathcal{A}}_m)$}
\label{sec:opPromotionCuts}
Imagine a generalized gauge scheme, call it amplitudes gauge, where application of Feynman rules for contacts strictly of multiplicitly less than $m$  to graph diagrammatics at $m$-points yielded precisely the form of $\slashed{A}$ calculated from the desired form minus the necessary novel contact information.  In such a case one can promote the novel contact term from amplitude data directly to the Lagrangian density without care for decisions made at earlier interactions and the gauge used for writing off-shell propagators.  Naturally decisions made in all generality can affect the necessary form of the contact required to achieve the target form of the cubic graph dressing at each multiplicity.  The functional automorphic form of kinematic numerators we favor here protects us from many unintentional glitches, but it is a matter of certainty that one must verify that written contacts yield precisely the desired form of the cubic dressings given the  conventions used in writing down earlier contacts.  Let us make the following clarifying example:  $(k_1+k_2)^2=2 k_1\cdot k_2$ when legs one and two are on-shell.  One will not have any problems  ignoring this distinction when writing the contribution to a four-point contact as $\oop{s}=\oop{k}_1{}^2 + \oop{k}_2{}^2 + 2\oop{k}_1\cdot\oop{k}_2$ as opposed to $\oop{s}=2\oop{k}_1\cdot\oop{k}_2$, but the Feynman rules will yield significantly different results in other contexts (at higher multiplicity or in the midst of loop-level diagrams) when $k_1$ and $k_2$ are off-shell. 

 As a matter of course for the $m$-point amplitude one should verify in any particular action that $ \OpPromotion{}( [ \slashed{\mathcal{A}}_m])= \OpPromotion{}( [ \slashed{\mathcal{A}}_m])_{\mathcal{L}}$.  The latter term represents the explicit Feynman construction of the amplitude using only lower-multiplicity interactions as written in the theory so far as well as the particular choice of propagator.   If it does not, reflecting features of previously made generalized gauge choices, then one must of course simply adjust the contact $\mathcal{L}_m$  by the difference accordingly.  This recipe embodies the pragmatic recognition that $\mathcal{L}_m = \OpPromotion{}(\mathcal{A}_m) - \OpPromotion{}( [ \slashed{\mathcal{A}}_m])_{\mathcal{L}}$.  
 
  A fair question is whether one should bother identifying the contact at the level of the amplitude to begin with, i.e.~why not simply promote the full amplitude and subtract the promotion of the partial amplitude generated from lower-multiplicity Feynman rules?   One motivation that has found success in repeated aspects of the amplitudes program is to distinguish the universal --- in this case the contact unambiguously identified from on-shell cut-data --- from the particular, i.e. the form of the contact established via ones gauge choices and propagator structure.  This may appear to be a minor or subtle technical point but it is a reality that in much of our lives we often find ourselves confronting in the present the repercussions of decisions we have made in the past.  
  
This is a point that is well emphasized in all approaches that lift gauge amplitudes to gravity contacts, from ref.~\cite{Bern:1999ji} through e.g.~refs.~\cite{Bern:2010yg,Tolotti:2013caa}

\section{Applications and Examples}
\label{sec:examples}

In this section, we illustrate the operator promotion framework across a range of theories. These examples are selected to emphasize different features: pedagogical clarity, connection to EFT classification, the power of resummation from string-inspired amplitudes, and gravitational double-copy consistency.

\subsection{Pedagogical Case:  Yang-Mills and Einstein-Hilbert Gravity}
\label{sect:YMops}

We begin simply with the Yang-Mills theory, where amplitudes are tractable and the color-kinematics structure is manifest on-shell for all low multiplicities requiring novel contact information. This serves as a clear entry point into the operator promotion method, highlighting how contact terms arise and how redundancies are automatically eliminated.

  The traditional representation for Yang-Mills actions modulo ghosts and gauge fixing terms is compactly given in terms of the field strength:
\begin{equation}
\mathcal{L}^{\text{YM}} = -\frac{1}{4} \text{Tr}( F^2)
\end{equation}
where we trace over the color indices.  For perturbative calculation,  this is often expanded out in terms of vector gauge fields, $A$, resulting in cubic and quartic contact interactions:
\begin{equation}
\mathcal{L}^{\text{YM}} = \Lag^{\text{YM}} _2 + g\, \Lag^{\text{YM}} _3 +  g^2\, \Lag^{\text{YM}} _4
\end{equation}
where $g$ is here taken to be the coupling $g^{\text{YM}}$, $\Lag_3$ can be written in position-space as proportional to $f^{abc} A^{a \mu} A^{b \nu} \partial_\mu A^c_\nu$, and $\Lag_4$ as proportional to $c_s (A^{a} \cdot A^c) ( A^{b} \cdot A^{d})$.    This is incredibly compact, completely obscuring the  tremendous factorial complexity in extracting predictions with traditional approaches.  That can be seen perhaps as a boon, certainly both gauge and Lorentz invariance are manifest.  One price we do pay is that this traditional form also obscures fundamental color-dual building blocks that, for example, Yang-Mills shares with gravitational theories.  It also obscures the well-known fact that these interactions are forced on us at lowest mass-dimension when we require massless vector interactions to be gauge invariant.

 Our approach as described in the previous two sections, is to uplift the  predictive kinematic weights of $m$-point graphs $n(g)$ directly to $m$-field operators $n(g) \to \oop{n}(g)$ and similarly with the propagator structure $d(g^m) \to \oop{d}(g^m)$.  As already discussed such a promotion involves an almost trivial transcription.  External polarizations are be replaced by contractions with explicit gauge fields and momentum invariants are  appropriately Fourier-transformed and clarified by introducing auxiliary spacetime coordinates constrained via delta functions.  
 
 Indeed this simple trick of introducing auxiliary spacetime coordinates will make double-copy structure trivially manifest at the operator level.  Carrying out the integration over those spatial delta-functions will obscure the relation to explicit prediction but can allow straightforward identification of these operators with more familiar representations.  If we do not explicitly carry out the integration over the auxiliary space-time coordinates, then the reverse operation when generating amplitudes is equally trivial $\oop{n}\to n$.   Having done the hard work of extracting predictions we can recycle that work for future excavation in the description of the theory itself.

Let us quote our earlier example 3 operators for Yang-Mills:
\begin{equation}
\oop{n}^\mu_3(x) =   -2 \left[ \left( \eta^{\mu_1 \mu_2} \right) \left(  \eta^{\mu_3 \rho} \partial_\rho{}_{x_2}) \right) + \text{cyclic}  \right]\,.\end{equation}
This is the only non-trivial kinematics allowed at three-points between massless vectors at this mass-dimension.  It is maximally antisymmetric which necessitates being dressed with an antisymmetric tensor, $f^{abc}$ to make a Bose-invariant amplitude.
As shown in \cref{3ptLymOp}, the traditional $\Lag^{\text{YM}} _3$ is completely equivalent to,
\begin{align}
\Lag^{\text{YM}} _3 &=- \frac{g}{6}  \int \mathcal{D}_3 f^{a_1 a_2 a_3} \, \oop{n}^\mu_3 \,\fieldsBold{A}{a}{\mu,3}{x}  \,.\\
  &=- g f^{a_1 a_2 a_3} A^{a_1\, \mu} \partial^\nu A^{a_2}_{\mu} A^{a_3}_{\nu} \, .
\end{align}

Only slightly more involved than the very simple operator for three-point Yang-Mills is the operator associated with the four-point color-dual Yang-Mills numerator at tree-level.  Assuming cubic dressings follow the isomorphism of graphs under relabeling there is a unique answer for how contacts are added in a way that satisfies linearized gauge invariance at four-points.   These contacts can always be absorbed to the dressings of cubic graphs by multiplying and dividing by propagators.  In other words if there is a contact proportional to $c_s$, say $X c_s$.  Then $n_s$ can absorb the contact by taking some non-contact $n_s' \to n_s = n_s' + s X$.  Gauge invariance can only occur if the color factors satisfy Jacobi relations, and any such representation which assigns contacts to cubic graphs must also satisfy the color-dual Jacobi relations.

 The numerator associated with the $s$ channel graph is simply given as the product of cubic numerators added then to a contact term:
\begin{align}
n^{\text{YM}}_s &= n_3(1,2,-k_{12}) n_3(3,4,k_{12}) +\\
 &~~~ -  s_{12} \left( \eta_{\alpha_1\alpha_2}\eta_{\alpha_3\alpha_4}+\eta_{\alpha_1\alpha_3}\eta_{\alpha_2\alpha_4}-  \eta_{\alpha_1\alpha_4}\eta_{\alpha_2\alpha_3} \right)
  \varepsilon^{\alpha_1}(k_1)  \varepsilon^{\alpha_2}(k_2)  \varepsilon^{\alpha_3}(k_3)  \varepsilon^{\alpha_4}(k_4)  \nonumber
\end{align}
with $k_{ij}=k_i+k_j$. The four-point amplitude is simply given as a sum over all three-factorization channel graphs: $\mathcal{A}_4 = \frac{c_s n_s}{s} +  \frac{c_t n_t}{t} +  \frac{c_u n_u}{u}$, where the additional channel dressings\footnote{  One could note that the conventions chosen here follow $c_s = c_t+c_u$, rather then the more symmetric $c_s+c_t+c_u=0$.  This is just a matter of convention of permutation order around vertices used to define what is labeled by $t$ and $u$ graphs. It does not matter at all as long as the numerators follow the same symmetry convention as is required by Bose invariance. }  are simply related to the $s$ channel by permuting labels.   

 The above numerator is promoted to an operator following the pervious section's conventions,
\begin{align}
\oop{n}^{\text{YM}, \boldsymbol{\mu}}_s &= \oop{n}^{\mu_1\mu_2\nu}_3(1,2,-k_{12})\cdot \oop{n}^{\mu_3\mu_4}_3{}_{\nu}(3,4,k_{12}) +\\
 &~~~ -\oop{s} \left(  \eta^{\mu_1\mu_2}\eta^{\mu_3\mu_4}+\eta^{\mu_1\mu_3}\eta^{\mu_2\mu_4}-  \eta^{\mu_1\mu_4}\eta^{\mu_2\mu_3} \right) \,. \nonumber
\end{align}

Locality of operators at higher multiplicity is made explicit by subtracting out the equivalent of cut-contributions as per the organization of the Method of Maximal cuts.  
\begin{equation}
\oop{\mathcal{A}}_4= \frac{1}{8}  \int \mathcal{D}_4 \frac{ c^{\boldsymbol{a}}_s \oop{n}^{\text{YM}, \boldsymbol{\mu}}_{s} }{\oop{s}}  \fieldsBold{A}{}{\mu,4}{x} \end{equation}
Where we subtract off everything but the contact contribution as described earlier.  We unpack the cut notation we introduced earlier,
\begin{equation}
\oop{\slashed{\mathcal{A}}}_4= \frac{1}{8}  \int \mathcal{D}_4  \frac{ c^{\boldsymbol{a}}_s  \oop{\slashed{n}}^{\text{YM}, \boldsymbol{\mu}}_{s} }{\oop{s}}    \fieldsBold{A}{a}{\mu,4}{x} \,.
\end{equation}
Here, following \cref{slashedDef} and identifying $n^{(1)}_s$ explicitly as the coefficient of $s$ in $n^{\text{YM}}_s$, we find,
\begin{align}
  \oop{\slashed{n}}^{\text{YM}, \boldsymbol{\mu}}_{s} &= \oop{n}^{\text{YM}, \boldsymbol{\mu}}_{s}  -  \oop{s} {\oop{ n}^{(1)}_s}{}^{\text{YM}, \boldsymbol{\mu}}\\
  &= \left[ \oop{n}^{\mu_1\mu_2\nu}_3(1,2,-k_{12})\cdot \oop{n}^{\mu_3\mu_4}_3{}_{\nu}(3,4,k_{12}) - \oop{s}\left(  \eta^{\mu_1\mu_2}\eta^{\mu_3\mu_4}+\eta^{\mu_1\mu_3}\eta^{\mu_2\mu_4}-  \eta^{\mu_1\mu_4}\eta^{\mu_2\mu_3} \right) \  \right] \nn\\
 &~~~ +  \oop{s} \left(  \eta^{\mu_1\mu_2}\eta^{\mu_3\mu_4}+\eta^{\mu_1\mu_3}\eta^{\mu_2\mu_4}-  \eta^{\mu_1\mu_4}\eta^{\mu_2\mu_3} \right) \ \nn \\
  &= \oop{n}^{\mu_1\mu_2\nu}_3(1,2,-k_{12})\cdot \oop{n}^{\mu_3\mu_4}_3{}_{\nu}(3,4,k_{12}) \,.
  \end{align}
 Clearly, the only surviving contribution to $\Lag^{\text{YM}}_4$ after subtracting off cuts is the usual contact term, here written in a way that makes the amplitude's color-dual nature manifest.  
 \begin{align}
\Lag^{\text{YM}}_4 &=  \frac{g^2}{8}  \int \mathcal{D}_4 \left(\frac{ c^{\boldsymbol{a}}_s \oop{n}^{\text{YM}, \boldsymbol{\mu}}_{s} }{\oop{s}}\fieldsBold{A}{a}{\mu,4}{x} \right)  -  \oop{\slashed{\mathcal{A}}}_4 \\
&= -\frac{g^2}{8}  \int \mathcal{D}_4   \frac{ c^{\boldsymbol{a}}_s  \oop{s}_{12} \left(  \eta_{\alpha_1\alpha_2}\eta_{\alpha_3\alpha_4}+\eta_{\alpha_1\alpha_3}\eta_{\alpha_2\alpha_4}-  \eta_{\alpha_1\alpha_4}\eta_{\alpha_2\alpha_3} \right)}{\oop{s}_{12}}   \fieldsBold{A}{}{\mu,4}{x}\\
&= -\frac{g^2}{8}  \int \mathcal{D}_4    c^{\boldsymbol{a}}_s \left(  \eta_{\alpha_1\alpha_2}\eta_{\alpha_3\alpha_4}+\eta_{\alpha_1\alpha_3}\eta_{\alpha_2\alpha_4}-  \eta_{\alpha_1\alpha_4}\eta_{\alpha_2\alpha_3} \right)   \fieldsBold{A}{}{\mu,4}{x}\\
&= -\frac{g^2}{8}    f^{a_1 a_2 b} f^{b a_3 a_4} \left( \eta_{\alpha_1\alpha_2}\eta_{\alpha_3\alpha_4}+\eta_{\alpha_1\alpha_3}\eta_{\alpha_2\alpha_4}-  \eta_{\alpha_1\alpha_4}\eta_{\alpha_2\alpha_3} \right)  A^{a_1}_{\mu_1}A^{a_2}_{\mu_2}A^{a_3}_{\mu_3}A^{a_4}_{\mu_4}\\
&= -\frac{g^2}{8}    f^{a_1 a_2 b} f^{b a_3 a_4} \left( A^{a_1} \cdot A^{a_2}A^{a_3} \cdot A^{a_4}+A^{a_1} \cdot A^{a_3}A^{a_2} \cdot A^{a_4} - A^{a_1} \cdot A^{a_4}A^{a_2} \cdot A^{a_3} \right)  \\
&=- \frac{g^2}{8}   ( f^{a_1 a_2 b} f^{b a_3 a_4} - f^{a_1 a_2 b} f^{b a_4 a_3}   )\left( A^{a_1} \cdot A^{a_3}A^{a_2} \cdot A^{a_4} \right) \\
&=- \frac{g^2}{4}   ( f^{a_1 a_2 b} f^{b a_3 a_4} )\left( A^{a_1} \cdot A^{a_3}A^{a_2} \cdot A^{a_4} \right)\,.
 \end{align} 
 Note that in the third to last line the  $A^{a_1} \cdot A^{a_2}A^{a_3} \cdot A^{a_4}$ term vanishes due to antisymmetry of color weights.
 
  The same procedure can be used to make every color dual numerator manifest at higher multiplicity, while explicitly only adding $0$ to the action.    
 
 We will carry out the five field contribution as an example, but perhaps first here it is important to emphasize the connection at four-points with the gravitational contact operator. Any term from the double-copy that survives the removal of cut terms contributes to the contact,
 \begin{align}
  \Lag^{\text{GR}}_4&=\left( \frac{\kappa}{2}\right)^2 \frac{1}{8}  \int \mathcal{D}_4 \frac{ \oop{n}^{\text{YM}, \boldsymbol{\mu}}_{4}  \oop{\tilde{n}}^{\text{YM}, \boldsymbol{\nu}}_{4} }{\oop{s}_{12} }    \fieldsBold{h}{}{\mu\nu,4}{x} - \oop{\slashed{\mathcal{A}}}^{\text{GR}}_4\,  \\
  &=\left( \frac{\kappa}{2}\right)^2 \frac{1}{8}  \int \mathcal{D}_4 \frac{ \oop{n}^{\text{YM}, \boldsymbol{\mu}}_{4}  \oop{\tilde{n}}^{\text{YM}, \boldsymbol{\nu}}_{4}  - \text{cut}(\oop{n}^{\text{YM}, \boldsymbol{\mu}}_{4}  \oop{\tilde{n}}^{\text{YM}, \boldsymbol{\nu}}_{4})}{\oop{s}_{12} }    \fieldsBold{h}{}{\mu\nu,4}{x}   \\
    &=\left( \frac{\kappa}{2}\right)^2\frac{1}{8}  \int \mathcal{D}_4 \frac{ \oop{n}^{\text{YM}, \boldsymbol{\mu}}_{4}  \oop{\tilde{n}}^{\text{YM}, \boldsymbol{\nu}}_{4}  
    -  \oop{\slashed{n}}^{\text{YM}, \boldsymbol{\mu}}_{s} \oop{\tilde{\slashed{n}}}^{\text{YM}, \boldsymbol{\nu}}_{s} }{\oop{s}_{12} }    \fieldsBold{h}{}{\mu\nu,4}{x}   \\
 &=\left( \frac{\kappa}{2}\right)^2 \frac{1}{8}  \int \mathcal{D}_4 \frac{
    \oop{s}_{12} \left(   
    \oop{\slashed{n}}^{\text{YM}, \boldsymbol{\mu}}_{s} \oop{\tilde{n}}^{(1)}_s{}^{\text{YM}, \boldsymbol{\nu} }+  
    \oop{n}^{(1)}_s{}^{\text{YM}, \boldsymbol{\mu}}  \oop{\tilde{\slashed{n}}}^{\text{YM}, \boldsymbol{\nu}}_{s} +      
    \oop{s}_{12} \oop{n}^{(1)}_s{}^{\text{YM}, \boldsymbol{\mu}} \oop{\tilde{n}}^{(1)}_s{}^{\text{YM},\boldsymbol{\nu}}  \right )}{\oop{s}_{12} }    \fieldsBold{h}{}{\mu\nu,4}{x}  \\
 &=\left( \frac{\kappa}{2}\right)^2 \frac{1}{8}  \int \mathcal{D}_4 \left(   
    \oop{\slashed{n}}^{\text{YM}, \boldsymbol{\mu}}_{s} \oop{\tilde{n}}^{(1)}_s{}^{\text{YM}, \boldsymbol{\nu} }+  
    \oop{n}^{(1)}_s{}^{\text{YM}, \boldsymbol{\mu}}  \oop{\tilde{\slashed{n}}}^{\text{YM}, \boldsymbol{\nu}}_{s} +      
    \oop{s}_{12} \oop{n}^{(1)}_s{}^{\text{YM}, \boldsymbol{\mu}} \oop{\tilde{n}}^{(1)}_s{}^{\text{YM},\boldsymbol{\nu}}  \right )    \fieldsBold{h}{}{\mu\nu,4}{x}  \,.
\end{align}
It is of course straightforward to carry out the delta-point integration to localize to a familiar representation of contact terms, but already here it is sufficient to see the main point which is that in the product of $n^{\text{YM}}_s\tilde{n}^{\text{YM}}_s$ that gives the gravitational numerator dressings is the contact term that arises from either side canceling the pole.  
 
 Continuing to higher multiplicity contacts is purely mechanical.  Color-kinematics and factorization uniquely solves linearized diffeomorphism for Yang-Mills, and  associated double-copy satisfies linearized diffeomorphism at each multiplicity.  Five points is an instructive example to see how this yields both no additional contact terms for Yang-Mills but non-trivial contact for gravitation.
 
 We can introduce the vanishing $\Lag^{\text{YM}} _5$, as follows.
 \begin{equation}
 \Lag^{\text{YM}}_5= \frac{1}{8}  \int \mathcal{D}_5 \frac{ c^{\boldsymbol{a}}_5 \oop{n}^{\text{YM}, \boldsymbol{\mu}}_{5} }{\oop{s}_{12} \oop{s}_{45} }    \fieldsBold{A}{a}{\mu,5}{x} - \oop{\slashed{\mathcal{A}}}^{\text{YM}}_5
 \end{equation}
 Because there is no five-point contact for Yang-Mills, we will find that $\Lag^{\text{YM}}_5=0$.  Let us see how that works out.  First  we promote the known color-dual five-point Yang-Mills graph numerator to operator form as previous examples.  We start by defining various (operator promotions of) sewings of lower multiplicity,
 \begin{align}
  \oop{n}^{{(3,3,3)},\boldsymbol{\mu}}_5&\equiv  \eta_{\nu\rho} \eta_{\sigma\tau} \oop{n}^{\mu_1\mu_2\nu}_3(1,2,-k_{12})  \oop{n}^{\rho \mu_3 \sigma}_3(k_{12},3,k_{45})  \oop{n}^{\tau \mu_4\mu_5}_3(-k_{45},4,5)\\
  \oop{n}^{(4,3),\boldsymbol{\mu}}_5& \equiv [\eta_{\nu\rho}   \oop{n}^{\text{YM},  \mu_1\mu_2\mu_3 \nu }_{s}(1,2,3,k_{45})  \oop{n}^{\rho \mu_4 \mu_5}_3(-k_{45},4,5)-  \oop{n}^{{(3,3,3)},\boldsymbol{\mu}}_5] +\\
       &~~~~~~ [\eta_{\nu\rho}  \oop{n}^{ \mu_1 \mu_2 \nu}_3(1,2,-k_{12})\oop{n}^{\rho \mu_3 \mu_4 \mu_5}_s(k_{12},3,4,5) -  \oop{n}^{{(3,3,3)},\boldsymbol{\mu}}_5] \, ,
       \nonumber
 \end{align}
and can then compactly write the color-dual five-point graph numerator operator as:
 \begin{align}
  \oop{{n}}^{\text{YM}, \boldsymbol{\mu}}_{5} &=  \oop{n}^{{(3,3,3)},\boldsymbol{\mu}}_5+\oop{n}^{(4,3),\boldsymbol{\mu}}_5\\
  &+\left[\left( s_{45} \eta^{\alpha _3 \alpha _4} \left(\eta^{\alpha _1 \alpha _2} k_1{}^{\alpha _5}-2 \eta^{\alpha _1 \alpha _5} k_1{}^{\alpha _2} - \left(1\leftrightarrow2\right)\right)  - \left(4\leftrightarrow 5 \right) \right)- \right.  \nonumber \\
 & \left. \left( \{45\}\leftrightarrow\{21\} \right) \right] \,. \nonumber
 \end{align}

 The cut contribution, following \eqn{slashedDef}, is given by 
 \begin{equation}
 \oop{\slashed{\mathcal{A}}}^{\text{YM}}_5 =\frac{1}{8}  \int \mathcal{D}_5 \frac{ c^{\boldsymbol{a}}_5 \oop{\slashed{n}}^{\text{YM}, \boldsymbol{\mu}}_{5} }{\oop{s}_{12} \oop{s}_{45} }    \fieldsBold{A}{a}{\mu,5}{x}
 \end{equation}
 with
 \begin{equation}
  \oop{\slashed{n}}^{\text{YM}, \boldsymbol{\mu}}_{5} = \oop{n}^{\text{YM}, \boldsymbol{\mu}}_{5} -   \oop{s}_{12} \oop{s}_{45} {\oop{ n}^{(2)}_{\Delta,5}}\,.
 \end{equation}
 But as there is no term in $\oop{{n}}^{\text{YM}, \boldsymbol{\mu}}_{5} $ proportional to the product $\Delta= \oop{s}_{12} \oop{s}_{45} $,  we have that
 \begin{equation}
 \oop{ n}^{(2)}_{\Delta,5}= 0 \, 
 \end{equation} 
 and so the cut contribution is the entire five-point numerator,  and thus $ \Lag^{\text{YM}}_5$ indeed vanishes.
 
 We note that the analogous gravity five field contribution does not vanish: 
  \begin{align}
  \label{fiveGravAmpGauge}
 \Lag^{\text{GR}}_5&= \frac{1}{8}  \int \mathcal{D}_5 \frac{ \oop{n}^{\text{YM}, \boldsymbol{\mu}}_{5}  \oop{n}^{\text{YM}, \boldsymbol{\nu}}_{5} }{\oop{s}_{12} \oop{s}_{45}  }    \fieldsBold{h}{}{\mu\nu,5}{x} -  \mathcal{O}(\oop{\slashed{\mathcal{A}}}^{\text{GR}}_5)_{\mathcal{L}}\,  \\
 &= \frac{1}{8}  \int \mathcal{D}_5 \left( \oop{n}{}^{(1), \boldsymbol{\mu}}_{(12)}   \, \oop{\tilde{n}} {}^{(1), \boldsymbol{\nu}}_{(45)}+ \oop{n} {}^{(1), \boldsymbol{\mu}}_{(45)} \,  \oop{\tilde{n}} {}^{(1), \boldsymbol{\nu}}_{(12)}\right)    \fieldsBold{h}{}{\mu\nu,5}{x}  + \text{gauge dep. terms} \, ,
 \label{eqn:gravFivePointContact}
 \end{align}
 as the product $ \oop{n}^{\text{YM}, \boldsymbol{\mu}}_{5}  \oop{n}^{\text{YM}, \boldsymbol{\nu}}_{5} $ does have terms proportional to $\Delta= \oop{s}_{12} \oop{s}_{45}$, as illustrated explicitly in \cref{sect:gravContact}.  Notice critically, from the first to second line, that no inverse-derivative contributions survive after subtracting the cut operator from the  non-local operator representing the full amplitude. The resulting operator is manifestly local.  We annotate here the cut operator $\mathcal{O}(\oop{\slashed{\mathcal{A}}}^{\text{GR}}_5)_{\mathcal{L}}$  to emphasize that the important difference is at the level of lower-multiplicity Feynman rules generated from operators already defined (yielding potential additional gauge-choice specific terms) as discussed in \cref{sec:opPromotionCuts}.  Through four-points we can be rather cavalier about this, but starting at five points lower multiplicity choices of operator feed into the form of the contact operator we must add to the action.  
 
 This mechanism of generating gravitational contact terms from products of lower-order effective structures within the constituent Yang-Mills numerators persists to all higher multiplicities. For instance, at six points, while the Yang-Mills action receives no new local 6-field operator, the double copy of 6-point color-dual Yang-Mills numerators (which must now include contributions from distinct cubic topologies beyond the half-ladder, related by Jacobi identities) will generate a non-vanishing $\mathcal{L}^{\text{GR}}_6$. The extraction of this term follows the same principles of promoting the full $n^{\text{YM}}_6 \tilde{n}^{\text{YM}}_6$ numerator and systematically subtracting all contributions reconstructible from 3-, 4-, and 5-point gravitational interactions via our cut-based procedure on both half-ladder and trimerous topologies at the level of Feynman rules.

Our primary goal, contrasting with previous efforts to make the duality between color and kinematics manifest at the level of the action, is the construction of a standard local effective action, achieved by isolating novel contact terms $\mathcal{L}_m = \OpPromotion(\mathcal{C}_m)$. However, it's noteworthy that the intermediate operators $\overline{\mathcal{L}}_m \equiv \OpPromotion(\mathcal{A}_m)$ (before subtracting the cut-constructible parts $\OpPromotion(\slashed{\mathcal{A}}_m)$) could themselves be viewed as an alternative specification of the predictions of theory. In such an approach, the $m$-point scattering amplitude would be generated solely by an $m$-point contact diagram dressed with the Feynman rule derived from $\overline{\mathcal{L}}_m$. This rule would, by construction, reproduce the full $\mathcal{A}_m$, including all its poles and residues, and if $\mathcal{A}_m$ was provided in a color-dual form, this vertex would directly yield that structure. While these $\overline{\mathcal{L}}_m$ operators are generally non-local (as they encode propagator structures), this perspective offers a direct map from full $m$-point amplitudes to $m$-point effective vertices that generate them in their entirety, conceptually aligning with efforts to find Lagrangians or Feynman rules that directly manifest color-kinematics duality for all graph contributions.  There is nothing surprising about this --- as we have endeavored to emphasize we are simply writing operators to correctly produce amplitudes from known amplitude data.  

\subsection{Higher Derivative Operators from Pointlike Fields to String Theory}
\label{sec:hdopAndStrings}
We now turn to one of the motivating points for our formalism --- the canonical identification of higher derivative operators.   As demonstrated in refs.~\cite{Carrasco:2019yyn,Carrasco:2021ptp}, the duality between color and kinematics allows for a particularly efficient way to construct and classify distinct amplitudes associated with higher derivative operators.  As such we have now a sharp mechanism for promoting these amplitudes to construct and classify the higher derivative operators themselves while maintaining the color-dual structure.

Recall that via field redefinition, color-zeroes, kinematic-zeros, and total derivatives there are an uncountably infinite number of ways of rewriting any given operator.    Consider the symmetrized four-field gauge operator responsible for generating the abelian four-point Born-Infeld amplitude.   The four-field operator is given simply as
\begin{equation}
\mathcal{O}_{\text{BI}}=\frac{g^2}{32} \left(  (F^2)^2+(F\tilde{F})^2 \right)\,.
\end{equation}
As many readers may appreciate this same operator (suitably supersymmetrized) results in a superamplitude that is a universal state-encoding prefactor for four-point scattering for the maximally supersymmetric Yang-Mills at any loop order, and squared is the universal prefactor for four-point scattering in the maximally supersymmetric supergravity amplitude.  The four point amplitude it produces is quite simply:
\begin{align}
\mathcal{A}_{\text{BI}} &= s t  A^{\text{YM}}_4(1234) \\
   &= n^{\text{YM}}_s t + n^{\text{YM}}_t s  \\
   &= n^{\text{YM}}_s  (t-u)/3 + n^{\text{YM}}_t (s-u)/3 + n^{\text{YM}}_u (t-s)/3 \\
   &= \frac{n^{\text{YM}}_s s (t-u)/3 }{s}  + \text{perms} \\
   &= \frac{n^{\text{YM}}_s n^{\pi}_s}{s} + \text{perms} 
\end{align}
In the third line we used that all four-point cubic graph representations of tree-level Yang-Mills satisfy Jacobi $n_u=n_s-n_t$ as well as conservation of momenta.  In the fifth line we identified the NLSM four-point kinematic numerator $n^{\pi}_s \propto s (t-u)$.  Here we see the  four-point demonstration that we can understand Born-Infeld as the double-copy between Yang-Mills and NLSM pions --- a pattern that explicitly holds to all multiplicity.  This extends to supersymmetric Dirac-Born-Infeld-Volkov-Akulov (DBI-VA) where the scalar and fermionic states are carried from supersymmetry on the Yang-Mills single-copy. 

We can clearly therefore write the Born-Infeld four-point operator. 
\begin{equation}
\mathcal{O}_{\text{BI}}= \mD{4}  \frac{1}{8} \frac{\oop{n}^{\pi}_s \oop{n}^{\text{YM},\boldsymbol{\mu}}_s}{\oop{s}}  \fieldsBold{A^\text{BI}}{}{\mu,4}{x} 
\end{equation}
The seeming non-locality is immediately canceled with the inverse-propagator present in the pion numerator.

As fantastic as this is for exposing the double-copy structure of BI operator by operator, our approach comes into its own in that that the entire tower of higher derivative gluonic operators contributing to the supersymmetric open string, whose predictions are described in  ref.~\cite{Carrasco:2019yyn}, are simply given by promoting the higher-derivative color-weights of \eqn{gaugeSoln} to operators.   Let's remind ourselves what the structure looks like.  To build these types of higher derivative operators we are considering all  higher derivative modifications to the color-weight of while respecting adjoint-type algebraic relations.   In general this means considering all color-dressings $o(g)$ that are adjoint compositions between graph dresings:
\begin{equation}
   o(g) = \compAdj{c}{n}(g)
\end{equation}
Where we differentiate between pure-scalar kinematic $n(g)$ and color weights, $c(g)$ that may contain higher-derivative contributions.  Clearly the mass-dimension of the LHS is the sum of the mass-dimensions of both $c(g)$ and $n(g)$.    As a matter of book-keeping we mod out any permutation invariants such that the contribution to a full amplitude of a particular mass dimension will always be given by:
\begin{equation}
\label{colorHDgen}
 o(g) = \sum_{ij} \alpha'^i  w_{j\vec{k}[j]}\, o_j(g)  \left(\prod_l  p_l(g)^{{k[j]}_l} \right) \,
 \end{equation}
 where $\alpha^i$ tracks mass-dimension,   $w_X$ are Wilson coefficients, and the sum of mass-dimensions of each permutation invariant $p_l$ weighted by $k[j]_l$ is equal to $i$ minus the mass-dimension of $o^{\text{HD}}_j$, i.e.,
 \begin{equation}
 i = \sum_l k_l\times[p_l] + [o^{\text{HD}}_j]\,.
 \end{equation}
 
 At four-points this is particularly simple as the number of permutation invariants for massless kinematics is incredibly small:
 \begin{align}
   \sigma_2 &= s^2+t^2+u^2\\
   \sigma_3 &= s t u
 \end{align}
Where $s+t+u=0$,  $s=(k_1+k_2)^2$,  and $t=(k_2+k_3)^2$. It turns out that every permutation invariant of $(s,t,u)$ of higher mass-dimension $m$ can be represented by:
\begin{align}
 p_m = \sum_{ij}  w_{ij} \sigma_2^i \sigma_3^j
\end{align}
with constant $w_{ij}$ and $2 i +3 j =m$.   If we restrict ourselves to four-point  $o^{\text{HD}}_j(g)$ that satisfy antisymmetry and Jacobi we have a very small number of building blocks then as per  ref.~\cite{Carrasco:2019yyn},:
\begin{align}
  o_1(g_s) &= c_s  \, \\
  o_2(g_s) &=  \compAdj{c}{n^{\text{ss}}}_s = c_t  (s-u) - c_u (t-s)  \,  \\
  o_3(g_s) &= d^{abcd} n^{\pi}_s \,.
\end{align}  
The first requires little discussion ---  higher-derivative contributions play a part only via the trivial product of permutation invariants that have been modded out.  The second arises from composing color weights with the covariantized free-scalar kinematic numerator.   The third's Jacobi satisfying properties arise from the color-dual pion kinematic weight $n^{\pi}_s \propto s (t-u)$, and the color-weights contribute through the four-point color permutation invariant.   Note the mass dimension of $[o_j]=2(j-1)$ for these three graph dressings.
 
So for the four-point case we have simply that all higher-derivative adjoint four-field color-weights are spanned by:
\begin{equation}
o(g) = \sum_{i=1}^\infty  \alpha'^i  \sum_{j=1}^3  \sum_{k_2,k_3} w_{j k_2 k_3}\sigma_2^{k_2} \sigma_3^{k_3} o_j(g)
\end{equation} 
where $\left( (j-1)+2 k_2 + 3 k_3 \right)= i$.  For $o(g)$ to be the result of a local (non-factorizing) operator we require that $k_3 \ge 1$ for $o_1$ and $o_2$.   The promotion to an operator is straightforward using:
\begin{align}
  \oop{o}^{\boldsymbol{a}}_1(g_s) &= c_s  \, \\
  \oop{o}^{\boldsymbol{a}}_2(g_s) &=  c_t  (\oop{s}-\oop{u}) - c_u (\oop{t}-\oop{s})    \,  \\
  \oop{o}^{\boldsymbol{a}}_3(g_s) &= d^{a_1 a_2 a_3 a_4} \oop{n}^{\pi}_s \,.\\
  \oop{\sigma_2} &= \oop{s}^2+\oop{t}^2+\oop{u}^2 \\
    \oop{\sigma_3} &= \oop{s} \oop{t} \oop{u} \,.
\end{align}  
Indeed we can write any four-field operator contributing to the gluonic sector of the $\alpha'$ expansion of the open superstring as:
\begin{equation}
 \mathcal{O}^{\text{OSS}}=  \int \mathcal{D}_4  \frac{1}{8} \frac{\oop{o}^{\boldsymbol{a}}_s \, \oop{n}^{\text{YM},\boldsymbol{\mu}}_s}{\oop{s}}  \fieldsBold{A}{a}{\mu,4}{x} \,.
 \label{oss}
 \end{equation}

Recall that the all multiplicty open superstring was notably expressed in terms of $m$-point super-Yang-Mills amplitudes~\cite{Mafra:2011nv,Mafra:2011nw}.   Remarkably the authors of ref.~\cite{Broedel:2013tta}, isolated the doubly-ordered disk integrals which carry all orders in $\alpha'$ corrections and double-copy with ordered Yang-Mills amplitudes to form color-ordered string theory amplitudes.  These integrals were later interpreted as doubly-ordered amplitudes in a very special bi-colored all-order effective field theory called $Z$-theory~\cite{Carrasco:2016ldy, Mafra:2016mcc, Carrasco:2016ygv}.  One ordering crucially obeys field theory KK and BCJ relations, and the other ordering obeys string monodromy relations.  When the monodromy ordering is dressed with Chan-Paton factors then we find color-dual field theory amplitudes.  The duality between color and kinematics appears to all orders in $\alpha'$ is precisely due to the all-order mixing of color and scalar kinematic weights as was clarified in~\cite{Carrasco:2019yyn,Carrasco:2021ptp}.

Indeed, from the structure of \cref{oss} one immediately  recognizes that that the four-point Z-theory $\alpha'$ expanded amplitude is encoded in 
\begin{equation}
 \mathcal{O}^{\text{Z}}_4=  \int \mathcal{D}_4  \frac{1}{8} \frac{\oop{o}^{\boldsymbol{a}}_s  \, c^{\widetilde{\boldsymbol{a} } }_s}{\oop{s}}  \fieldsBold{\phi}{a \tilde{a}}{4}{x} \,.
\end{equation}
By construction this doubly-colored theory obeys field theory relations on its $\tilde{a}$ color-ordered amplitudes.  Actual $Z$ theory amplitudes are reproduced with appropriate choice of Wilson coefficients which also imposes string theory monodromy relations on its $a$ color-ordered amplitudes.  We include in \tab{aCoeffsMD13} the necessary Wilson coefficients through mass dimension 26 that reproduce the low energy expansions of Z and open super-string amplitudes.

\begin{table*}
\begin{center}
\caption{Values of the $w^{[\text{MD}]/2}_{j,k_2,k_3 }$ coefficients appearing in $\oop{o}$ that match $\mathcal{O}^{\text{OSS}}$ to the low energy expansion of the open superstring amplitude through mass dimension 26. }
\label{aCoeffsMD13}{\tiny
\begin{equation*}
\begin{array}{cc}
w_{1,0,0}^{[0]}= 1 & w_{3,0,0}^{[2]}= 2 \zeta_2 \\
 w_{1,0,1}^{[3]}= -\zeta_3 & w_{2,0,1}^{[4]}= \frac{\zeta_4}{4} \\
 w_{3,1,0}^{[4]}= \frac{5 \zeta_4}{4} & w_{3,0,1}^{[5]}= -2 \zeta_2\zeta_3 \\
 w_{1,1,1}^{[5]}= -\frac{\zeta_5}{2} & w_{2,1,1}^{[6]}= \frac{5\zeta_6}{32} \\
 w_{3,2,0}^{[6]}= \frac{21 \zeta_6}{32} & w_{1,0,2}^{[6]}=\frac{1}{16} \left(8 \zeta_3^2+5 \zeta_6\right) \\
 w_{2,0,2}^{[7]}= -\frac{1}{4} \zeta_3 \zeta_4 & w_{3,1,1}^{[7]}=-\frac{5}{4} \zeta_3 \zeta_4-\zeta_2 \zeta_5 \\
 w_{1,2,1}^{[7]}= -\frac{\zeta_7}{4} & w_{2,2,1}^{[8]}= \frac{21\zeta_8}{256} \\
 w_{3,3,0}^{[8]}= \frac{85 \zeta_8}{256} & w_{3,0,2}^{[8]}= \zeta_2\zeta_3^2+\frac{155 \zeta_8}{96} \\
 w_{1,1,2}^{[8]}= \frac{\zeta_3 \zeta_5}{2}+\frac{49 \zeta_8}{128} &w_{2,1,2}^{[9]}= \frac{1}{32} \left(-4 \zeta_4 \zeta_5-5 \zeta
_3 \zeta_6\right) \\
 w_{3,2,1}^{[9]}= \frac{1}{32} \left(-20 \zeta_4 \zeta_5-21 \zeta_3\zeta_6-16 \zeta_2 \zeta_7\right) & w_{1,0,3}^{[9]}= \frac{1}{48}
\left(-8 \zeta_3^3-15 \zeta_3 \zeta_6-16 \zeta_9\right) \\
 w_{1,3,1}^{[9]}= -\frac{\zeta_9}{8} & w_{2,3,1}^{[10]}= \frac{85\zeta _{10}}{2048} \\
 w_{3,4,0}^{[10]}= \frac{341 \zeta _{10}}{2048} & w_{2,0,3}^{[10]}=\frac{1}{8} \zeta_3^2 \zeta_4+\frac{35 \zeta _{10}}{256}
\\
 w_{3,1,2}^{[10]}= \frac{5}{8} \zeta_3^2 \zeta_4+\zeta_2 \zeta_3\zeta_5+\frac{737 \zeta _{10}}{320} & w_{1,2,2}^{[10]}= \frac{1}{8}
\left(\zeta_5^2+2 \zeta_3 \zeta_7\right)+\frac{321 \zeta _{10}}{1024}\\
 w_{2,2,2}^{[11]}= \frac{1}{256} \left(-20 \zeta_5 \zeta_6-16 \zeta_4\zeta_7-21 \zeta_3 \zeta_8\right) & w_{3,0,3}^{[11]}=
-\frac{155}{96} \zeta_3 \zeta_8-\frac{1}{3} \zeta_2 \left(\zeta_3^3+2\zeta_9\right) \\
 w_{3,3,1}^{[11]}= \frac{1}{256} \left(-84 \zeta_5 \zeta_6-80 \zeta_4\zeta_7-85 \zeta_3 \zeta_8-64 \zeta_2 \zeta_9\right) &
w_{1,1,3}^{[11]}= \frac{1}{128} \left(-4 \zeta_5 \left(8 \zeta_3^2+5\zeta_6\right)-49 \zeta_3 \zeta_8-64 \zeta _{11}\right) \\
 w_{1,4,1}^{[11]}= -\frac{\zeta _{11}}{16} & w_{2,4,1}^{[12]}=\frac{341 \zeta _{12}}{16384} \\
 w_{3,5,0}^{[12]}= \frac{1365 \zeta _{12}}{16384} & w_{2,1,3}^{[12]}=\frac{1}{64} \zeta_3 \left(8 \zeta_4 \zeta_5+5 \zeta
_3 \zeta_6\right)+\frac{152911 \zeta _{12}}{707584} \\
 w_{3,2,2}^{[12]}= \frac{1}{4} \zeta_2 \zeta_5^2+\frac{21}{64}\zeta_3^2 \zeta_6+\frac{1}{8} \zeta_3 \left(5 \zeta_4 \zeta_5+4
\zeta_2 \zeta_7\right)+\frac{3162705 \zeta _{12}}{1415168} &w_{1,0,4}^{[12]}= \frac{1}{96} \zeta_3 \left(4 \zeta_3^3+15 \zeta_3\zeta_6+32
\zeta_9\right)+\frac{199881 \zeta _{12}}{707584} \\
 w_{1,3,2}^{[12]}= \frac{1}{8} \left(\zeta_5 \zeta_7+\zeta_3 \zeta_9\right)+\frac{1793 \zeta _{12}}{8192} & w_{2,3,2}^{[13]}= \frac{-80
\zeta_6 \zeta_7-84 \zeta_5 \zeta_8-64 \zeta_4 \zeta_9-85 \zeta_3\zeta _{10}}{2048} \\
 w_{2,0,4}^{[13]}= -\frac{1}{24} \zeta_4 \left(\zeta_3^3+2 \zeta_9\right)-\frac{35 \zeta_3 \zeta _{10}}{256} & w_{3,1,3}^{[13]}=
\frac{-5 \left(40 \zeta_3^3 \zeta_4+96 \zeta_2 \zeta_3^2 \zeta_5+155\zeta_5 \zeta_8+80 \zeta_4 \zeta_9\right)-2211 \zeta
_3 \zeta _{10}}{960} -\zeta_2 \zeta _{11} \\
 w_{3,4,1}^{[13]}= \frac{-336 \zeta_6 \zeta_7-340 \zeta_5 \zeta_8-320\zeta_4 \zeta_9-341 \zeta_3 \zeta _{10}-256 \zeta_2 \zeta
_{11}}{2048} & w_{1,2,3}^{[13]}= \frac{-4 \left(32 \zeta_3\zeta_5^2+32 \zeta_3^2 \zeta_7+20 \zeta_6 \zeta_7+49 \zeta_5 \zeta_8\right)-321
\zeta_3 \zeta _{10}-512 \zeta _{13}}{1024} \\
 w_{1,5,1}^{[13]}= -\frac{\zeta _{13}}{32} & \text{} \\
\end{array}
 \end{equation*}}
\end{center}
\end{table*}

Given the ease of the series expansion one can wonder if it is possible to write the resummed open string and $Z$ theory operators.  At four-points this is straightforward to do in closed form as  the disc integrals are known in terms of Euler Gamma functions.  One can simply write the s-channel $Z$-theory numerator as,
\begin{equation}
n^{\text{Z}}_s =  \frac{1}{3 \sigma_3} s ( t - u) [ s t A^{\text{Z}}(s,t) ] \,
\end{equation}
where the Chan-Paton dressed permutation invariant  $s t A^{\text{Z}}(s,t)$ is given in ref.~\cite{Carrasco:2019yyn}, and we quote here,
\begin{multline}
  \left [s t A^{Z}(s,t) \right] =  \frac{\pi^2}{\ap} \frac{\csc(\pi \ap s) \csc(\pi \ap t) \csc(\pi \ap u)}{\Gamma({-\ap s}) \Gamma({-\ap t}) \Gamma({-\ap u})} \times\\
 \left( c_s z_s + c_t z_t + c_u z_u + d^{a_1 a_2 a_3 a_4}\, 2 \, \left [ \sin({ \pi \ap s})+ \sin({ \pi \ap t})+ \sin({ \pi \ap u}) \right] \right) \,,
\end{multline}
where $z_s= (\sin( \pi \ap u) - \sin(\pi \ap t))/3$,  $c_s=f^{a_1a_2 e}f^{e a_3 a_4}$, and the other channel $c_g$ follow from simple relabeling.  Note the $z_g$ satisfy $z_s=z_t+z_u$ in concordance with $c_s = c_t + c_u$.

The promotion of kinematic invariants $s_{ij}$ to operators $\oop{s}_{ij}$ is applied directly within these expressions. This defines operators that compactly encode the all-orders $\alpha'$ behavior. It is worth noting that the resulting position-space operators, such as $\Gamma(-\alpha'\oop{s})$, are necessarily non-polynomial in derivatives and reflect the rich analytic structure of the underlying string amplitudes. While their formal properties as differential operators can be intricate, their definition and utility within our framework are anchored by their ability to generate the correct, known string S-matrix elements. This S-matrix-centric perspective allows us to construct action-level counterparts for theories whose amplitudes may not conform to the strictest notions of point-particle locality or perturbative unitarity (as is the case for full string theory or certain exotic theories like conformal supergravity, respectively). The primary goal is a faithful operator encoding of the on-shell physics.

If one is happy to promote unevaluated disk integrals one can use the technology of virtuous trees to write Jacobi satisfying numerators at any multiplicity.    The $m$-point Chan-Paton dressed, but field theory ordered Z-theory amplitude is given by summing the doubly-ordered $Z$ integrals over the integration domain weighted by Chan-Paton traces.
\begin{equation}
Z(q_1,\ldots,q_m)=\alpha'^{m-3}\sum_{\rho \in S_{m-1}(2,\ldots,m)} \Tr[1\rho] \int\limits_{D(1\rho)} \frac{dz_1 dz_2\cdots dz_m}{\text{vol}(SL(2,\mathbb{R}))} 
\frac{\prod_{i<j}^m |z_{ij}|^{\alpha' s_{ij}}}{z_{q_1 q_2} z_{q_2 q_3} \cdots z_{q_m q_1}}
\end{equation}
Where the trace $\Tr[1\rho]$ is shorthand for the appropriate color trace on Chan-Paton indices $\Tr[T^{a_1} T^{a_{\rho_2}}\cdots T^{a_{\rho_m}}]$. 
Notice all kinematics appear in terms of Mandelstam invariants which are trivially promoted to differential operators as we will demonstrate.  One simply needs to map from ordered amplitudes to functional kinematic numerators, which can be accomplished via the virtuous tree representation introduced by Broedel and Carrasco in \cite{Broedel:2011pd} and can be constructively built at any multiplicity by symmetrizing over the KLT kernel as per Fu, Du, and Feng in ref.~\cite{Fu:2014pya} and Naculich in ref.~\cite{Naculich:2014rta}.

As an example let us consider the five-point Chan-paton dressed, field theory ordered $Z$-theory amplitude which satisfies the KK and BCJ relations on the field-theory ordering,
\begin{equation}
\label{z5}
Z(12345) = \alpha'^{2}\sum_{\rho \in S_{4}(2,\ldots,5)} \Tr[1\rho] \int\limits_{D(1\rho)} \frac{dz_1 dz_2\cdots dz_5}{\text{vol}(SL(2,\mathbb{R}))} 
\frac{\prod_{i<j}^5 |z_{ij}|^{\alpha' s_{ij}}}{z_{q_1 q_2} z_{q_2 q_3} \cdots z_{q_5 q_1}}
\end{equation}

Conceptually, one could imagine promoting $s_{ij} \to \oop{s}_{ij}$ directly within the Koba-Nielsen integrand, e.g., in terms like $|z_{ab}|^{\alpha' s_{ab}} \to |z_{ab}|^{\alpha' \oop{s}_{ab}}$. Such an object would formally define an operator whose coefficients are given by integrals of operator-valued functions over the string worldsheet coordinates $z_i$. While a full exploration of such operators is beyond our present scope, their formal series expansion in $\alpha'$ would correspond to an infinite tower of local higher-derivative operators whose coefficients are the standard string integrals. If this bothers discerning analytic readers of taste, feel free to read the prescription as promoting $s_{ij} \to \oop{s}_{ij}$ in the expressions for numerators that are already expressed as functions of Mandelstam invariants (which may themselves be the result of evaluating string integrals or their expansions).

The low energy expansion of \eqn{z5} has been verified~\cite{Carrasco:2021ptp} to be spanned by a constructive compositional ansatz through $\alpha'^9$ (mass dimension 18). Llet us see how we can build a color-dual numerator in terms of the unexpanded disk integrals.  
A manifestly Jacobi satisfying, functional, color-dual numerator of the five-point half-ladder graph~\cite{Broedel:2011pd} is given by:
\begin{multline}
n^{Z}_5=
\frac{1}{30} \Bigg(
\Big[s_{12}s_{45}(Z_{12345}-Z_{12354} -Z_{21345}+Z_{21354})\Big]\\
\!\!\!\!\!\!\!\!\!\!\!\!\!\!\!\!\!\!\!\!+ \Big[s_{12}(s_{34}-s_{35})(Z_{1435 2}+Z_{15342})\\
+
s_{45}(s_{13}-s_{23})(Z_{51324}-Z_{41325})\Big]\\
\!\!\!\!\!\!\!\!\!\!\!\!\!\!\!\!\!\!\!\!+\Big[(s_{12}s_{34}-s_{12}s_{35})Z_{14352}+(s_{12}s_{34}-s_{12}s_{35})Z_{15342}\\
+(-s_{15}s_{23}-s_{25}s_{34})Z_{14325}+(s_{14}s_{23}+s_{24}s_{35})Z_{15324}\\
~~~~~+(s_{13}s_{24}+s_{14}s_{25})Z_{41352}+ (-s_{13}s_{25}-s_{15}s_{24})Z_{51342}\Big] \Bigg)\,.
\end{multline}
Here we use the shorthand $Z_X=Z(X)$. The $s_{ij}$ are trivially promoted to differential operators $\oop{s}_{ij}$, including in the numerator of the disk integrals $|z_{ij}|^{s_{ij}}$, yielding a five-point operator which reproduces the entire gluonic sector of the tree-level open superstring  by:
\begin{equation}
 \mathcal{O}^{\text{OSS}}_5 = \int \mathcal{D}_5  \frac{\oop{n}^{Z,\boldsymbol{a}}_5  n^{\text{YM},\boldsymbol{\mu} }_s}{\oop{s}}  \fieldsBold{A}{a}{\mu,5}{x} \,.
\end{equation}
Promoting $n^{\text{YM}}$ to its supernumerator using on-shell superspace naturally reproduces the entire multiplet.

The virtuous (tree-encoded, color-dual, functionally symmetric) numerator for the half-ladder graph labeled $\sigma_1\dots \sigma_m$ is given in closed form in terms of virtuous trace-kinematic graph weights $\tau$ as follows.
\begin{equation}
    n_\sigma = \tau_{\sigma_1 [ \sigma_2, [ \sigma_3, [\ldots, m]\cdots ]]}\,.
\end{equation}
Here the brackets in the kinematic trace $\tau$ signify an antisymmetric combination, following BernDennen.  i.e. 
\begin{equation}
\tau_{1[2,[3,4]]} = \tau_{1234} - \tau_{1243} - \left(\tau_{1342} - \tau_{1432} \right)\,.
\end{equation}

The approach to generating such $\tau$ is perhaps obvious from KLT construction, they come easily from ordered amplitudes.  Given that  full amplitudes are bose-symmetric, one can average the KLT expression over all permutations of leg labels and simply read off the cyclc $\tau$,
\begin{equation}
\tau_\sigma =- \frac{\partial}{\partial \tilde{A}(\sigma)} \frac{1}{m!} \sum_{S_m} \frac{1}{2} \left( \sum_{\rho, \tau \in S_{m-3}} \tilde{A}(1,\rho, m,m-1) S_{\rho|\tau} A(1,\tau,m-1,m)  + \tilde{A} \leftrightarrow A  \right) 
\end{equation}
where the outer sum is over all permutations of leg labels. The derivative is taken with the understanding that all orderings of trees are cyclically identified.  Amusingly the definition of the antisymmetric Jacobi-satisfying $n_\sigma$ in terms of shuffle operations on $\tau$ can be seen as equivalent of re-expressing the permutation summed KLT expression in a Kleiss-Kujif basis for the $\tilde{A}$,
\begin{equation}
n_\sigma =- \frac{\partial}{\partial \tilde{A}(\sigma)} \left. \left(  \frac{1}{m!} \sum_{S_m} \left[  \frac{1}{2} \sum_{\rho, \tau \in S_{m-3}}\tilde{A}(1,\rho, m,m-1) S_{\rho|\tau} A(1,\tau,m-1,m)  + \tilde{A} \leftrightarrow A  \right] \right|_{\tilde{A}(\sigma_1\beta \sigma_n)}\right)\,.
\end{equation}

We see here an all-order constructive form of color-dual numerators.  If the ordered amplitudes are Yang-Mills then the double-copy is to gravitation, and the novel contact information of the $n$-point graviton is encoded as the expression in the double-copy that survives subtraction of all cuts.

\section{Path to Quantum Gravity via Double Copy}
\label{sec:doublecopyGrav}

Having established the operator promotion framework, 
we now explore its implications for understanding quantum gravity 
as a double copy of Yang-Mills theory, both at the level of the 
effective action and the quantum states.

\subsection{From Yang-Mills amplitudes to the Einstein-Hilbert action}

Our framework offers a constructive route\footnote{We should emphasize, as noted in our introduction, that we are not the first to do make this constructive argument.  Notably using only the existence of KLT relations --- Bern and Grant first demonstrated this was towards generating spectacularly compact  representations of higher-multiplicity gravitational operators in ref.~\cite{Bern:1999ji}.} to the full expansion of the Einstein-Hilbert action, $\sqrt{-g}R$, written in terms of the fluctuation $h_{\mu\nu} = g_{\mu\nu} - \eta_{\mu\nu}$. While it is well known that gravity amplitudes arise as the double copy of gauge theory, we go further: we construct each individual interaction term in $\sqrt{-g}R$ from local operator contributions derived directly from gauge theory amplitudes.  

This relies on two established facts:

\begin{enumerate}
\item Gravitational amplitudes are a double copy of Yang-Mills. At tree level and all multiplicities, the gravitational $m$-point amplitude takes the form
\begin{equation}
\mathcal{A}_m^{\text{GR}} = \sum_{g \in \Gamma^m_3} \frac{n^{\text{YM}}_m(g)\, \tilde{n}^{\text{YM}}_m(g)}{d_g} \,,
\end{equation}
where $n^{\text{YM}}_m(g)$ and $\tilde{n}^{\text{YM}}_m(g)$ are color-dual numerators for each cubic graph $g$, and $d_g$ is the product of graph propagators.

\item Color-dual numerators exist at all multiplicity. A variety of constructions provide explicit expressions for $n^{\text{YM}}_m(g)$ that satisfy the kinematic Jacobi relations and generate the full Yang-Mills S-matrix.
\end{enumerate}

Given these facts, our formalism promotes each gravitational amplitude to a local operator in the action. We extract the novel $m$-point contact term by subtracting contributions reconstructible from lower-point factorizations,
\begin{equation}
\mathcal{C}^{\text{GR}}_m = \mathcal{A}^{\text{GR}}_m - \slashed{\mathcal{A}}^{\text{GR}}_m\,,
\end{equation}
and define the corresponding operator as
\begin{equation}
\mathcal{L}^{\text{GR}}_m = \OpPromotion(\mathcal{C}^{\text{GR}}_m)\,.
\end{equation}
Each $\mathcal{L}_m$ is manifestly local and required for the interacting action to be invariant under linearized diffeomorphism invariance and arises entirely from graph-level double-copy structure.

Crucially, Yang-Mills theory contains no fundamental contact terms beyond four points. In our framework, this means $\mathcal{L}^{\text{YM}}_m = 0$ for $m > 4$, and all higher contact structure in gravity must arise from cross terms in the double copy. That is, products of pole-canceling terms from each gauge-theory copy can combine to produce contact terms in gravity. We make this mechanism explicit at five points in Section~\ref{sect:YMops} and more generally in Section~\ref{sect:gravContact}.

For example, the Yang-Mills half-ladder at five points contains separate terms one proportional to $s_{12}$ and another proportional to $s_{45}$.  Upon double copy between two such numerator factors these terms appear together yielding a contact term proportional to $s_{12}s_{45}$, precisely the form required for the five-point expansion of $\sqrt{-g}R$. This structure persists at six points and beyond. The gravitational contact terms $\mathcal{L}^{\text{GR}}_m$ are nonvanishing for all $m \geq 3,$ and for $m >4$ are not physically present in either gauge-theory factor alone.

Summing over all multiplicities gives the complete expansion of the gravitational action,
\begin{equation}
\mathcal{S}^{\text{GR}}_{\text{tree}} = \int d^D x \sum_{m \ge 3} \mathcal{L}^{\text{GR}}_m\,.
\end{equation}
Each term $\mathcal{L}_m$ generates the correct $m$-point amplitude, and thus the sum matches the standard expansion of the Einstein-Hilbert action up to field redefinitions and total derivatives. This provides a direct, structured map from gauge theory to gravity that makes manifest  the double-copy origin of each interaction.

\subsection{State level description}

We have shown how local operators in double-copy theories can be expressed in terms of factorized operators acting on constituent gauge-theory-like copies. Let us now outline the path toward a state-level encoding of this structure, focusing on quantum gravity emerging from Yang-Mills theory.

To begin, recall the conceptual path from Maxwell theory to Yang-Mills. Free photon states encode linearized gauge invariance via $D-2$ physical polarizations. On-shell three-vector interactions, constrained by mass dimension and gauge invariance, uniquely fix the kinematic numerator to be the maximally antisymmetric structure:
\begin{equation}
   n_3(k_1,k_2,k_3) = (\varepsilon_1 \cdot \varepsilon_2)(\varepsilon_3 \cdot (k_1 - k_2)) + \text{cyclic}.
\end{equation}
Consistency with Bose symmetry then demands an accompanying maximally antisymmetric color factor, the first manifestation of the duality between color and kinematics. Imposing gauge invariance at four points necessitates a local contact term such that both color and kinematic numerators (assigned to cubic graphs) obey Jacobi identities. Promoting this amplitude structure to field-level operators yields Yang-Mills theory.

In constructing gravity, we proceed in parallel. Free graviton states must be consistent with spin-2 gauge symmetry, i.e., invariance under linearized diffeomorphisms: $h_{\mu\nu} \to h_{\mu\nu} + \partial_\mu \xi_\nu + \partial_\nu \xi_\mu$. In our amplitude-centric formulation, $h_{\mu\nu}$ (defined as $g_{\mu\nu} - \eta_{\mu\nu}$ without restriction on magnitude) primarily encodes asymptotic on-shell data. Each free graviton state corresponds to a symmetric-traceless (ST) polarization tensor, which precisely captures the $D(D-3)/2$ physical polarizations. This ST structure is realized as a double copy of two gauge-theory polarization vectors:
\begin{equation}
\label{eq:graviton_polarization}
\varepsilon^{\mu\nu}_{\text{GR}} = \varepsilon^{(\mu} \tilde{\varepsilon}^{\nu)} - \frac{1}{D} \eta^{\mu\nu} (\varepsilon \cdot \tilde{\varepsilon}).
\end{equation}
This definition inherently enforces the requirements of linearized diffeomorphism invariance at the level of asymptotic states. 

Interactions must preserve this symmetry at every multiplicity. This is guaranteed when gravitational amplitudes are expressed as a double copy of color-dual Yang-Mills amplitudes. The operator promotion framework developed in this paper provides an all-multiplicity algorithm for expressing every gravitational operator in a factorized form, $\oop{\mathcal{O}}_{GR} \sim \oop{\mathcal{O}}_{YM} \otimes \oop{\mathcal{O}}_{\widetilde{YM}}$, suitable for acting on a correspondingly structured Hilbert space.

Let us make this state-level structure concrete. A free Yang-Mills state of momentum $k$, helicity $\lambda$, and adjoint color $a$ is:
\begin{equation}
\ket{\text{YM}(k)^a_{\lambda}{}} = \ket{k} \otimes \ket{\varepsilon_\lambda} \otimes \ket{a}_G.
\end{equation}
A free graviton state, built as a double copy, then takes the form:
\begin{equation}
\label{eq:graviton_state_simple}
\ket{\text{GR}(k)_{\lambda,\tilde{\lambda}}} = \ket{k} \otimes \ket{\varepsilon_{\text{GR}; \lambda,\tilde{\lambda}}},
\end{equation}
where $\varepsilon^{\mu\nu}_{\text{GR}; \lambda,\tilde{\lambda}}(k)$ can be expressed as the symmetric traceless (ST) tensor from Eq.~\eqref{eq:graviton_polarization} formed from $\varepsilon_\lambda(k)$ and $\tilde{\varepsilon}_{\tilde{\lambda}}(k)$. This defines a state-level double copy:
\begin{equation}
\label{eq:state_double_copy_conceptual}
\ket{\text{GR}(k)_{\lambda,\tilde{\lambda}}} 
= \left( \ket{\text{YM}(k)^{\slashed{a}}_{\lambda}} \otimes_{\text{ST}} \ket{\widetilde{\text{YM}}(k)^{\tilde{\slashed{b}}}_{\tilde{\lambda}}} \right)_{\text{color-singlet}}.
\end{equation}
The slashed color indices $\slashed{a}, \tilde{\slashed{b}}$ signify that color degrees of freedom are traced over, yielding a color-singlet state essential for diffeomorphism invariance of observables. The $\otimes_{\text{ST}}$ denotes that the tensor product of the kinematic parts (momentum and polarization kets) is projected onto the symmetric-traceless representation for the graviton. More explicitly, one can construct this state from constituent polarization kets as:
\begin{equation}
\label{eq:state_double_copy_explicit}
\ket{\text{GR}(k)_{\lambda,\tilde{\lambda}}} = \frac{1}{\sqrt{N_c N_{\tilde{c}}}} \sum_{c,\tilde{c}} \int \! d\tilde{k} \, \delta^D(k \!-\! \tilde{k}) \left[  \ket{k} \otimes  P_{ST} \left(\ket{\varepsilon_\lambda} \otimes \ket{\tilde{\varepsilon}_{\tilde{\lambda}}} \right) \otimes \ket{c}_G \otimes \ket{\tilde{c}}_{\tilde{G}} \right]_{\text{projected to singlet}}.
\end{equation}
Here, the sum and normalization factor ensure a color-singlet state if $\ket{c}_G$ and $\ket{\tilde{c}}_{\tilde{G}}$ are from identified color spaces; for distinct groups, the singlet projection is trivial.  

Representation theory dictates the decomposition of the tensor product of two vector ($\mathbf{1}$) representations:
$
\mathbf{1} \otimes \mathbf{1} = \mathbf{2}_{\text{ST}} \oplus \mathbf{A}_{\text{AS}} \oplus \mathbf{S}_{\text{Tr}},
$
corresponding to the spin-2 graviton (symmetric-traceless), a 2-form (antisymmetric), and a dilaton (scalar trace), respectively. Explicitly, if $\ket{T^{\mu\nu}} \equiv \ket{\varepsilon^\mu} \otimes \ket{\tilde{\varepsilon}^\nu}$, then
$
\ket{T^{\mu\nu}} = \ket{h^{\mu\nu}_{\text{ST}}} + \ket{B^{\mu\nu}_{\text{AS}}} + \eta^{\mu\nu} \ket{\phi_{\text{Tr}}}.
$
Our construction of Einstein-Hilbert gravity, by promoting amplitudes that solely describe interacting gravitons, effectively projects onto the $\ket{h^{\mu\nu}_{\text{ST}}}$ sector. This projection is crucial\footnote{The alternative lands on the states of the fat graviton~\cite{Luna:2016hge} which include states of the Kalb-Ramond antisymmetric two-form and dilaton.}, particularly when considering interactions with matter or loop-level effects, where unphysical propagation of the full $\mathbf{A}$ or $\mathbf{S}$ sectors must be avoided for pure gravity.  The method of maximal cuts, as employed in our contact term extraction and in advanced loop computations~\cite{Carrasco:2021bmu}, provides a systematic way to enforce such projections, ensuring that only the desired physical degrees of freedom contribute.

The operators $\mathcal{L}_m = \oop{\mathcal{O}}(\mathcal{A}_m) - \oop{\mathcal{O}}(\slashed{\mathcal{A}}_m)$ derived in this paper manifestly factorize into gauge-theory-like operators. These are designed to act on the constituent kets within the tensor product structure of states like Eq.~\eqref{eq:state_double_copy_conceptual}, ensuring that interactions respect the double-copy inheritance. The construction of all such gravitational operators from their gauge theory counterparts via our systematic procedure is thus, in principle, an algorithmic task to all multiplicities.

At the level of quantum states, we remain agnostic about whether the color trace implies a true partial trace over entangled subsystems in a quantum information sense, or if it is a formal projection onto the $G \times \tilde{G}$ invariant subspace of $\mathcal{H}_{\text{YM}} \otimes \widetilde{\mathcal{H}}_{\text{YM}}$. What is essential is that the resulting gravitational Hilbert space sector is built from these gauge-invariant (color-singlet) tensor factor combinations, correctly capturing the physical graviton polarizations and their interactions.

While this state-level construction is most transparent in flat spacetime, it is not fundamentally limited. If the background geometry itself arises from a double copy of classical gauge field configurations (as in Kerr-Schild metrics or generalized classical double-copy solutions), graviton fluctuations $\delta h_{\mu\nu}$ can still be expressed as ST tensor products of gauge-field fluctuations $\delta A_\mu \otimes \delta \tilde{A}_\nu$ around these backgrounds. The projection to physical degrees of freedom remains locally well-defined, extending the applicability of this framework to a broad class of curved spacetimes.

Our construction of single-graviton states as specific projections of tensor products of Yang-Mills state structures, $\ket{\text{GR}} \sim (\ket{\text{YM}} \otimes_{\text{ST}} \ket{\widetilde{\text{YM}}})_{\text{color-singlet}}$, provides a Fock-space basis for gravitational theories built from double-copy principles. This perspective, where the double copy is realized at the level of individual particle states and the operators that act upon them, complements other approaches that explore state-level manifestations of the double copy, notably c.f.~the work of Cheung and Remmen~\cite{Cheung:2020uts}. which explored $N$-graviton dynamics by introducing an entanglement ansatz between $2N$-gluon states under the same SU($N_c$) gauge group.

\section{Outlook} 
\label{sec:outlook}
In this paper we introduced a method to promote color-dual amplitudes directly to operators at the level of the action which make manifest their double-copy structure.  We introduce a generalization of the method of maximal cuts that lets us uniquely and algorithmically identify contact contributions,
\begin{equation}
\mathcal{L}_m = \OpPromotion(\mathcal{A}_m - \slashed{\mathcal{A}}_m)\,,
\end{equation}  
and apply this to modding out redundancy that can occur when writing entire amplitudes as non-local operators.  We introduce a transparent operator promotion that allows for quantum field operators to look like the momentum-space amplitude expressions they generate. We demonstrate the utility by for the first time, giving an action level expression that correctly generates the massless vector contributions to the open superstring theory at five-points --- encoding an infinite number of higher-dimensional operators as a disk integral that can be lifted directly from the predictions of the two-dimensional CFT. 

This has immediate applications.  For EFT construction it offers a  alternative to ansatz methods, especially for complex high-erivative operators or theories with many fields.  The procedure is straightforward and there is non-trivial potential for automation and database engagement with physical content defining both actions and predictions simultaneously.

It should be noted that the  efficacy of the double-copy prescription, evident in scattering amplitudes across a wide range of theories, has naturally spurred investigations into its underlying mathematical foundations. Significant progress has been made in understanding these structures from the perspective of homotopy algebras, such as $L_\infty$-algebras and the Batalin-Vilkovisky formalism (e.g., \cite{Borsten:2023ned} and references therein).  There has also been progress on the kinematic Hopf algebra relevant to double-copy~\cite{Brandhuber:2021bsf,Fu:2025jpp}. These approaches aim to provide a rigorous algebraic basis for color-kinematics duality and the double-copy operation itself, often by identifying kinematic algebras as part of these sophisticated mathematical structures.

Our work, while starting from the S-matrix and focusing on a direct, constructive path to actions, is complementary to these formal algebraic investigations. By providing an explicit `amplitude-to-action' map that preserves double-copy structures, we offer a concrete realization of these principles at the level of effective Lagrangians and quantum states. It would be interesting future work to explore the precise connections between our constructed operators and their counterparts in the language of homotopy algebras.

The construction presented here explicitly applies the method of maximal cuts to identify and isolate the novel contribution of every local gravitational operator. This framework is not restricted to tree level: it builds upon the same unitarity-compatible bookkeeping used in multiloop computations, where maximal-cut techniques have been extensively employed to project out unwanted scalar and two-form contributions. In fact, this approach was recently formalized precisely as a maximal-cut-based operation in~\cite{Carrasco:2021bmu}, where it was used to extract the Einstein–Hilbert predictions at loop level from the gravitational double copy.

Crucially, the same method also resolves subtleties that arise already at tree level when coupling to massive matter. In such cases, intermediate states in the double copy can include unphysical scalar components unless projected out via symmetric-traceless projections. The method of maximal cuts provides a systematic and physically meaningful way to enforce this projection channel-by-channel, ensuring that only the gravitational degrees of freedom propagate. As such, the operator construction presented here is not only valid at the level of free states but is fully compatible with loop-level unitarity and matter couplings at tree level.

More broadly, the double-copy principle extends far beyond field-theoretic S-matrix constructions, with both open and closed string theories admitting fully consistent, all-orders-in-$\alpha'$ double-copy formulations. The open superstring (OSS) amplitude, for instance, can be expressed as a field-theory-level double copy involving Z-theory and super Yang-Mills (sYM)~\cite{Broedel:2013tta, Carrasco:2016ldy,Carrasco:2016ygv, Mafra:2016mcc}.  We exploited this earlier to give the open-superstring operator at tree-level for five-points.  Z-theory acts as a single-copy theory capturing all $\alpha'$ dependence and satisfying string monodromy relations (associated here with capital Latin indices, e.g., $A, B$). When its Chan-Paton stripped, ordered amplitudes $Z_{Ab}$ (where $b$ is a field-theory-like index) are appropriately combined with ordered sYM amplitudes $\text{SYM}_{\tilde{b}}$, they yield the complete ordered OSS amplitude, denoted $\text{OSS}_A = Z_a \cdot \text{sYM} \equiv Z_{Ab} \otimes^{b\tilde{b}} \text{sYM}_{\tilde{b}}$, not just its low-energy expansion. Here, lower-case indices obey standard field-theory KK and BCJ relations.

The closed superstring (CSS), in turn, emerges from a KLT-like relation where this string KLT kernel, $\stringKLT{}^{AB}$ (which implements a single-valued projection at the level of multiple zeta values), acts on these OSS structures:
\begin{align}
\text{CSS} &= \text{OSS}_A \stringKLT{}^{AB} \text{OSS}_B \\
&= (\text{sYM}\cdot Z_{A}) \stringKLT{}^{AB} (\text{sYM} \cdot Z_{B}) \label{eq:css_substituted_oss} \\ 
&= \text{sYM} \cdot \left( (Z_{A} \stringKLT{}^{AB} Z_{B}) \cdot \text{sYM} \right) \label{eq:css_regrouped} \\
&\equiv \text{sYM}\cdot \text{sv}(\text{sYM})\,. \label{eq:css_final_form_placeholder_indices}
\end{align}
Critically note both open and closed superstring theory amplitudes are expressed ultimately in terms of field-theory double-copies in terms of objects that respect field theory rules (color-kinematics  duality / KK-BCJ relations), extending naturally  to bosonic and heterotic string amplitudes~\cite{Azevedo:2018dgo,Carrasco:2022lbm}.

A potential concern regarding operator constructions based on the duality between color and kinematics is whether such representations exist at arbitrary multiplicity and loop order. However, from an S-matrix forward perspective this is a red herring: all multiloop integrands can be systematically constructed from products of tree-level amplitudes via generalized unitarity. If color-kinematics duality holds at tree leve ( the bare minimum required to play in the web of color-dual theories)  then any loop-level prediction can be built from tree data that already satisfies the required duality.  This point was emphasized to spectacular effect in the constructive work of ref.~\cite{Bern:2024vqs} . In this sense, the universality of the operator construction is inherited from the tree-level data, and no additional off-shell extension of color-kinematics duality is required.

Moreover, the method of maximal cuts explicitly selects the contributing operator structures from on-shell configurations, which are fully determined by consistent Jacobi-satisfying tree-level numerators. The operator mapping defined in this work thus provides a constructive algorithm for building a gravitational operator basis compatible with the double-copy structure at all multiplicity and loop order, without requiring an off-shell or Lagrangian-level duality.

One of the defining strengths of the double-copy framework is its constructive nature: it does not presuppose a simple or perturbative gravitational background. Even highly nontrivial gravitational configurations --- including those far from asymptotic flatness --- can be realized as double copies of appropriately structured gauge-theory field configurations. In this sense, the complexity of the gravitational field \( h_{\mu\nu} \) is not a limitation but a challenge to be matched by equally intricate single-copy data. This flexible, bottom-up construction is a feature, not a flaw, of the double-copy program.

The true promise of establishing the double copy as an action-and-state-level duality extends far beyond perturbative calculations. It offers a concrete strategy for confronting notoriously difficult  gravitational phenomena by reframing them in the language of Yang-Mills theory. While non-perturbative Yang-Mills dynamics, such as those governing instanton-mediated processes, are themselves challenging, they are fundamentally rooted in a well-understood, unitary quantum field theory on flat spacetime. 

This perspective invites a paradigm shift: questions about Hawking radiation, the black hole information paradox, or even eternal inflation, need not be solely pursued within the often conceptually fraught arena of (semi-classical) quantum gravity on curved backgrounds. Instead, the double copy provides a map to potentially more tractable (though still intricate)  problems in their dual gauge theories. The path to understanding these profound gravitational puzzles may therefore lie in leveraging our robust toolkit for flat-space Yang-Mills theory. Indeed, with structure that simplifies the S-matrix, and with explicit operator connections now established, the challenge becomes one of technical execution within these ``simpler'' copies. We hope that this doubled motivation can inspire new innovations around the still-significant challenges within Yang-Mills theory, such as confinement, with the promise that their resolution will also pay dividends in our understanding of the evolution of time and space.

\acknowledgments
We especially thank Zvi Bern, Marco Chiodaroli, Radu Roiban, and Henrik Johansson for conversations, comments, and related collaboration.  We would like to also thank members of the Amplitudes and Insight group at Northwestern University, Sai Sasank Chava, Yaxi Chen, Alex Edison, Nic Pavao, Aslan Seifi, Cong Shen, Theodore Wecker, and John Zhang, for important conversations and feedback on an earlier versions.
SZ is grateful for the support from Northwestern University’s
Presidential Fellowship.
This work was supported by the DOE under contract DE-SC0015910, by the
Alfred P. Sloan Foundation, and by Northwestern University via the
Amplitudes and Insight Group, Department of Physics and Astronomy, and
Weinberg College of Arts and Sciences.
\appendix
\bibliographystyle{JHEP}
\bibliography{someCitations}

\providecommand{\href}[2]{#2}\begingroup\raggedright\begin{thebibliography}{10}

\bibitem{Carrasco:2019yyn}
J.J.M.~Carrasco, L.~Rodina, Z.~Yin and S.~Zekioglu, \emph{{Simple encoding of
  higher derivative gauge and gravity counterterms}},
  \href{https://doi.org/10.1103/PhysRevLett.125.251602}{\emph{Phys. Rev. Lett.}
  {\bfseries 125} (2020) 251602}
  [\href{https://arxiv.org/abs/1910.12850}{{\ttfamily 1910.12850}}].

\bibitem{Carrasco:2021ptp}
J.J.M.~Carrasco, L.~Rodina and S.~Zekioglu, \emph{{Composing effective
  prediction at five points}},
  \href{https://doi.org/10.1007/JHEP06(2021)169}{\emph{JHEP} {\bfseries 06}
  (2021) 169} [\href{https://arxiv.org/abs/2104.08370}{{\ttfamily
  2104.08370}}].

\bibitem{Carrasco:2023wib}
J.J.M.~Carrasco and N.H.~Pavao, \emph{{UV massive resonance from IR double copy
  consistency}}, \href{https://doi.org/10.1103/PhysRevD.109.065006}{\emph{Phys.
  Rev. D} {\bfseries 109} (2024) 065006}
  [\href{https://arxiv.org/abs/2310.06316}{{\ttfamily 2310.06316}}].

\bibitem{Bern:2010yg}
Z.~Bern, T.~Dennen, Y.-t.~Huang and M.~Kiermaier, \emph{{Gravity as the Square
  of Gauge Theory}},
  \href{https://doi.org/10.1103/PhysRevD.82.065003}{\emph{Phys. Rev. D}
  {\bfseries 82} (2010) 065003}
  [\href{https://arxiv.org/abs/1004.0693}{{\ttfamily 1004.0693}}].

\bibitem{Tolotti:2013caa}
M.~Tolotti and S.~Weinzierl, \emph{{Construction of an effective Yang-Mills
  Lagrangian with manifest BCJ duality}},
  \href{https://doi.org/10.1007/JHEP07(2013)111}{\emph{JHEP} {\bfseries 07}
  (2013) 111} [\href{https://arxiv.org/abs/1306.2975}{{\ttfamily 1306.2975}}].

\bibitem{Cheung:2016prv}
C.~Cheung and C.-H.~Shen, \emph{{Symmetry for Flavor-Kinematics Duality from an
  Action}}, \href{https://doi.org/10.1103/PhysRevLett.118.121601}{\emph{Phys.
  Rev. Lett.} {\bfseries 118} (2017) 121601}
  [\href{https://arxiv.org/abs/1612.00868}{{\ttfamily 1612.00868}}].

\bibitem{Ben-Shahar:2021doh}
M.~Ben-Shahar and M.~Guillen, \emph{{10D super-Yang-Mills scattering amplitudes
  from its pure spinor action}},
  \href{https://doi.org/10.1007/JHEP12(2021)014}{\emph{JHEP} {\bfseries 12}
  (2021) 014} [\href{https://arxiv.org/abs/2108.11708}{{\ttfamily
  2108.11708}}].

\bibitem{Ben-Shahar:2021zww}
M.~Ben-Shahar and H.~Johansson, \emph{{Off-shell color-kinematics duality for
  Chern-Simons}}, \href{https://doi.org/10.1007/JHEP08(2022)035}{\emph{JHEP}
  {\bfseries 08} (2022) 035}
  [\href{https://arxiv.org/abs/2112.11452}{{\ttfamily 2112.11452}}].

\bibitem{Bern:1999ji}
Z.~Bern and A.K.~Grant, \emph{{Perturbative gravity from QCD amplitudes}},
  \href{https://doi.org/10.1016/S0370-2693(99)00524-9}{\emph{Phys. Lett. B}
  {\bfseries 457} (1999) 23}
  [\href{https://arxiv.org/abs/hep-th/9904026}{{\ttfamily hep-th/9904026}}].

\bibitem{Kawai:1985xq}
H.~Kawai, D.C.~Lewellen and S.H.H.~Tye, \emph{{A Relation Between Tree
  Amplitudes of Closed and Open Strings}},
  \href{https://doi.org/10.1016/0550-3213(86)90362-7}{\emph{Nucl. Phys. B}
  {\bfseries 269} (1986) 1}.

\bibitem{Bern:1998sv}
Z.~Bern, L.J.~Dixon, M.~Perelstein and J.S.~Rozowsky, \emph{{Multileg one loop
  gravity amplitudes from gauge theory}},
  \href{https://doi.org/10.1016/S0550-3213(99)00029-2}{\emph{Nucl. Phys. B}
  {\bfseries 546} (1999) 423}
  [\href{https://arxiv.org/abs/hep-th/9811140}{{\ttfamily hep-th/9811140}}].

\bibitem{Cheung:2016say}
C.~Cheung and G.N.~Remmen, \emph{{Twofold Symmetries of the Pure Gravity
  Action}}, \href{https://doi.org/10.1007/JHEP01(2017)104}{\emph{JHEP}
  {\bfseries 01} (2017) 104}
  [\href{https://arxiv.org/abs/1612.03927}{{\ttfamily 1612.03927}}].

\bibitem{Bern:2008qj}
Z.~Bern, J.J.M.~Carrasco and H.~Johansson, \emph{{New Relations for
  Gauge-Theory Amplitudes}},
  \href{https://doi.org/10.1103/PhysRevD.78.085011}{\emph{Phys. Rev. D}
  {\bfseries 78} (2008) 085011}
  [\href{https://arxiv.org/abs/0805.3993}{{\ttfamily 0805.3993}}].

\bibitem{Bern:2010ue}
Z.~Bern, J.J.M.~Carrasco and H.~Johansson, \emph{{Perturbative Quantum Gravity
  as a Double Copy of Gauge Theory}},
  \href{https://doi.org/10.1103/PhysRevLett.105.061602}{\emph{Phys. Rev. Lett.}
  {\bfseries 105} (2010) 061602}
  [\href{https://arxiv.org/abs/1004.0476}{{\ttfamily 1004.0476}}].

\bibitem{Monteiro:2014cda}
R.~Monteiro, D.~O'Connell and C.D.~White, \emph{{Black holes and the double
  copy}}, \href{https://doi.org/10.1007/JHEP12(2014)056}{\emph{JHEP} {\bfseries
  12} (2014) 056} [\href{https://arxiv.org/abs/1410.0239}{{\ttfamily
  1410.0239}}].

\bibitem{Luna:2015paa}
A.~Luna, R.~Monteiro, D.~O'Connell and C.D.~White, \emph{{The classical double
  copy for Taub\textendash{}NUT spacetime}},
  \href{https://doi.org/10.1016/j.physletb.2015.09.021}{\emph{Phys. Lett. B}
  {\bfseries 750} (2015) 272}
  [\href{https://arxiv.org/abs/1507.01869}{{\ttfamily 1507.01869}}].

\bibitem{Luna:2016due}
A.~Luna, R.~Monteiro, I.~Nicholson, D.~O'Connell and C.D.~White, \emph{{The
  double copy: Bremsstrahlung and accelerating black holes}},
  \href{https://doi.org/10.1007/JHEP06(2016)023}{\emph{JHEP} {\bfseries 06}
  (2016) 023} [\href{https://arxiv.org/abs/1603.05737}{{\ttfamily
  1603.05737}}].

\bibitem{Kosower:2018adc}
D.A.~Kosower, B.~Maybee and D.~O'Connell, \emph{{Amplitudes, observables, and
  classical scattering}},
  \href{https://doi.org/10.1007/JHEP02(2019)137}{\emph{JHEP} {\bfseries 02}
  (2019) 137} [\href{https://arxiv.org/abs/1811.10950}{{\ttfamily
  1811.10950}}].

\bibitem{Elvang:2013cua}
H.~Elvang and Y.-t.~Huang, \emph{{Scattering Amplitudes}},
  \href{https://arxiv.org/abs/1308.1697}{{\ttfamily 1308.1697}}.

\bibitem{Carrasco:2015iwa}
J.J.M.~Carrasco, \emph{{Gauge and Gravity Amplitude Relations}},  in
  \emph{{Theoretical Advanced Study Institute in Elementary Particle Physics}:
  {Journeys Through the Precision Frontier: Amplitudes for Colliders}},
  pp.~477--557, WSP, 2015,
  \href{https://doi.org/10.1142/9789814678766_0011}{DOI}
  [\href{https://arxiv.org/abs/1506.00974}{{\ttfamily 1506.00974}}].

\bibitem{Bern:2019prr}
Z.~Bern, J.J.~Carrasco, M.~Chiodaroli, H.~Johansson and R.~Roiban, \emph{{The
  duality between color and kinematics and its applications}},
  \href{https://doi.org/10.1088/1751-8121/ad5fd0}{\emph{J. Phys. A} {\bfseries
  57} (2024) 333002} [\href{https://arxiv.org/abs/1909.01358}{{\ttfamily
  1909.01358}}].

\bibitem{Borsten:2020bgv}
L.~Borsten, \emph{{Gravity as the square of gauge theory: a review}},
  \href{https://doi.org/10.1007/s40766-020-00003-6}{\emph{Riv. Nuovo Cim.}
  {\bfseries 43} (2020) 97}.

\bibitem{Adamo:2022dcm}
T.~Adamo, J.J.M.~Carrasco, M.~Carrillo-Gonz\'alez, M.~Chiodaroli, H.~Elvang,
  H.~Johansson et~al., \emph{{Snowmass White Paper: the Double Copy and its
  Applications}},  in \emph{{Snowmass 2021}}, 4, 2022
  [\href{https://arxiv.org/abs/2204.06547}{{\ttfamily 2204.06547}}].

\bibitem{Bern:2022wqg}
Z.~Bern, J.J.~Carrasco, M.~Chiodaroli, H.~Johansson and R.~Roiban, \emph{{The
  SAGEX review on scattering amplitudes Chapter 2: An invitation to
  color-kinematics duality and the double copy}},
  \href{https://doi.org/10.1088/1751-8121/ac93cf}{\emph{J. Phys. A} {\bfseries
  55} (2022) 443003} [\href{https://arxiv.org/abs/2203.13013}{{\ttfamily
  2203.13013}}].

\bibitem{Bern:2023zkg}
Z.~Bern, J.J.M.~Carrasco, M.~Chiodaroli, H.~Johansson and R.~Roiban,
  \emph{{Supergravity Amplitudes, the Double Copy, and Ultraviolet Behavior}},
  (2023), \href{https://doi.org/10.1007/978-981-19-3079-9_49-1}{DOI}
  [\href{https://arxiv.org/abs/2304.07392}{{\ttfamily 2304.07392}}].

\bibitem{Carrasco:2022jxn}
J.J.M.~Carrasco and N.H.~Pavao, \emph{{Virtues of a symmetric-structure double
  copy}}, \href{https://doi.org/10.1103/PhysRevD.107.065005}{\emph{Phys. Rev.
  D} {\bfseries 107} (2023) 065005}
  [\href{https://arxiv.org/abs/2211.04431}{{\ttfamily 2211.04431}}].

\bibitem{rafi2013}
K.~Rafi and J.~Tao, \emph{The diameter of the thick part of moduli space and
  simultaneous whitehead moves},
  \href{https://doi.org/10.1215/00127094-2323128}{\emph{Duke Math. J.}
  {\bfseries 162} (2013) 1833}.

\bibitem{Carrasco:2023vjg}
J.J.M.~Carrasco and A.~Seifi, \emph{{Loop-level double-copy for massive
  fermions in the fundamental}},
  \href{https://doi.org/10.1007/JHEP05(2023)217}{\emph{JHEP} {\bfseries 05}
  (2023) 217} [\href{https://arxiv.org/abs/2302.14861}{{\ttfamily
  2302.14861}}].

\bibitem{Bern:2007ct}
Z.~Bern, J.J.M.~Carrasco, H.~Johansson and D.A.~Kosower, \emph{{Maximally
  supersymmetric planar Yang-Mills amplitudes at five loops}},
  \href{https://doi.org/10.1103/PhysRevD.76.125020}{\emph{Phys. Rev. D}
  {\bfseries 76} (2007) 125020}
  [\href{https://arxiv.org/abs/0705.1864}{{\ttfamily 0705.1864}}].

\bibitem{Bern:1994zx}
Z.~Bern, L.J.~Dixon, D.C.~Dunbar and D.A.~Kosower, \emph{{One loop n point
  gauge theory amplitudes, unitarity and collinear limits}},
  \href{https://doi.org/10.1016/0550-3213(94)90179-1}{\emph{Nucl. Phys. B}
  {\bfseries 425} (1994) 217}
  [\href{https://arxiv.org/abs/hep-ph/9403226}{{\ttfamily hep-ph/9403226}}].

\bibitem{Bern:1994cg}
Z.~Bern, L.J.~Dixon, D.C.~Dunbar and D.A.~Kosower, \emph{{Fusing gauge theory
  tree amplitudes into loop amplitudes}},
  \href{https://doi.org/10.1016/0550-3213(94)00488-Z}{\emph{Nucl. Phys. B}
  {\bfseries 435} (1995) 59}
  [\href{https://arxiv.org/abs/hep-ph/9409265}{{\ttfamily hep-ph/9409265}}].

\bibitem{Bern:1995db}
Z.~Bern and A.G.~Morgan, \emph{{Massive loop amplitudes from unitarity}},
  \href{https://doi.org/10.1016/0550-3213(96)00078-8}{\emph{Nucl. Phys. B}
  {\bfseries 467} (1996) 479}
  [\href{https://arxiv.org/abs/hep-ph/9511336}{{\ttfamily hep-ph/9511336}}].

\bibitem{Bern:1997sc}
Z.~Bern, L.J.~Dixon and D.A.~Kosower, \emph{{One loop amplitudes for e+ e- to
  four partons}},
  \href{https://doi.org/10.1016/S0550-3213(97)00703-7}{\emph{Nucl. Phys. B}
  {\bfseries 513} (1998) 3}
  [\href{https://arxiv.org/abs/hep-ph/9708239}{{\ttfamily hep-ph/9708239}}].

\bibitem{Britto:2004nc}
R.~Britto, F.~Cachazo and B.~Feng, \emph{{Generalized unitarity and one-loop
  amplitudes in N=4 super-Yang-Mills}},
  \href{https://doi.org/10.1016/j.nuclphysb.2005.07.014}{\emph{Nucl. Phys. B}
  {\bfseries 725} (2005) 275}
  [\href{https://arxiv.org/abs/hep-th/0412103}{{\ttfamily hep-th/0412103}}].

\bibitem{Bern:2008pv}
Z.~Bern, J.J.M.~Carrasco, L.J.~Dixon, H.~Johansson and R.~Roiban,
  \emph{{Manifest Ultraviolet Behavior for the Three-Loop Four-Point Amplitude
  of N=8 Supergravity}},
  \href{https://doi.org/10.1103/PhysRevD.78.105019}{\emph{Phys. Rev. D}
  {\bfseries 78} (2008) 105019}
  [\href{https://arxiv.org/abs/0808.4112}{{\ttfamily 0808.4112}}].

\bibitem{Bern:2010tq}
Z.~Bern, J.J.M.~Carrasco, L.J.~Dixon, H.~Johansson and R.~Roiban, \emph{{The
  Complete Four-Loop Four-Point Amplitude in N=4 Super-Yang-Mills Theory}},
  \href{https://doi.org/10.1103/PhysRevD.82.125040}{\emph{Phys. Rev. D}
  {\bfseries 82} (2010) 125040}
  [\href{https://arxiv.org/abs/1008.3327}{{\ttfamily 1008.3327}}].

\bibitem{Carrasco:2011hw}
J.J.M.~Carrasco and H.~Johansson, \emph{{Generic multiloop methods and
  application to N=4 super-Yang-Mills}},
  \href{https://doi.org/10.1088/1751-8113/44/45/454004}{\emph{J. Phys. A}
  {\bfseries 44} (2011) 454004}
  [\href{https://arxiv.org/abs/1103.3298}{{\ttfamily 1103.3298}}].

\bibitem{Broedel:2011pd}
J.~Broedel and J.J.M.~Carrasco, \emph{{Virtuous Trees at Five and Six Points
  for Yang-Mills and Gravity}},
  \href{https://doi.org/10.1103/PhysRevD.84.085009}{\emph{Phys. Rev. D}
  {\bfseries 84} (2011) 085009}
  [\href{https://arxiv.org/abs/1107.4802}{{\ttfamily 1107.4802}}].

\bibitem{deRoo:2003xv}
M.~de~Roo and M.G.C.~Eenink, \emph{{The Effective action for the four point
  functions in Abelian open superstring theory}},
  \href{https://doi.org/10.1088/1126-6708/2003/08/036}{\emph{JHEP} {\bfseries
  08} (2003) 036} [\href{https://arxiv.org/abs/hep-th/0307211}{{\ttfamily
  hep-th/0307211}}].

\bibitem{Mafra:2011nv}
C.R.~Mafra, O.~Schlotterer and S.~Stieberger, \emph{{Complete N-Point
  Superstring Disk Amplitude I. Pure Spinor Computation}},
  \href{https://doi.org/10.1016/j.nuclphysb.2013.04.023}{\emph{Nucl. Phys. B}
  {\bfseries 873} (2013) 419}
  [\href{https://arxiv.org/abs/1106.2645}{{\ttfamily 1106.2645}}].

\bibitem{Mafra:2011nw}
C.R.~Mafra, O.~Schlotterer and S.~Stieberger, \emph{{Complete N-Point
  Superstring Disk Amplitude II. Amplitude and Hypergeometric Function
  Structure}},
  \href{https://doi.org/10.1016/j.nuclphysb.2013.04.022}{\emph{Nucl. Phys. B}
  {\bfseries 873} (2013) 461}
  [\href{https://arxiv.org/abs/1106.2646}{{\ttfamily 1106.2646}}].

\bibitem{Broedel:2013tta}
J.~Broedel, O.~Schlotterer and S.~Stieberger, \emph{{Polylogarithms, Multiple
  Zeta Values and Superstring Amplitudes}},
  \href{https://doi.org/10.1002/prop.201300019}{\emph{Fortsch. Phys.}
  {\bfseries 61} (2013) 812} [\href{https://arxiv.org/abs/1304.7267}{{\ttfamily
  1304.7267}}].

\bibitem{Carrasco:2016ldy}
J.J.M.~Carrasco, C.R.~Mafra and O.~Schlotterer, \emph{{Abelian Z-theory: NLSM
  amplitudes and $\alpha$'-corrections from the open string}},
  \href{https://doi.org/10.1007/JHEP06(2017)093}{\emph{JHEP} {\bfseries 06}
  (2017) 093} [\href{https://arxiv.org/abs/1608.02569}{{\ttfamily
  1608.02569}}].

\bibitem{Mafra:2016mcc}
C.R.~Mafra and O.~Schlotterer, \emph{{Non-abelian $Z$-theory: Berends-Giele
  recursion for the $\alpha'$-expansion of disk integrals}},
  \href{https://doi.org/10.1007/JHEP01(2017)031}{\emph{JHEP} {\bfseries 01}
  (2017) 031} [\href{https://arxiv.org/abs/1609.07078}{{\ttfamily
  1609.07078}}].

\bibitem{Carrasco:2016ygv}
J.J.M.~Carrasco, C.R.~Mafra and O.~Schlotterer, \emph{{Semi-abelian Z-theory:
  NLSM$+\phi^{3}$ from the open string}},
  \href{https://doi.org/10.1007/JHEP08(2017)135}{\emph{JHEP} {\bfseries 08}
  (2017) 135} [\href{https://arxiv.org/abs/1612.06446}{{\ttfamily
  1612.06446}}].

\bibitem{Fu:2014pya}
C.-H.~Fu, Y.-J.~Du and B.~Feng, \emph{{Note on symmetric BCJ numerator}},
  \href{https://doi.org/10.1007/JHEP08(2014)098}{\emph{JHEP} {\bfseries 08}
  (2014) 098} [\href{https://arxiv.org/abs/1403.6262}{{\ttfamily 1403.6262}}].

\bibitem{Naculich:2014rta}
S.G.~Naculich, \emph{{Scattering equations and virtuous kinematic numerators
  and dual-trace functions}},
  \href{https://doi.org/10.1007/JHEP07(2014)143}{\emph{JHEP} {\bfseries 07}
  (2014) 143} [\href{https://arxiv.org/abs/1404.7141}{{\ttfamily 1404.7141}}].

\bibitem{Luna:2016hge}
A.~Luna, R.~Monteiro, I.~Nicholson, A.~Ochirov, D.~O'Connell, N.~Westerberg
  et~al., \emph{{Perturbative spacetimes from Yang-Mills theory}},
  \href{https://doi.org/10.1007/JHEP04(2017)069}{\emph{JHEP} {\bfseries 04}
  (2017) 069} [\href{https://arxiv.org/abs/1611.07508}{{\ttfamily
  1611.07508}}].

\bibitem{Carrasco:2021bmu}
J.J.M.~Carrasco and I.A.~Vazquez-Holm, \emph{{Extracting Einstein from the
  loop-level double-copy}},
  \href{https://doi.org/10.1007/JHEP11(2021)088}{\emph{JHEP} {\bfseries 11}
  (2021) 088} [\href{https://arxiv.org/abs/2108.06798}{{\ttfamily
  2108.06798}}].

\bibitem{Cheung:2020uts}
C.~Cheung and G.N.~Remmen, \emph{{Entanglement and the double copy}},
  \href{https://doi.org/10.1007/JHEP05(2020)100}{\emph{JHEP} {\bfseries 05}
  (2020) 100} [\href{https://arxiv.org/abs/2002.10470}{{\ttfamily
  2002.10470}}].

\bibitem{Borsten:2023ned}
L.~Borsten, B.~Jurco, H.~Kim, T.~Macrelli, C.~Saemann and M.~Wolf,
  \emph{{Double Copy From Tensor Products of Metric
  BV\ensuremath{\blacksquare}-Algebras}},
  \href{https://doi.org/10.1002/prop.202300270}{\emph{Fortsch. Phys.}
  {\bfseries 73} (2025) 2300270}
  [\href{https://arxiv.org/abs/2307.02563}{{\ttfamily 2307.02563}}].

\bibitem{Brandhuber:2021bsf}
A.~Brandhuber, G.~Chen, H.~Johansson, G.~Travaglini and C.~Wen,
  \emph{{Kinematic Hopf Algebra for Bern-Carrasco-Johansson Numerators in
  Heavy-Mass Effective Field Theory and Yang-Mills Theory}},
  \href{https://doi.org/10.1103/PhysRevLett.128.121601}{\emph{Phys. Rev. Lett.}
  {\bfseries 128} (2022) 121601}
  [\href{https://arxiv.org/abs/2111.15649}{{\ttfamily 2111.15649}}].

\bibitem{Fu:2025jpp}
C.-H.~Fu, P.~Vanhove and Y.~Wang, \emph{{HEFT numerators from kinematic
  algebra}}, \href{https://doi.org/10.1007/JHEP06(2025)248}{\emph{JHEP}
  {\bfseries 06} (2025) 248}
  [\href{https://arxiv.org/abs/2501.14523}{{\ttfamily 2501.14523}}].

\bibitem{Azevedo:2018dgo}
T.~Azevedo, M.~Chiodaroli, H.~Johansson and O.~Schlotterer, \emph{{Heterotic
  and bosonic string amplitudes via field theory}},
  \href{https://doi.org/10.1007/JHEP10(2018)012}{\emph{JHEP} {\bfseries 10}
  (2018) 012} [\href{https://arxiv.org/abs/1803.05452}{{\ttfamily
  1803.05452}}].

\bibitem{Carrasco:2022lbm}
J.J.M.~Carrasco, M.~Lewandowski and N.H.~Pavao, \emph{{Color-Dual Fates of F3,
  R3, and N=4 Supergravity}},
  \href{https://doi.org/10.1103/PhysRevLett.131.051601}{\emph{Phys. Rev. Lett.}
  {\bfseries 131} (2023) 051601}
  [\href{https://arxiv.org/abs/2203.03592}{{\ttfamily 2203.03592}}].

\bibitem{Bern:2024vqs}
Z.~Bern, E.~Herrmann, R.~Roiban, M.S.~Ruf and M.~Zeng, \emph{{Global bases for
  nonplanar loop integrands, generalized unitarity, and the double copy to all
  loop orders}}, \href{https://doi.org/10.1007/JHEP06(2025)115}{\emph{JHEP}
  {\bfseries 06} (2025) 115}
  [\href{https://arxiv.org/abs/2408.06686}{{\ttfamily 2408.06686}}].

\end{thebibliography}\endgroup
\end{document}